\documentclass[traditabstract]{aa}

\usepackage{natbib}
\usepackage{graphicx}
\usepackage{txfonts}

\begin{document}

\title{The XXL Survey\thanks{Based on observations obtained with \emph{XMM-Newton}, an ESA science mission with instruments and contributions directly funded by ESA Member States and NASA.} \\ %
XIII. Baryon content of the bright cluster sample}

\author{D. Eckert\inst{1,2} \and S. Ettori\inst{3,4} \and J. Coupon\inst{1} \and F. Gastaldello\inst{2} \and M. Pierre\inst{5} \and J.-B. Melin\inst{6} \and A. M. C. Le Brun\inst{5,7} \and I. G. McCarthy\inst{7} \and C. Adami\inst{8} \and L. Chiappetti\inst{2} \and L. Faccioli\inst{5} \and P. Giles\inst{9} \and S. Lavoie\inst{10}  \and J. P. Lef\`evre\inst{11} \and M. Lieu\inst{12} \and A. Mantz \inst{13} \and B. Maughan\inst{9} \and S. McGee\inst{12} \and F. Pacaud\inst{14} \and S. Paltani\inst{1} \and T. Sadibekova\inst{5} \and G. P. Smith\inst{12} \and F. Ziparo\inst{12}  }
\institute{
Department of Astronomy, University of Geneva, ch. d'Ecogia 16, 1290 Versoix, Switzerland\\
\email{Dominique.Eckert@unige.ch}
\and
INAF - IASF-Milano, Via E. Bassini 15, 20133 Milano, Italy
\and
INAF - Osservatorio Astronomico di Bologna, Via Ranzani 1, 40127 Bologna, Italy
\and
INFN, Sezione di Bologna, viale Berti Pichat 6/2, 40127 Bologna, Italy
\and 
Laboratoire AIM, IRFU/Service d'Astrophysique -- CEA/DSM -- CNRS -- Universit\'e Paris Diderot, B\^at. 709, CEA-Saclay, 91191 Gif-sur-Yvette Cedex, France
\and
DSM/Irfu/SPP, CEA-Saclay, 91191 Gif-sur-Yvette Cedex, France
\and
Astrophysics Research Institute, Liverpool John Moores University, 146 Brownlow Hill, Liverpool L3 5RF, UK
\and
Universit\' e Aix Marseille, CNRS, LAM (Laboratoire d'Astrophysique de Marseille), UMR 7326, 13388, Marseille, France
\and
Department of Physics, H.H. Wills Physics Laboratory, University of Bristol, Tyndall Avenue, Bristol, BS8 1TL, UK
\and
Department of Physics and Astronomy, University of Victoria, 3800 Finnerty Road, Victoria, BC, V8P 1A1, Canada
\and
SEDI CEA Saclay, France
\and
School of Physics and Astronomy, University of Birmingham, Edgbaston, Birmingham, B15 2TT, UK
\and
Department of Astronomy and Astrophysics, University of Chicago, 5640 South Ellis Avenue, Chicago, IL 60637, USA
\and 
Argelander-Institut f\"ur Astronomie, University of Bonn, Auf dem H\"ugel 71, 53121 Bonn, Germany
}

\abstract{Traditionally, galaxy clusters have been  expected to retain all the material accreted since their formation epoch. For this reason, their matter content should be representative of the Universe as a whole, and thus their baryon fraction should be close to the Universal baryon fraction $\Omega_b/\Omega_m$. We make use of the sample of the 100 brightest galaxy clusters discovered in the XXL Survey to investigate the fraction of baryons in the form of hot gas and stars in the cluster population. Since it spans a wide range of mass ($10^{13}-10^{15}M_\odot$) and redshift ($0.05-1.1$) and benefits from a large set of multiwavelength data, the XXL-100-GC sample is ideal for measuring the global baryon budget of massive halos. We measure the gas masses of the detected halos and use a mass--temperature relation directly calibrated using weak-lensing measurements for a subset of XXL clusters to estimate the halo mass. We find that the weak-lensing calibrated gas fraction of XXL-100-GC clusters is substantially lower than  was found in previous studies using hydrostatic masses. Our best-fit relation between gas fraction and mass reads $f_{\rm gas,500}=0.055_{-0.006}^{+0.007}\left(M_{\rm 500}/10^{14}M_\odot\right)^{0.21_{-0.10}^{+0.11}}$. The baryon budget of galaxy clusters therefore falls short of the Universal baryon fraction by about a factor of two at $r_{\rm 500,MT}$. Our measurements require a hydrostatic bias $1-b=M_X/M_{\rm WL}=0.72_{-0.07}^{+0.08}$ to match the gas fraction obtained using lensing and hydrostatic equilibrium, which holds independently of the instrument considered. Comparing our gas fraction measurements with the expectations from numerical simulations, we find that our results favour an extreme feedback scheme in which a significant fraction of the baryons are expelled from the cores of halos. This model is, however, in contrast with the thermodynamical properties of observed halos, which might suggest that weak-lensing masses are overestimated. In light of these results, we note that a mass bias $1-b=0.58$ as required to reconcile \emph{Planck} CMB and cluster counts should translate into an even lower baryon fraction, which poses a major challenge to our current understanding of galaxy clusters.}

\maketitle

\section{Introduction}

Recent observations of the cosmic microwave background with \emph{Planck} were able to measure the relative amount of baryons and dark matter in the Universe with very high precision, indicating that baryons account for $(15.6\pm0.3)\%$ of the  total matter content of the Universe \citep{planck15_13}. Because of their deep gravitational wells, galaxy clusters are traditionally expected to have retained most of the material accreted since the formation epoch \citep{white93,eke98}. For this reason, the relative amount of baryons and dark matter in galaxy clusters should be close to the Universal value, provided that the measurement has been performed over a sufficiently large volume inside which the effects of baryonic physics can be neglected \citep{evrard97,ettori03}. Recently, \citet{sembolini15} presented a comparison between 12 different non-radiative hydrodynamical codes, showing that in all cases within an overdensity of 500 compared to the critical density the overall baryon budget of galaxy clusters is close to the Universal value. 

In recent times, numerical simulations including baryonic physics \citep[cooling, star formation, and feedback from supernovae and active galactic nuclei (AGNs), e.g.][]{planelles13,battaglia13,lebrun14} have shown that energy injection around the cluster formation epoch may be able to expel some of the gas from the cores of halos \citep{mccarthy11}, leading to a depletion of baryons in the central regions. In particular, \citet{planelles13} has found that the \emph{depletion factor} $Y_b=f_{\rm bar}/(\Omega_b/\Omega_m)$, where $f_{\rm bar}$ denotes the cluster baryon fraction, reaches the value $Y_b=0.85$ at $r_{500}$. \citet{lebrun14} has shown that different AGN feedback implementations have an impact on the depletion factor, i.e. models with strong feedback tend to produce a lower baryon fraction, with a depletion factor ranging from 0.7 to 1.0 for $M_{500}=10^{15}M_\odot$ depending on the adopted setup. Therefore, robust observational constraints on the baryon fraction and its mass dependence are crucial to calibrating the implementation of baryonic physics in cosmological simulations.

In the most massive halos, the majority of the baryons resides in the hot, ionised intracluster medium (ICM), which accounts for $80-90\%$ of the baryons at this mass scale \citep[e.g.][]{arnaudlxt,ettori03b,vikhlinin06,pratt09}. The observed gas fraction is generally found to increase with radius because of non-gravitational energy input \citep{allen04,vikhlinin06,eckert13b} and it converges toward the value of $\sim13\%$ at $r_{500}$. The stellar content of cluster galaxies and intracluster light generally represents 10-20\% of the total baryonic mass \citep{lin04,gonzalez07,andreon10,andreon15,mulroy14}. The overall baryon content of galaxy clusters is therefore found to be close to the Universal value \citep{lin03,giodini09,lagana13}. Interestingly, these studies observed a general trend of increasing gas fraction \citep{sun09,pratt09,lovisari15} and decreasing stellar fraction \citep[e.g.][]{behroozi10,leauthaud12,coupon15} with increasing halo mass, which indicates a mass dependence of the star formation efficiency, although the strength of this effect is subject to some debate \citep[e.g.][]{gonzalez07,budzynski14}.

It must be noted, however, that most of the studies measuring the hot gas fraction assumed that the ICM is in hydrostatic equilibrium to derive the total cluster mass, which may be biased low in the presence of a significant non-thermal pressure contribution \citep[e.g.][]{rasia04,nagai07}. Recent studies comparing X-ray and weak-lensing mass estimates have found large discrepancies between the results obtained with the two methods, although significant differences are found between the various studies \citep[see][and references therein]{sereno15}. In light of these results, a careful assessment of the cluster baryon budget in a sample spanning a wide mass range is required.
 
The  XXL Survey \citep{xxl1} is the largest observing programme undertaken by \emph{XMM-Newton}. It covers two distinct sky areas (XXL-North and XXL-South) for a total of 50 square degrees down to a sensitivity of $5\times10^{-15}$ ergs cm$^{-2}$ s$^{-1}$ for point-like sources ([0.5-2] keV band). Thanks to this combination of area and depth, the survey is ideally suited to measuring the overall baryon content of massive halos. Indeed, the detected clusters cover a wide range of nearly two decades in cluster mass \citep[hereafter Paper III]{xxl3}. In addition, a wealth of high-quality optical and near-infrared data are available in the survey area (e.g. CFHTLS, WIRCam), which provides a robust measurement of the total and stellar mass of the detected clusters. 

In this paper, we exploit the combination of multiwavelength data for the brightest XXL clusters to perform a comprehensive census of the baryon fraction of dark-matter halos in the range $10^{13}-10^{15}M_\odot$. The paper is organised as follows. In Sect. \ref{data} we present the available \emph{XMM-Newton} data and the method developed to estimate the gas mass. In Sect. \ref{sec:hod} we describe the method used to estimate the stellar fraction. Our main results are presented in Sect. \ref{sec:results} and discussed in Sect. \ref{sec:disc}.

Throughout the paper we assume a WMAP9 cosmology \citep{wmap9} with $\Omega_m=0.28$, $\Omega_\Lambda=0.72$, and $H_o=70$ km/s/Mpc. In this cosmology the cosmic baryon fraction is $\Omega_b/\Omega_m=0.166\pm0.006$. The uncertainties are given at the $1\sigma$ level. The quantities indicated with the subscript 500 refer to quantities integrated within an overdensity of 500 with respect to the critical density.

\section{Data analysis}
\label{data}
\subsection{Cluster sample}

The XXL-100-GC cluster sample \citep[hereafter Paper II]{xxl2} is the sample of the 100 brightest clusters detected in the XXL Survey\footnote{XXL-100-GC data are available in computer readable form via the XXL Master Catalogue browser \url{http://cosmosdb.iasf-milano.inaf.it/XXL} and via the XMM XXL DataBase \url{http://xmm-lss.in2p3.fr}}. It is particularly well suited for the study conducted here, as it benefits from a well-defined selection function (see Paper II) and spans a broad range of mass ($2\times10^{13}-8\times10^{14}M_\odot$) and redshift ($0.05-1.1$). The definition and properties of the sample are described in detail in Paper II. Temperatures and luminosities within a fixed radius of 300 kpc were measured from the survey data and presented in Paper III. Halo masses were measured directly through weak lensing for 38 $z<0.6$ clusters coinciding with the CFHTLS survey \citep[hereafter Paper IV; see Sect. \ref{sec:mwltx}]{xxl4}, which allowed us to directly calibrate the relation between weak-lensing mass and X-ray temperature  using a subset of the XXL-100-GC. The mass of each cluster was estimated using the $M-T$ relation (see Paper IV), which allowed us to calculate an estimate for $r_{500}$ (hereafter $r_{\rm 500,MT}$). To validate this approach, in Appendix \ref{app:r500} we compare the values of $r_{\rm 500,MT}$ with those obtained from the weak-lensing measurements .

\subsection{\emph{XMM-Newton} data and processing}

We processed the XXL data using the XMMSAS package and calibration files v10.0.2 and the data reduction pipeline {\sc Xamin} \citep{pacaud06} to obtain cleaned event files for each observation (see Paper II for details on the data reduction scheme). We extracted photon images in the [0.5-2.0] keV band for each EPIC instrument and created co-added EPIC images by summing the images obtained for each detector. We then created exposure maps using the XMMSAS task \texttt{eexpmap} for each EPIC detector and summed them, each weighted by its respective effective area. 

The non-X-ray background (NXB) was estimated for each observation by measuring the count rate in several energy bands in the unexposed corners of the three EPIC instruments. A template image of the NXB was created using a collection of closed-filter observations and scaled to match the count rates recorded in the corners. For the details of this procedure, see \citet{lm08}. We neglected the contribution of residual soft protons and fitted this contribution together with the sky background components since the soft protons are funneled through the \emph{XMM-Newton} telescopes.

\subsection{Emission-measure profiles}

Surface brightness profiles were extracted for each cluster using the {\sc Proffit} package \citep{ccbias1}. Specifically, we defined annular regions of width 8 arcsec and accumulated the sky and NXB count rates in each annulus, taking vignetting and CCD gaps into account. The NXB was then subtracted from the observed profile. The profiles were centred on the best-fit coordinates provided by {\sc Xamin}. 

To estimate the local cosmic X-ray background and Galactic foreground emission, we fitted  the surface-brightness profiles beyond $1.2r_{\rm 500,MT}$\footnote{In principle the cluster emission might extend beyond $1.2r_{\rm 500,MT}$. However, in the survey data no significant emission is detected beyond $r_{\rm 500,MT}$, and we verified that the background measurements do not change significantly when a larger radius is used.} from the X-ray peak with a constant and subtracted the resulting value from the surface-brightness profiles, propagating the uncertainties in the sky background and the NXB to the source profile. As a result, a source-only profile was obtained for each cluster. In the cases for which the estimated value of $r_{\rm 500,MT}$ exceeds half of the \emph{XMM-Newton} field of view ($\sim15$ arcmin radius), the estimation of the local background may be affected by systematic uncertainties. For this reason, we excluded the corresponding clusters (mainly low-redshift systems) from the analysis.

To convert the observed surface brightness profiles into emission measure profiles, we used the temperature measured within 300 kpc from Paper III and simulated a single-temperature absorbed thin-plasma model using the APEC code in {\sc Xspec} \citep{apec}. The Galactic absorption was fixed to the 21cm value as measured by \citet{kalberla}. The mean column density is $2.2\times10^{20}$ cm$^{-2}$ in the XXL-North field and $1.2\times10^{20}$ cm$^{-2}$ in the XXL-South field. The metal abundance was fixed to the value of $0.3Z_\odot$ using the \citet{ag89} solar abundances \citep{lm08b}, and the cluster redshift was set to the spectroscopic value (see Paper II). This allowed us to compute the conversion between count rate in the [0.5-2.0] keV range and APEC norm, which is proportional to the emission measure,

\begin{equation}Norm=\frac{10^{-14}}{4\pi d_A^2(1+z)^2}\int n_en_H\,dV,\end{equation}

\noindent assuming constant temperature and metallicity. We note that this conversion is largely insensitive to the temperature and metallicity as long as the temperature exceeds $\sim1.5$ keV, thus in most cases any uncertainty associated with the temperature measurement propagates very weakly to the emission measure. For the least massive systems ($\sim10\%$ of the sample), the X-ray emissivity depends strongly on line emission and on the position of the bremsstrahlung cut-off. For these systems we thus expect a systematic uncertainty of $\sim30\%$ on the recovered emission measure (see Sect. \ref{sec:syst}).

\subsection{Gas density and gas mass}
\label{sec:method_mgas} 

To estimate the gas density and gas mass profiles from the emission measure, we proceeded in two steps: \emph{i)} deprojecting the source emission-measure profiles assuming spherical symmetry and \emph{ii)} converting the resulting profile into gas density assuming constant density inside each shell. We assumed an electron-to-proton ratio of 1.21 in a fully ionised astrophysical plasma. The gas mass within a radius $R$ could then be obtained by integrating the resulting gas density profiles over the volume,
\begin{equation}M_{\rm gas}(<r) = \mu m_p \int_{0}^r n_{\rm gas}(r^\prime)\,4\pi r^{\prime2}\, dr^\prime, \label{eq:mgas}\end{equation}
\noindent with $n_{\rm gas}=n_e+n_H=2.21n_H$ the particle number density, $\mu=0.61$ the mean molecular weight, and $m_p$ the proton mass. 

\begin{figure}
\resizebox{\hsize}{!}{\includegraphics{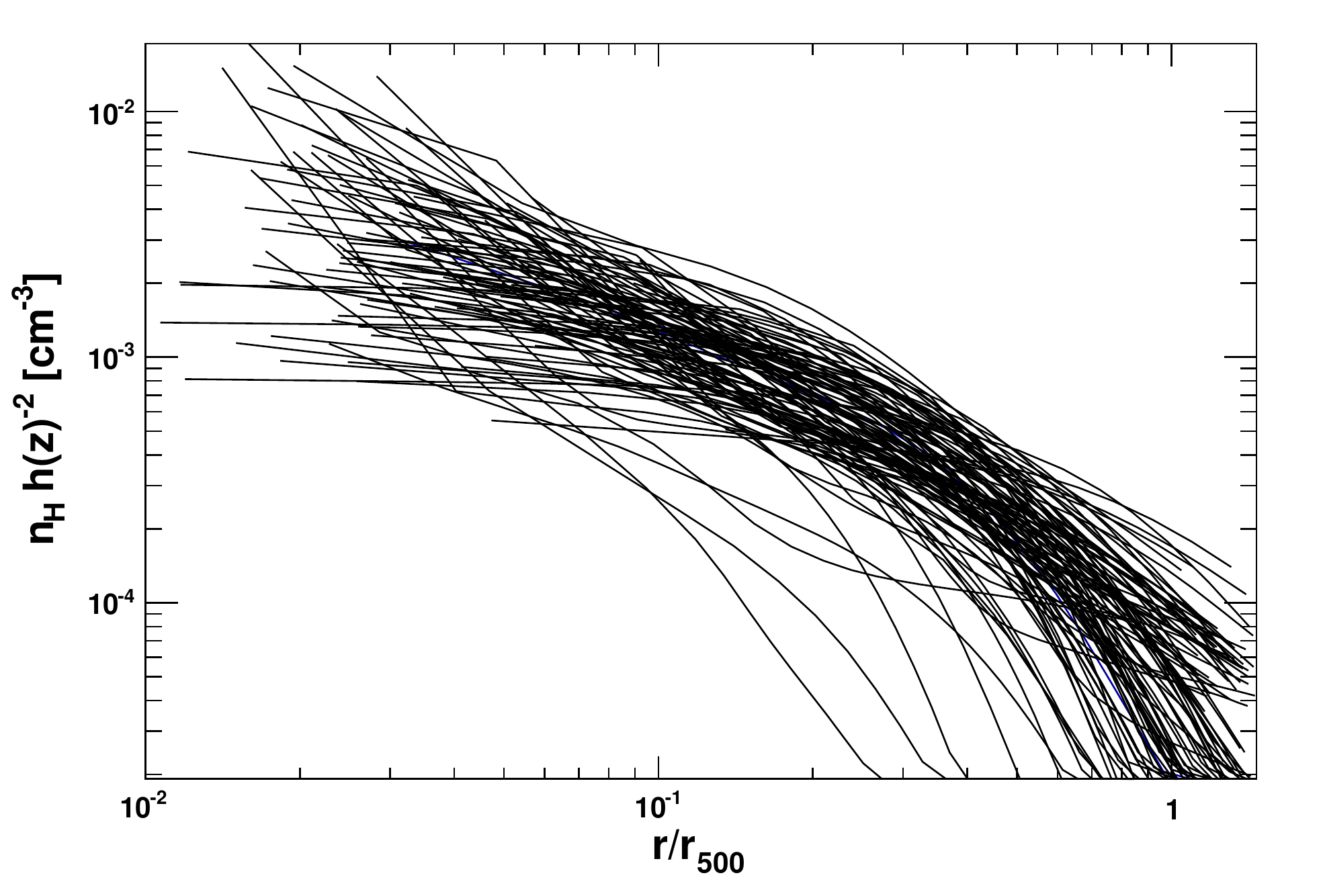}}
\caption{Self-similarly scaled gas density profiles for the XXL-100-GC clusters.}
\label{fig:all_ngas}
\end{figure}

Since the emission region is optically thin, the observed X-ray emissivity is the result of the projection of the 3D emissivity along the line of sight. Therefore, the observed emission-measure profiles must be deprojected to estimate the 3D gas density profile and integrate Eq. \ref{eq:mgas}. To deproject the profiles, we used a multiscale forward-fitting approach in which the signal is decomposed into a sum of King functions, which are then individually deprojected assuming spherical symmetry. The method is described in greater detail in Appendix \ref{app:deproj}. To take the effects of the \emph{XMM-Newton} point spread function (PSF) into account, the model was convolved with a PSF convolution matrix drawn from the latest calibration files (see Appendix \ref{app:psf}). A maximum likelihood algorithm was then applied to fit the emission-measure profiles, and the best-fit parameters were used to compute the 3D density profile. In Fig. \ref{fig:all_ngas} we show the gas density profiles recovered using this technique for the entire sample, scaled by their expected self-similar evolution $n(z)\sim E(z)^2\equiv\Omega_m(1+z)^3+\Omega_\Lambda$ \citep{bryan98}.

To sample the parameter space, we used the Markov chain Monte Carlo (MCMC) code \texttt{emcee} \citep{foreman-mackey13}. We then drew the posterior distributions of gas density and gas mass from the resulting Markov chains. The uncertainties on the values of $r_{\rm 500,MT}$ were propagated to the posterior $M_{\rm gas}$ distributions by  randomly drawing a value of $r_{\rm 500,MT}$ for each MCMC step according to the uncertainties and computing the gas mass within the corresponding radius.

In Table \ref{tab:data} we provide the gas masses recovered using this technique for the XXL-100-GC sample together with the basic properties of the sample. We also provide the measurements of $Y_{X,500}=T_{\rm 300 kpc}\times M_{\rm gas,500}$ and $f_{\rm gas,500}=M_{\rm gas,500}/M_{\rm 500,MT}$, where $M_{\rm 500,MT}$ was computed from the $M_{\rm 500,WL}-T$ relation presented in Paper IV. For the remainder of the paper, we restrict ourselves to the systems for which the estimated value of $r_{\rm 500,MT}$ does not exceed 8 arcmin, which affords a robust estimation of the local background and hence of the gas mass. Our final sample comprises 95 clusters in the redshift range 0.05-1.1.

\subsection{Validation using hydrodynamical simulations}
\label{sec:mgas_test}

To validate the method, we used mock X-ray images created from the OverWhelmingly Large cosmological Simulation \citep[cosmo-OWLS,][]{schaye10,mccarthy10,lebrun14} using the AGN-8.0 model (see Sect. \ref{sec:agn}) in a field comparable to the XXL Survey (Faccioli et al., in prep.). The mock images were then folded through the \emph{XMM-Newton} response and a realistic background was added to obtain simulated data that were as close as possible to real XXL observations \citep{valtchanov01}. 

\begin{figure}
\resizebox{\hsize}{!}{\includegraphics{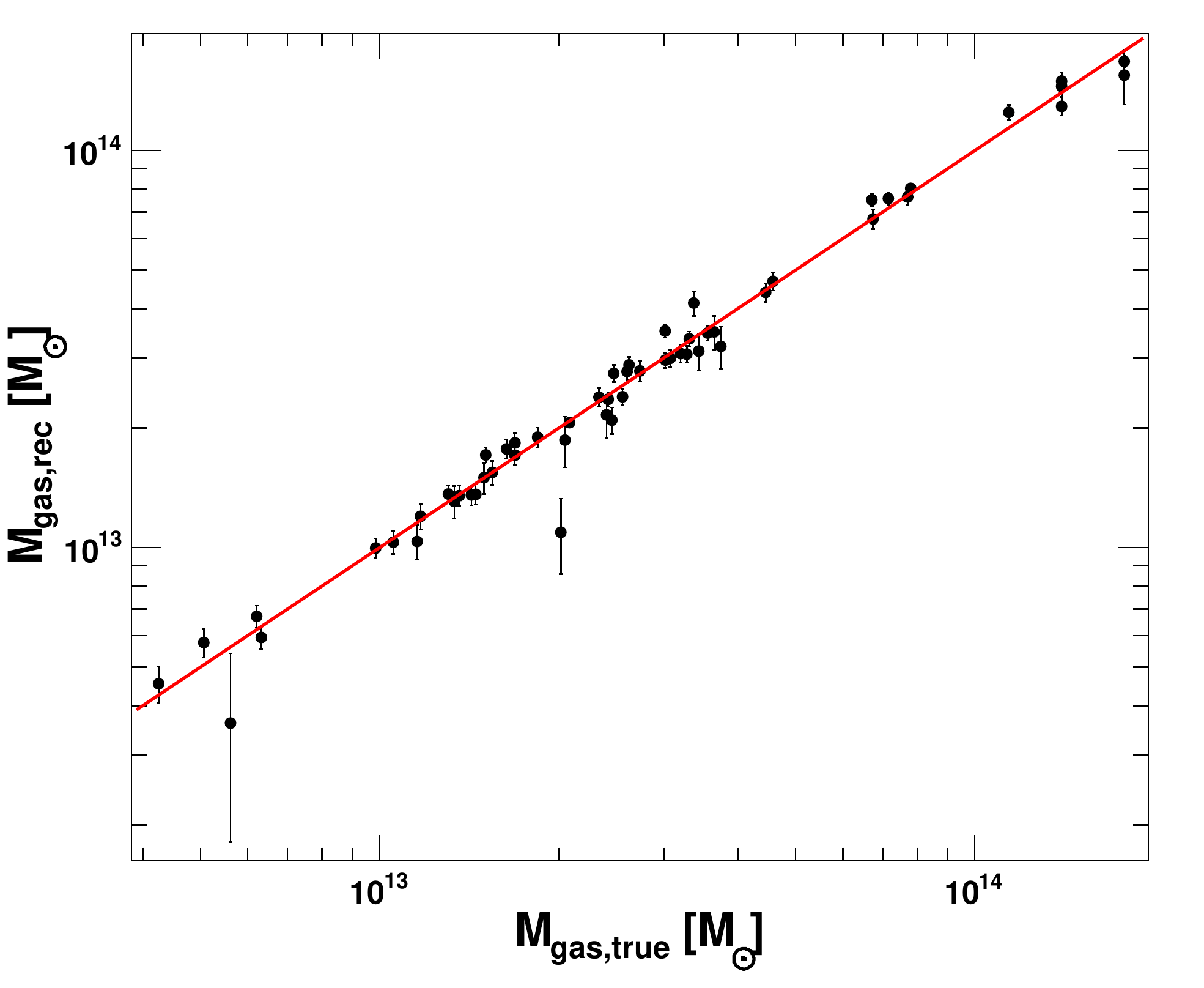}}
\caption{Gas mass within $r_{\rm 500,MT}$ reconstructed from mock \emph{XMM-Newton} observations as a function of the true 3D gas mass in cosmo-OWLS simulations. The red line shows the 1-1 relation.}
\label{fig:mock}
\end{figure}

We selected  a sample of 61 bright clusters from these mock observations and applied the method described above to reconstruct the gas mass. The true value of $r_{\rm 500}$ was used to integrate the gas density profile. In Fig. \ref{fig:mock} we show the reconstructed gas mass within $r_{\rm 500}$ compared to the true 3D gas mass within the same radius. The reconstructed quantity traces remarkably well the true value with no measurable bias ($M_{\rm gas,rec}/M_{\rm gas,true}=0.99\pm0.01$) and a low intrinsic scatter of 7\%. The good agreement between the true and reconstructed gas masses demonstrates the reliability of the method used here and was also found by other similar studies based on hydrodynamical simulations \citep[e.g.][]{nagai07,rasia11}.

\section{Stellar fraction}
\label{sec:hod}

To estimate the total stellar content of the XXL-100-GC clusters, we adopted a non-parametric approach to measure the mean number of galaxies as a function of halo mass. Unlike parametric halo occupation distribution (HOD) techniques, here no assumption on the statistical distribution of host halos is required as the hot gas is a direct tracer of the dark matter halo positions. We probed the distribution of satellites by means of the stacked galaxy overdensity as a function of distance from the central galaxy, and compared this value to the field galaxy density.

The stellar fractions are measured over 15~deg$^2$ of the XXL-North field (corresponding to 34 systems in the XXL-100-GC sample) where galaxy photometric redshifts and stellar masses are taken from the sample of \citet[and references therein]{coupon15}. In brief, this sample is composed of deep optical photometry from the Canada-France-Hawaii Telescope Legacy Survey (CFHTLS, $u,g,r,i,z_{\rm AB} \sim 25$) complemented by medium-deep $K$-band photometry ($K_{\rm AB} \sim 22$) from the WIRCam camera at CFHT. Calibrated with $60\,000$ spectroscopic redshifts from the VIPERS \citep{guzzo14} and VVDS surveys \citep{lefevre05}, these photometric redshifts reach a precision of $0.03\times(1+z)$ with less than $2\%$ catastrophic failures up to $z\sim1$. Stellar masses are computed with the initial mass function (IMF) of \citet{chabrier03} truncated at 0.1 and 100 $M_\odot$, and the stellar population synthesis templates from \citet{bruzual03}. For the details of the procedure we refer the reader to \citet{arnouts13}.

To measure the projected overdensity of satellite galaxies in halos, we use the two-point cross-correlation function between the bright central galaxy (BCG) positions of the XXL-100-GC clusters from the sample of \citet{xxl15}, and galaxies from the full photometric sample. To account for masking and edge effects, we used the \citet{landy93} estimator to measure the projected two-point correlation function $w(R)$; the mean number of galaxies within a radius $R$ away from the centre writes
 \begin{equation}
   N_{\rm gal} (r) = 2 \pi \, \eta \, \int_{0}^{r} (1+w(r')) \, r' \, d r' \, ,
\end{equation}
where $R$ is the physical transverse distance from the central galaxy (at the cluster redshift) and $\eta$ is the mean density of galaxies in the field. Galaxies in the photometric sample are selected using their photometric redshift estimates within the redshift range spanned by the cluster sample, extended at low and high redshift ($\pm0.1$) to account for photometric redshift uncertainties. 
Here we cross-correlate the full cluster sample with the full photometric sample. To increase the signal, galaxies can   also be selected  in a thin redshift slice around each cluster redshift; however, as the background estimation becomes significantly noisier in consequence, no gain in signal-to-noise is observed.

As the central galaxy does not correlate with the field galaxies outside of the halo, the number of satellites is simply expressed by the {over}density within $r_{\rm 500,MT}$ compared to the field,
\begin{equation}
   N_{\rm sat} = 2 \pi \, \eta \, \int_{0}^{r_{\rm 500,MT}} w(r') \, r' \, d r' \, ,
\end{equation}
where $r_{\rm 500,MT}$ is derived from the $M-T$ relation from Paper IV. This method provides a measurement of the projected galaxy overdensity,  which might lead to a slight overestimation of the enclosed stellar mass because of projection effects, since the cluster actually extends significantly beyond $r_{500,MT}$. Assuming that the distribution of stellar mass follows that of the dark matter, we used an NFW profile \citep{nfw97} to estimate the fraction of the projected stellar mass located beyond $r_{500,MT}$. We found that this fraction is of the order of 15\% for typical NFW parameters.

The process is repeated for each galaxy sample selected by stellar mass in eight stellar mass bins in the range $10^{9}-10^{12}M_\odot$, and the total stellar mass is obtained by integrating the number of satellites over their stellar masses, and adding the stellar mass of the central galaxy. The errors are estimated from a jackknife resampling of 64 sub-volumes. This accounts for Poisson error, cosmic variance and intrinsic halo-to-satellite number dispersion.

We note that this method differs from those relying on individual satellite identification using redshift probability distributions \citep[e.g.][]{george11}. Given our photometric redshift statistical uncertainties, our cross-correlation estimator provides a more robust (although noisier) satellite number estimate.

\section{Results}
\label{sec:results}

\subsection{Weak-lensing halo masses}
\label{sec:mwltx}

In a companion paper (Paper IV), we used weak-lensing shear measurements for a subsample of 38 XXL-100-GC clusters to derive the relation between X-ray temperature and halo mass in XXL clusters. The relation was complemented with weak-lensing mass measurements from the COSMOS \citep{kettula13} and CCCP \citep{hoekstra15} samples to increase the sample size to 95 systems and the dynamical range to 1-15 keV. The best-fit relation reads
\begin{equation}\log\left(\frac{E(z)M_{\rm WL}}{h_{70}^{-1}M_\odot}\right)=a+b\log\left(\frac{T_{\rm 300 kpc}}{\rm 1\, keV}\right),\label{eq:mwltx}\end{equation}
\noindent with $a=13.57_{-0.09}^{+0.09}$ and  $b=1.67_{-0.10}^{+0.14}$, i.e. slightly steeper than the self-similar expectation. For the details on the mass measurement method and the derivation of the $M-T$ relation, we refer to Paper IV.

The spectroscopic temperatures were measured for the entire XXL-100-GC within a fixed radius of 300 kpc and the halo mass for each cluster was estimated using Eq. \ref{eq:mwltx}, taking the scatter in the relation into account (see Paper III for details).   Using the 38 objects with measured weak-lensing signal, we verified that the masses and radii measured using the $M-T$ relation match the values expected from the lensing measurements (see Appendix \ref{app:r500}).

\subsection{$M_{\rm gas}-T$ relation and gas mass fraction}
\label{sec:fgas_mtot}

\begin{figure}
\resizebox{\hsize}{!}{\includegraphics{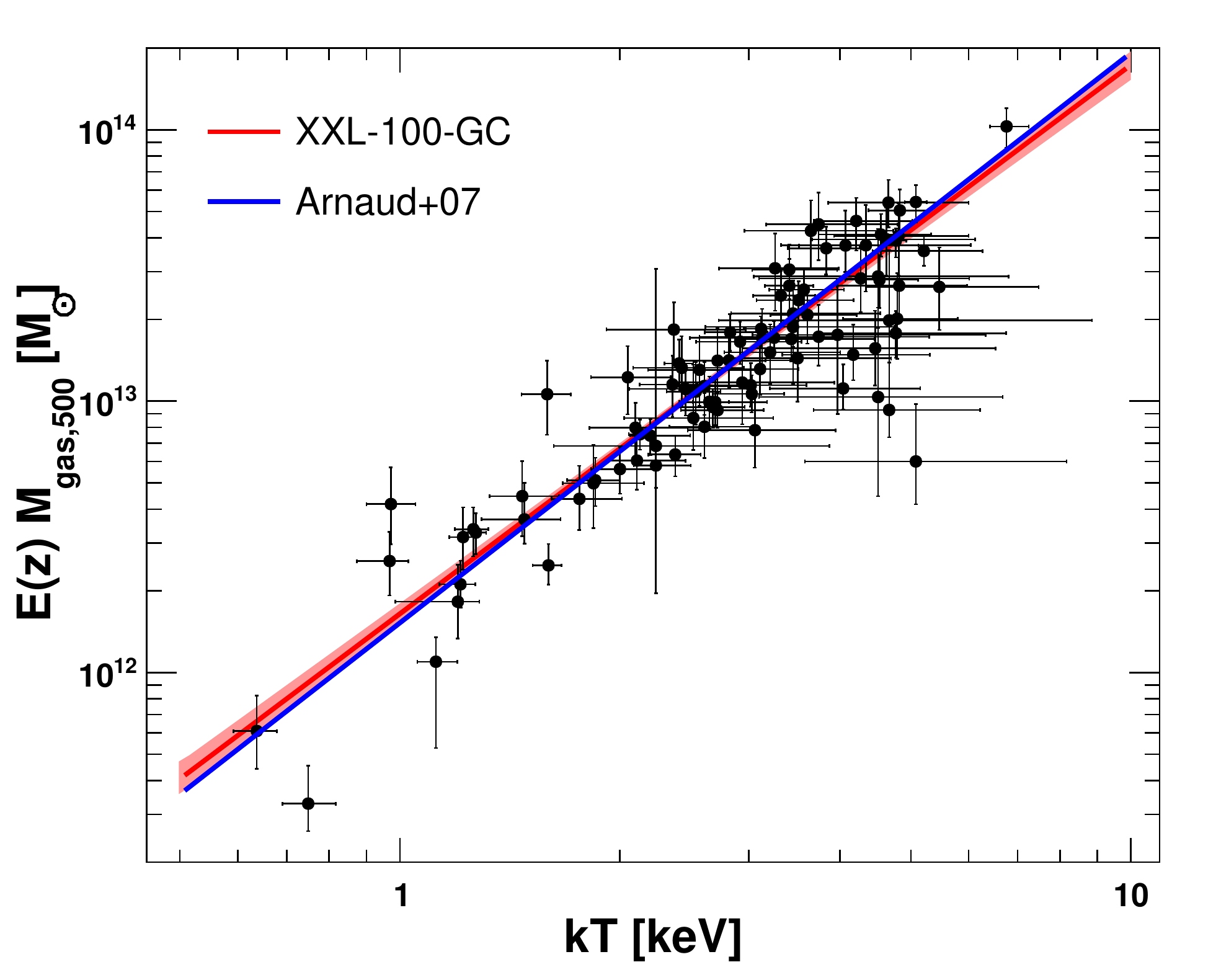}}
\caption{Gas mass within $r_{\rm 500,MT}$ for the XXL-100-GC sample as a function of their temperature within a fixed aperture of 300 kpc. The red line and the red shaded area show the best-fit relation and its uncertainty. The blue curve represents the relation of \citet{arnaud07}. }
\label{fig:mgas_tx}
\end{figure}

To estimate the average gas fraction and the relation between $f_{\rm gas,500}$ and $M_{\rm WL}$, we fitted the $M_{\rm gas}-T$ relation with the relation
\begin{equation}\log\left(\frac{E(z)M_{\rm gas,500}}{h_{70}^{-5/2}M_\odot}\right)=\log N+\alpha\log\left(\frac{T_{\rm 300 kpc}}{\rm 1\, keV}\right).\label{eq:mgastx}\end{equation}
As we did for the $M_{\rm WL}-T$ relation (see Paper IV), we fitted the data using the Bayesian regression code of \citet{kelly07} and the Gibbs MCMC sampler. In Fig. \ref{fig:mgas_tx} we show the relation between $M_{\rm gas,500}$ and $T_{\rm 300 kpc}$ for the XXL-100-GC clusters together with their best-fitting relation. We measured $\log N=12.22\pm0.04$ and $\alpha=2.02_{-0.09}^{+0.08}$. The relation is very tight, with a measured intrinsic scatter $\sigma_{\rm int}=0.06\pm0.03$ dex;  however,  the use of the temperature to estimate $r_{\rm 500,MT}$ introduces a small level of covariance between the two quantities, which might lead to an underestimation of the intrinsic scatter. As shown in Fig. \ref{fig:mgas_tx}, our best-fit relation agrees very well with the relation derived by \citet{arnaud07} using ten nearby relaxed clusters, which yields a slope of $2.10\pm0.05$ fully consistent with ours\footnote{In \citet{arnaud07} the temperatures were measured in the range $[0.15-0.75]r_{500}$, i.e. excluding the core. See Paper III for a discussion of the effects of excising (or not) the core region. \citet{arnaud07} integrated the gas mass within a hydrostatic-based $r_{500}$; since $M_{\rm gas}\propto M_{\rm tot}^{1/3}$ (see Sect. 5.2) the relation should depend mildly on the adopted value of $r_{500}$.}. The slope of the relation is steeper than the self-similar expectation \citep[1.5; e.g.][]{bryan98} at more than $5\sigma$ and steeper than the slope of the $M-T$ relation, which indicates a dependence of the gas fraction on cluster mass. 

We searched for breaks in the relation by splitting our sample into several temperature ranges and performing  the fitting procedure again. When considering only the systems with $T_{\rm 300 kpc}>2.5$ keV we measured a slope $\alpha_{T>2.5}=1.91\pm0.16$, compared to $\alpha_{T<2.5}=2.09\pm0.22$ below 2.5 keV. Thus, we find no evidence for a strong break in the relation. 

\begin{figure*}
\resizebox{\hsize}{!}{\hbox{\includegraphics{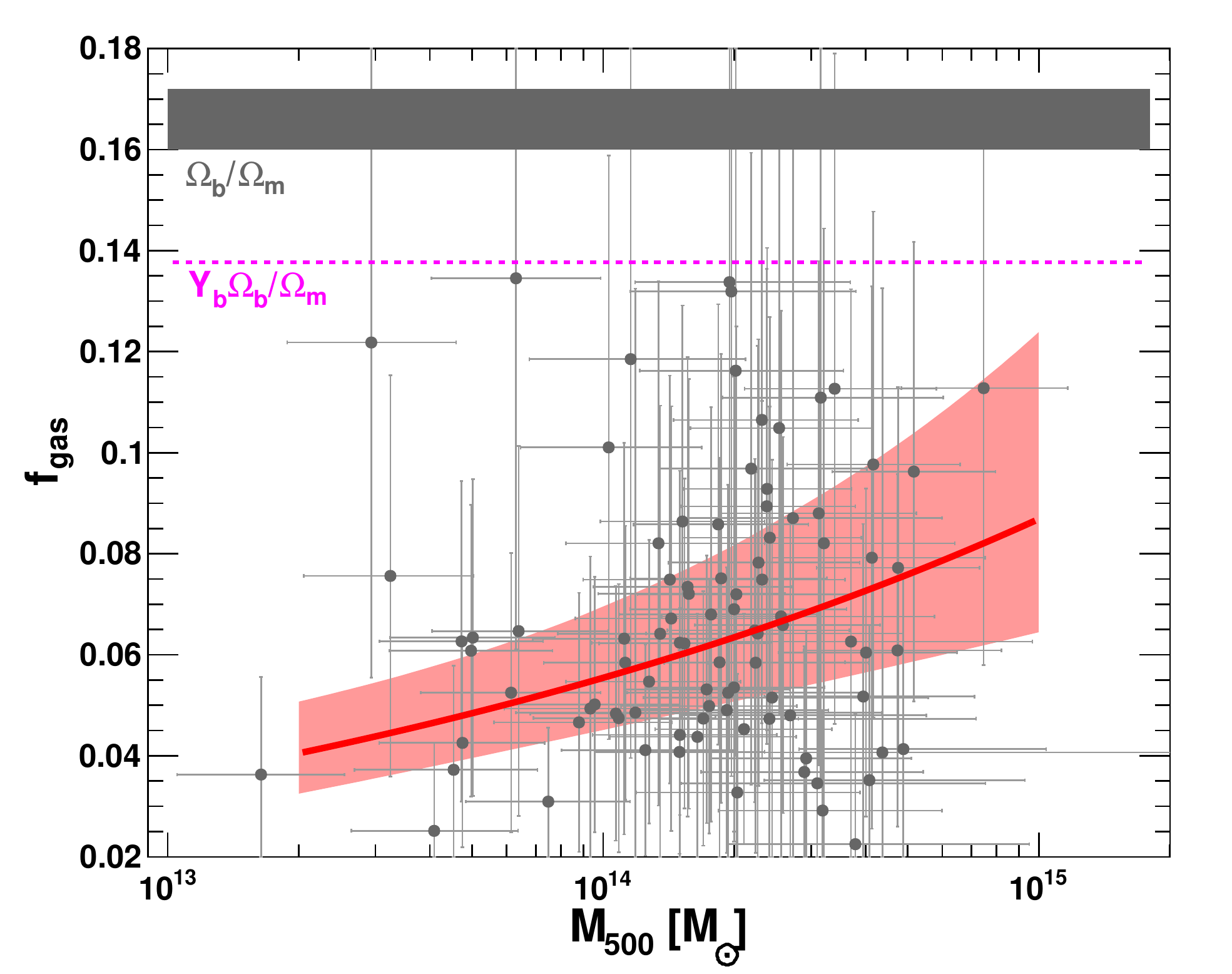}\includegraphics{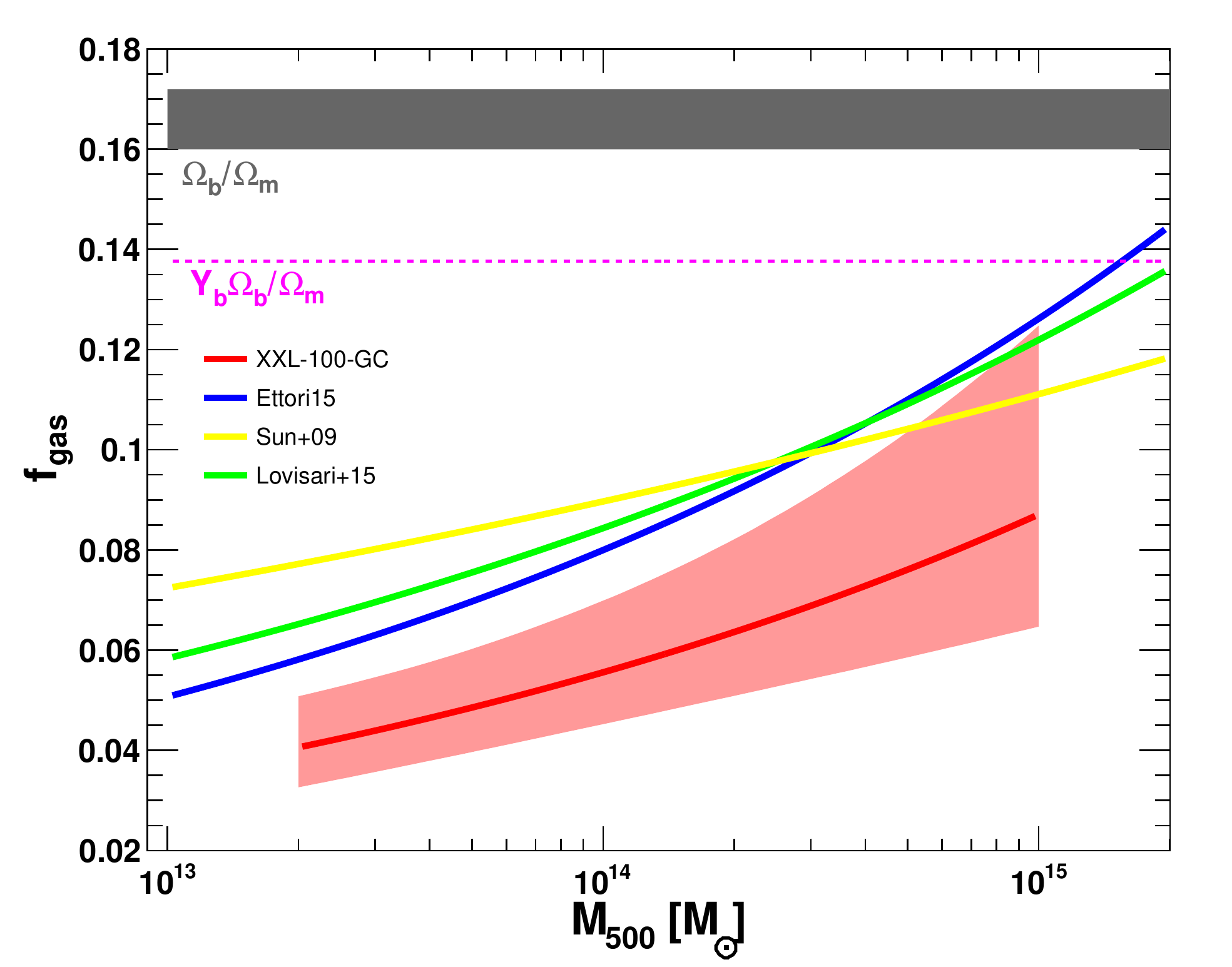}}}
\caption{Relation between gas fraction and halo mass within $r_{\rm 500,MT}$ for the XXL-100-GC sample. \emph{Left:} The red line and the red shaded area show the best-fit relation and its uncertainty. The data points show the individual $f_{gas}$ estimates obtained using the $M-T$ relation. The WMAP9 cosmic baryon fraction is displayed in the grey shaded area, whereas the dashed magenta line indicates the cosmic baryon fraction corrected by the depletion factor $Y_b=0.85$ at $r_{500}$ \citep{planelles13}. \emph{Right:} Same as in the left panel. For comparison, the solid curves show similar relations obtained using hydrostatic masses (yellow, \citet{sun09}; blue, \citet{ettori15}; green, \citet{lovisari15}). }
\label{fig:fgas_mtot}
\end{figure*}

To infer the relation between $f_{\rm gas,500}$ and $M_{\rm WL}$, we combined the best-fit $M-T$ and $M_{\rm gas}-T$ relations. This method is preferable to directly fitting  the $f_{\rm gas}-M$ relation, since the $M-T$ relation is affected by a significant scatter which is propagated to individual $f_{\rm gas,500}$ measurements. To recover the relation between $f_{\rm gas}$ and mass, we selected 10,000 MCMC realisations of the two relations and computed the resulting $f_{\rm gas}-M$ relation in each case. We then calculated the median and dispersion of the values of $f_{\rm gas,500}$ for a grid of $M_{WL}$ values. The relation obtained in this way reads
\begin{equation}h_{70}^{-3/2}f_{\rm gas,500}=0.055_{-0.006}^{+0.007}\times\left(\frac{M_{\rm 500,MT}}{10^{14}h_{70}^{-1}M_\odot}\right)^{0.21_{-0.10}^{+0.11}}.\label{eq:fgas_m}\end{equation}
In the left panel of Fig. \ref{fig:fgas_mtot} we show the recovered relation and its uncertainties, together with the individual $f_{\rm gas,500}$ values. In the right panel of Fig. \ref{fig:fgas_mtot} we compare our results with three state-of-the-art $f_{\rm gas}-M$ relations obtained for group and cluster-scale systems assuming that the ICM is in hydrostatic equilibrium: \citet[\emph{Chandra}]{sun09}, \citet[\emph{XMM-Newton}]{lovisari15}, and \citet[a large sample of published results]{ettori15}. We note a clear offset between our weak-lensing calibrated $f_{\rm gas}-M$ relation and the hydrostatic relations, for all mass scales. The relation also falls short of the cosmic baryon fraction $\Omega_b/\Omega_m$ \citep{planck15_13} by almost a factor of three at $10^{14}M_\odot$.

\begin{figure}
\resizebox{0.92\hsize}{!}{\includegraphics{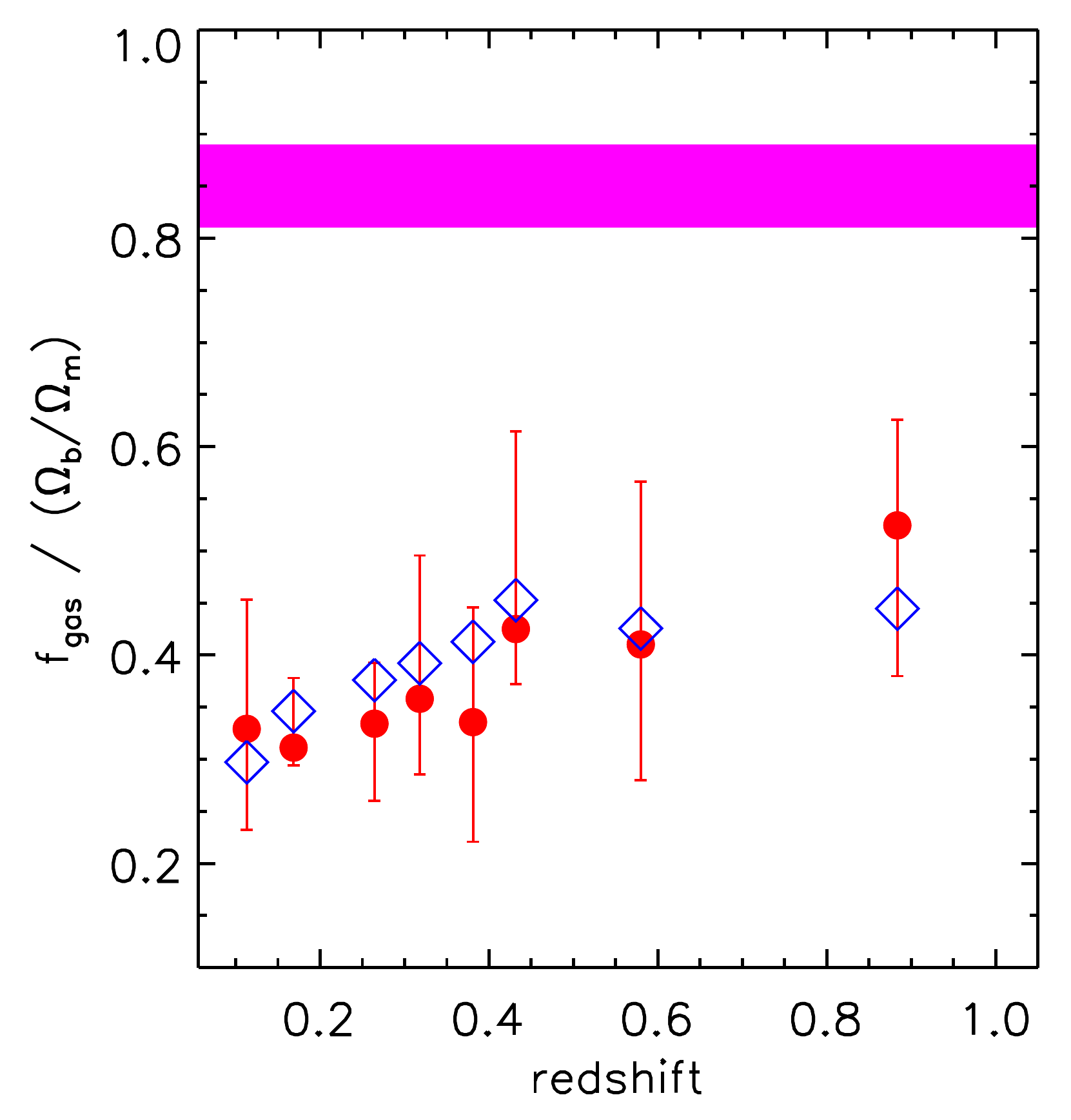}}
\caption{Mean gas fraction and inter-quartile ranges (red points) for the XXL-100-GC clusters in eight redshift bins normalised to the Universal baryon fraction $\Omega_b/\Omega_m$. The blue diamonds show the gas fraction expected from Eq. \ref{eq:fgas_m} at the median mass of each redshift bin in our sample. The magenta shaded area shows the value expected at $r_{500}$ from the simulations of \citet{planelles13}.}
\label{fig:fgas_z}
\end{figure}

In addition, we investigated the redshift dependence of our measurements. Fitting the $M_{\rm gas}-T$ relation only for the closest systems ($z<0.2$), we found a slope $\alpha_{z<0.2}=2.11\pm0.10$, in agreement with the results obtained for the entire data set. In Fig. \ref{fig:fgas_z} we show the mean gas fraction and inter-quartile ranges in eight redshift bins, normalised to the cosmic value. In all cases, a deficit of hot gas is observed compared to the expected gas fraction. We note a trend of increasing gas fraction with redshift. Since the median mass of the selected clusters increases with redshift, this effect can be explained by the mass dependence of the gas fraction. To confirm this statement, we calculated the gas fraction expected from Eq. \ref{eq:fgas_m} at the median mass of each redshift bin and compared with our measurements. This test closely reproduces the observed trend of increasing gas fraction with redshift. Therefore, we find no evidence for an evolution of the gas fraction with redshift.

The discrepancy reported here between X-ray and weak-lensing measurements of the gas fraction is independent of the instrument used for hydrostatic-based measurements. Recent calibration works reported discrepancies at the level of $\sim15\%$ between the temperatures measured with \emph{XMM-Newton} and \emph{Chandra} \citep{nevalainen10,schellenberger15}. However, this systematic issue was observed mainly for high-temperature systems, for which the energy of the bremsstrahlung cut-off is hard to calibrate. The bulk of the clusters in the XXL-100-GC sample have observed temperatures in the range 2-5 keV, where calibration uncertainties are much less important. 

\subsection{Stellar fraction}

To measure the stellar fraction, we applied the method presented in Sect.~\ref{sec:hod} to the subset of clusters in the XXL-100-GC northern sample where CFHTLS and WIRCam photometry is available, and within the redshift range $0.0 < z < 0.7$ to guarantee that most of the stellar mass is accounted for given our near-infrared completeness \citep[$10^{10} M_{\odot}$; see][]{coupon15}: this threshold corresponds to the limit in stellar mass above which the vast majority of the total stellar mass is encapsulated in massive clusters \citep[see also][]{vdb14}.

\begin{figure}  
  \resizebox{\hsize}{!}{\includegraphics[width=\textwidth]{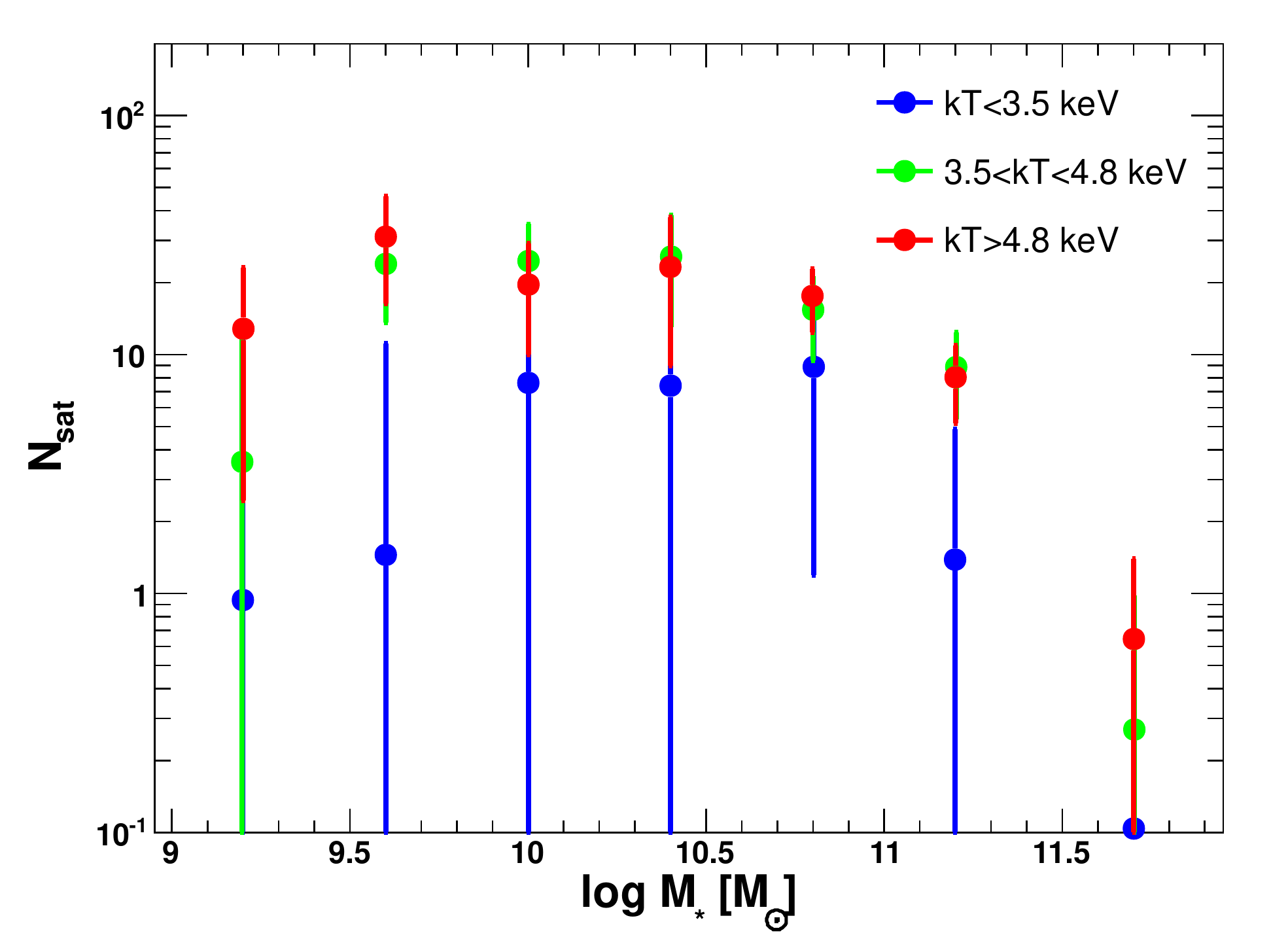} }
  \caption{Non-parametric halo occupation distribution of satellite galaxies in XXL-100-GC clusters in three temperature bins as a function of the galaxy's stellar mass for three temperature bins ($T_{\rm 300 kpc}<3.5$ keV, blue; $3.5<T_{\rm 300 kpc}<4.8$ keV, green; $T_{\rm 300 kpc}>4.8$ keV, red). }
  \label{fig:HOD}       
\end{figure}

This selection yields a subsample of 34 clusters for this analysis (marked with an asterisk in Table~\ref{tab:data}), further subdivided into three temperature bins with approximately similar signal-to-noise ($T_{\rm 300 kpc}<3.5$ keV, $3.5<T_{\rm 300 kpc}<4.8$ keV, and $T_{\rm 300 kpc}>4.8$ keV) to compute the mean satellite number as a function of halo mass. We cut the galaxy sample into stellar mass bins and measured the projected two-point correlation function in the range $0 < R < r_{\rm 500,MT}$ for each stellar mass and temperature bins. In Fig.~\ref{fig:HOD} we show the number of satellite galaxies as a function of stellar mass for the three temperature bins. The mean stellar mass for each temperature bin was then obtained by integrating the satellite numbers multiplied by their respective stellar mass in each bin, over the full stellar mass range. Although our sample is not complete below $10^{10}M_\odot$, low-mass galaxies contribute very little to the total stellar mass \citep{vdb14}. We used the mean temperature in each bin and the $M-T$ relation to estimate the corresponding halo mass. The results of this analysis are provided in Table \ref{tab:fstar}.

\begin{table*}
\caption{\label{tab:fstar} Results of the non-parametric halo occupation distribution analysis.}
\begin{center}
\begin{tabular}{cccccc}
\hline
$\langle T_{\rm 300 kpc}\rangle$ [keV] & $N_{\rm obj}$ & $M_{*} [M_\odot]$ & $M_{\rm 500,MT} [M_\odot]$ & $M_{\rm gas,500} [M_\odot]$ &  $f_{*}$ \\
\hline
\hline
$2.35\pm0.65$ &  21 & $(1.30\pm0.82)\times10^{12}$ & $1.4\times10^{14}$ & $(9.77\pm0.75)\times10^{12}$ & $0.0095\pm0.0061$ \\
$4.50\pm0.35$ & 7 & $(3.86\pm0.86)\times10^{12}$ & $3.9\times10^{14}$ & $(1.96\pm0.16)\times10^{13}$ & $0.0099\pm0.0022$ \\
$5.13\pm0.21$ & 6 & $(3.86\pm0.79)\times10^{12}$ & $4.8\times10^{14}$ & $(2.90\pm0.23)\times10^{13}$ & $0.0080\pm0.0017$ \\
\hline
\end{tabular}
\end{center}
\textbf{Column description:} 1. Mean temperature and standard deviation in the considered temperature bin. 2. Number of clusters per bin. 3. Mean stellar mass per cluster. 4. Mean halo mass estimated using the $M_{\rm WL}-T$ relation. 5. Mean gas mass per cluster (calculated for the same systems). 6. Mean stellar fraction.
\end{table*}

\begin{figure}
\resizebox{\hsize}{!}{\includegraphics{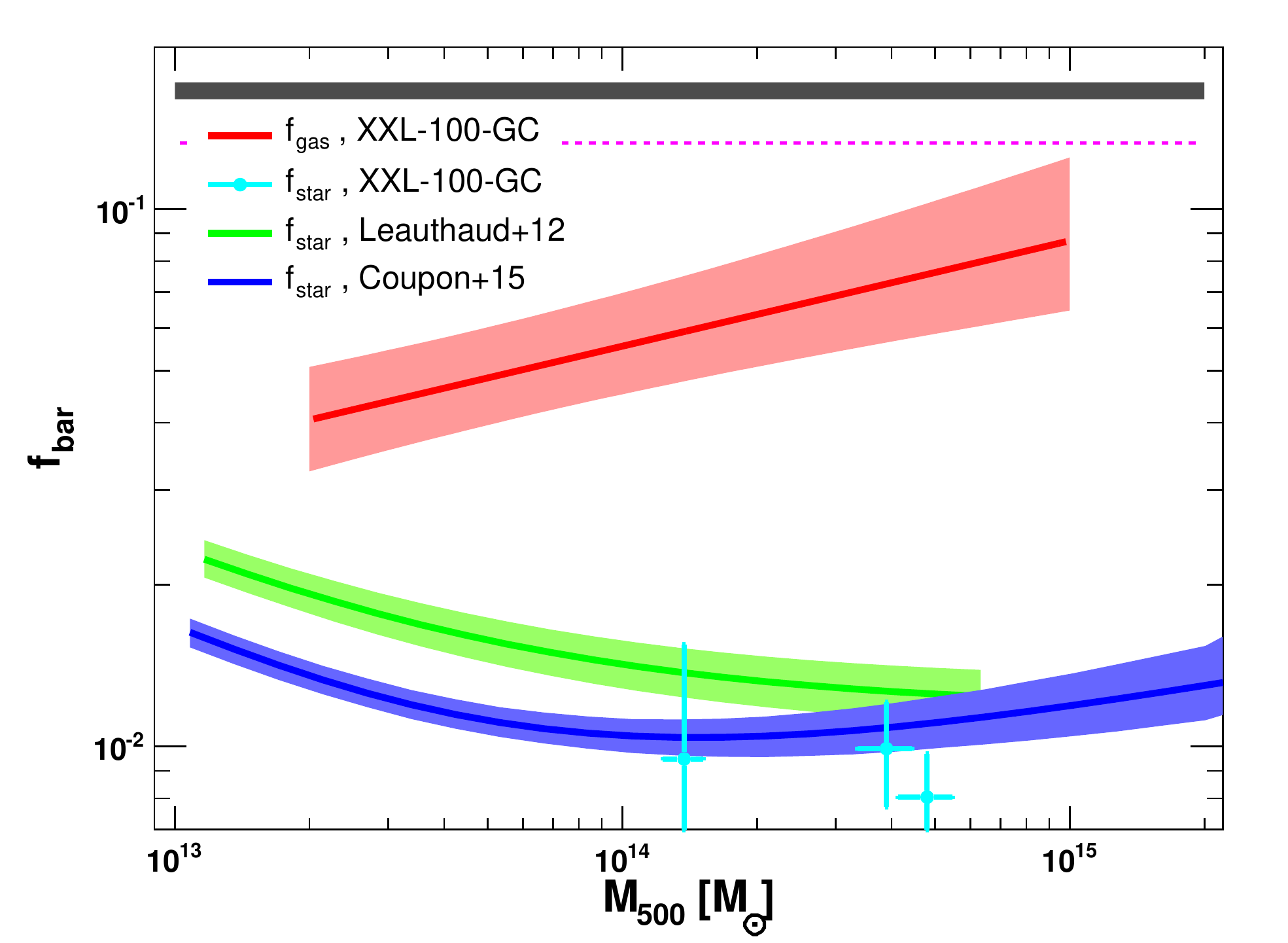}}
\caption{Baryon fraction in the form of hot gas (red, this work) and stars. The cyan data points show the measurements obtained for the XXL-100-GC sample in three temperature bins (see Table \ref{tab:fstar}), compared to literature measurements at different redshifts ($z\sim0.3$, \citet[green]{leauthaud12}; $z\sim0.8$, \citet[blue]{coupon15}). The WMAP9 cosmic baryon fraction is displayed in the grey shaded area, whereas the dashed magenta line indicates the cosmic baryon fraction corrected by the depletion factor $Y_b=0.85$ at $r_{500}$ \citep{planelles13}. }
\label{fig:fbar}
\end{figure}

In Fig.~\ref{fig:fbar} we show the measured stellar fraction and the gas fraction as a function of halo mass. Our results are compared to two literature studies combining galaxy clustering and galaxy-galaxy lensing in COSMOS \citep[$z\sim0.3$,][]{leauthaud12} and CFHTLenS/VIPERS \citep[$z\sim0.8$,][]{coupon15}. Our measurements are in good agreement with the latter study, whereas a slight discrepancy is observed with the former, most probably explained by stellar mass measurement systematics compared to \citet{leauthaud12}, as described in Sect. 5.3.1 of \citet{coupon15}.

Our measurements yield a stellar-to-halo mass fraction of about 1\% with little dependence on halo mass. This is clearly insufficient to bridge the gap between the measured hot gas fraction and the cosmic value. Combining the fraction of baryons in the form of hot gas and stars, we found $f_{\rm bar, 500}=0.067\pm0.008$ for $10^{14}M_\odot$ halos, which is discrepant with the cosmic baryon fraction corrected by the depletion factor at a confidence level of $6.9\sigma$.

We note that this estimate of the stellar fraction neglects the contribution of intracluster light (ICL) to the stellar content of dark-matter halos. Intracluster light is usually found to contribute to a fraction of the total stellar mass of galaxy clusters \citep[$20-30\%$, e.g.][]{zibetti05,gonzalez07,budzynski14}, so including this component would not significantly change the baryon budget. Therefore, the baryon fraction of XXL clusters only represents about half of the expected value, even when the total stellar content is accounted for.

\subsection{Gas distribution}

We investigated the distribution of intracluster gas using the PSF-corrected gas density profiles extracted following the method described in Sect. \ref{sec:method_mgas} (see Fig. \ref{fig:all_ngas}). We divided our sample into four temperature (and hence, mass) bins ($T_{\rm 300 kpc}<2$ keV; $2<T_{\rm 300 kpc}<3$ keV; $3<T_{\rm 300 kpc}<4$ keV; $T_{\rm 300 kpc}>4$ keV) and computed the mean gas density profile in each bin. To estimate the uncertainties in the mean profiles, for each subsample we performed $10^4$ bootstrap samplings of the population and computed the mean and standard deviation of the bootstrap samples. In addition to the statistical uncertainties, this method takes the intrinsic scatter in the population into account.

\begin{figure}
\resizebox{\hsize}{!}{\includegraphics{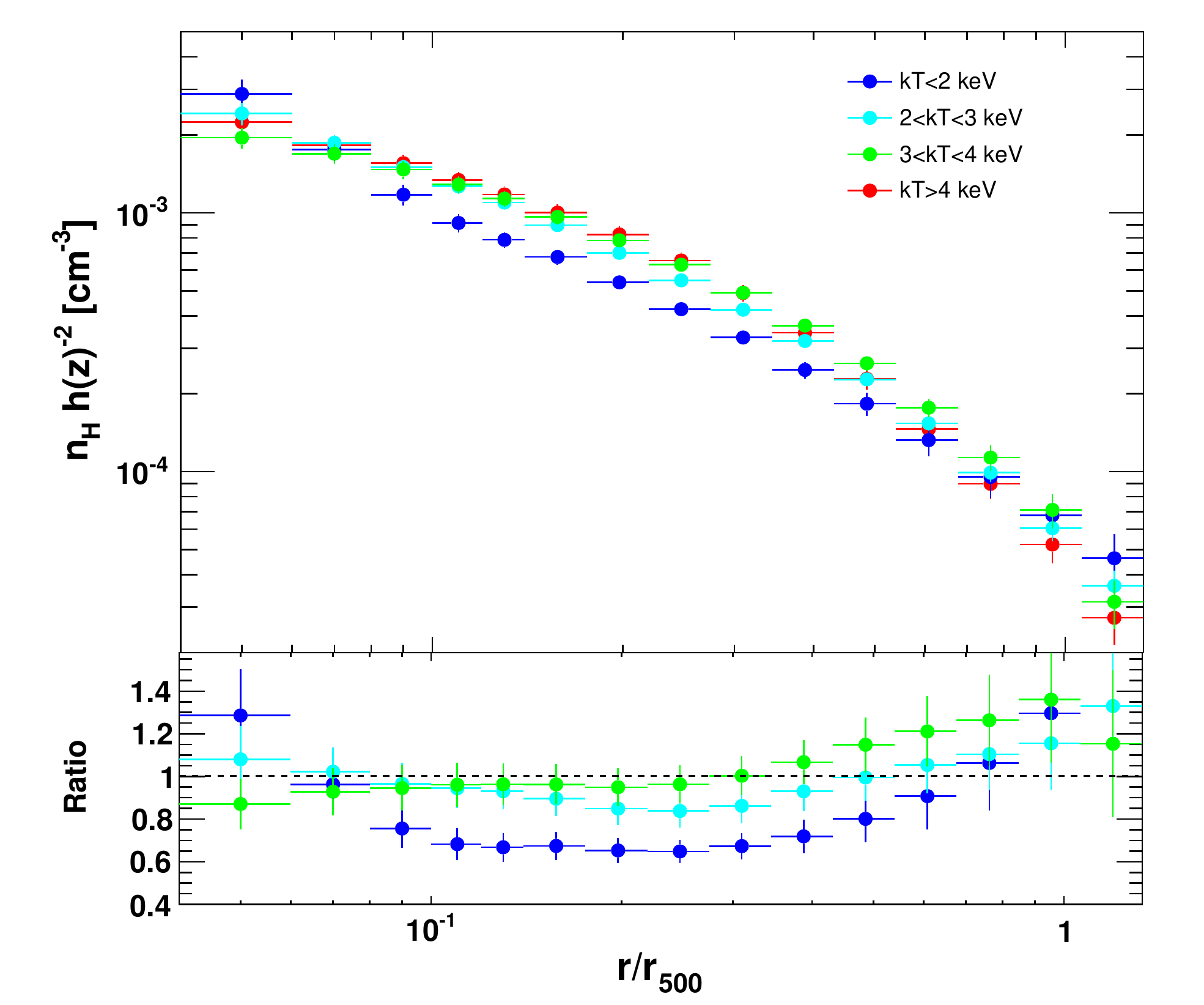}}
\caption{Mean self-similar scaled ion density profiles for the XXL-100-GC sample for four different temperature bins ($T_{\rm 300 kpc}<2$ keV, blue; $2<T_{\rm 300 kpc}<3$ keV, yellow; $3<T_{\rm 300 kpc}<4$ keV, green; $T_{\rm 300 kpc}>4$ keV, red). The uncertainties in the mean were computed using $10^4$ bootstrap samplings of the populations. The bottom panel shows the ratio of the mean gas density profiles to the high-temperature bin.}
\label{fig:mean_ngas}
\end{figure}

The mean gas density profiles are shown in Fig. \ref{fig:mean_ngas}, scaled according to the self-similar expectation. We observe significant differences between the various temperature bins, with a general trend of higher gas density in the central regions for massive clusters than for groups. This trend provides a clear confirmation of the dependence of the gas fraction on mass, since the gas fraction can be computed directly by integrating the self-similar scaled gas density profile \citep[see Appendix D in][]{e12}. 

We note clear differences in the shape of the gas density profiles between the various samples. While at high mass the mean profile shows a cored behaviour and a relatively steep outer gradient, as usually inferred for massive clusters \citep[see e.g.][]{mohr99,ettori98,e12}, galaxy groups exhibit a cuspy profile and a flatter outer slope, in agreement with previous studies \citep[e.g.][]{helsdon00,sun09}. Interestingly,  the gas density at $r_{\rm 500,MT}$ appears to be roughly independent of cluster mass \citep[see also][]{sun12}. This effect could indicate a mass dependence of entropy injection by AGN feedback, which would lead to a more pronounced expansion of the gas atmosphere in low-mass systems \citep[e.g.][]{ponman99}. Such differences would gradually disappear in cluster outskirts.

\section{Discussion}
\label{sec:disc}

As shown in Sect. \ref{sec:fgas_mtot} and highlighted in Fig. \ref{fig:fbar}, our results are in substantial tension with the Universal baryon fraction within $r_{\rm 500,MT}$. Here we review the possible explanations for this result, both physical and instrumental.

\subsection{Systematic uncertainties}
\label{sec:syst}

\subsubsection{X-ray measurements}

We investigated whether the subtraction of the background could have a significant impact on our gas mass measurements. Indeed, a systematic overestimation of the \emph{XMM-Newton} background level would lower the observed surface brightness and bias the recovered gas mass low. In the soft band, residual soft protons can typically affect the measured background by $\sim20\%$ \citep{lm08,kuntz08}. To investigate the impact of such an uncertainty on our measurements, we decreased the measured background level by 20\% when computing the emission measure and recalculated the gas masses. The resulting gas masses are only mildly affected by such a change, with typical differences at the 5\% level. 

A further source of systematic uncertainty is the conversion between count rate and emission measure, which depends on the temperature and metallicity of the plasma. For gas temperatures exceeding 1.5 keV (83 systems out of 95), the soft-band emissivity depends very weakly on temperature ($\sim3\%$ uncertainty) and the overall emissivity is dominated by the continuum, thus the metallicity does not play an important role. For the coolest systems, however, line emission and the position of the bremsstrahlung cut-off have a significant impact on the soft-band emissivity \citep[see the discussion by][]{lovisari15}. We estimate from the dependence of the cooling function on temperature and metal abundance that this effect can have an impact as large as 30\% on the soft-band emissivity, which translates into an uncertainty of 15\% on the gas density and gas mass. To estimate the impact of this systematic uncertainty on the gas fraction, we varied the gas masses of the systems with $T_{\rm 300 kpc}<1.5$ keV by $\pm15\%$ and fitted again the $M_{\rm gas}-T$ relation. We found that the slope of the relation varies in the range 1.92-2.11. The effect on the gas fraction is less than the statistical uncertainties in the entire mass range.

Our test of the gas mass measurement method using the cosmo-OWLS simulations (Sect. \ref{sec:mgas_test}) shows that our method is able to reconstruct the 3D gas mass with no bias and little intrinsic scatter (see Fig. \ref{fig:mock}), provided that the value of $r_{500}$ used to integrate the gas density profile is accurate. Since the uncertainties in $r_{\rm 500,MT}$ were propagated to the gas mass (see Sect. \ref{sec:method_mgas}), we conclude that potential systematics in the gas mass cannot reconcile our $f_{\rm gas} $ measurements with the cosmological value. In addition, the excellent agreement observed between the $M_{\rm gas}-T$ relation measured here and results from the literature (see Fig. \ref{fig:mgas_tx}) indicates that our measurements of the gas temperature are in line with other samples. 

\subsubsection{Comparison between X-ray and Sunyaev-Zeldovich measurements}
\label{sec:planck}

To cross-check our \emph{XMM-Newton} $Y_{X,500}$ measurements, we extracted measurements of the Sunyaev-Zeldovich (SZ) flux $Y_{500}$ from \emph{Planck}. This constitutes a powerful further check, since X-ray and SZ provide a measurement of the same physical quantity using two completely independent techniques. We extracted $Y_{500}$ for each object of our sample following the method used in~\citet{PlanckMCXC,PlanckMaxBCG,PlanckLBG}. Namely, we adopted the SZ profile from~\citet{arnaud2010} and applied multifrequency matched filters~\citep{melin2006} scaled to the radii $r_{\rm 500,MT}$ given in Table~\ref{tab:data} at the positions of the clusters. 

Unfortunately, only four clusters were detected individually by Planck at $S/N>3$ (XLSSC 003, XLSSC 060, XLSSC 091, XLSSC 509). We thus computed the inverse-variance weighted average of our X-ray and SZ signals in 8 regularly log-spaced $Y_{X,500}$ bins. For each $Y_{X,500}$, the individual variance is taken as the mean of the square of the upper and lower errors given in Table~\ref{tab:data}. The $Y_{X,500}$ values are then converted into ${\rm Mpc}^2$ adopting the factor $C_{\rm XSZ}=1.416\times10^{-19} \, {{\rm Mpc}^2 \over M_\odot {\rm keV}}$ \citep{arnaud2010}. We averaged the SZ flux similarly in the same $Y_{X,500}$ bins. In Table~\ref{tab:binned_yx_ysz} we give the resulting binned $Y_{X,500}$ and $Y_{500}$, and associated errors.

\begin{table}
\caption{\label{tab:binned_yx_ysz}Binned values for \emph{XMM-Newton} $Y_{X,500}=M_{\rm gas,500} \times T_{\rm 300 kpc}$ and \emph{Planck} SZ $Y_{500}$.}
\begin{center}
\begin{tabular}{cccc}
\hline
$Y_{X,500}$ & $\sigma_{Y_{X,500}}$ & $Y_{500}$ & $\sigma_{Y_{500}}$\\ 
\hline\hline
1.67 & 0.35 & -0.67 & 9.93\\
3.08 & 0.22 & 2.74 & 5.60\\
5.50 & 0.33 & -2.51 & 7.94\\
13.4 & 0.89 & -5.30 & 10.2\\
24.6 & 1.01 & 22.0 & 11.5\\
53.5 & 2.12 & 53.4 & 13.5\\
111 & 5.79 & 31.7 & 27.4\\
239 & 12.2 & 258 & 32.3\\
\hline
\end{tabular}
\end{center}
\textbf{Column description:} 1. Mean $Y_{X,500}$ value; 2. Error on $Y_{X,500}$; 3. Stacked SZ flux $Y_{500}$; 4. Error on $Y_{500}$. All measurements are in units of $10^{-7} {\rm Mpc}^2$.
\end{table}

\begin{figure}
\resizebox{\hsize}{!}{\includegraphics{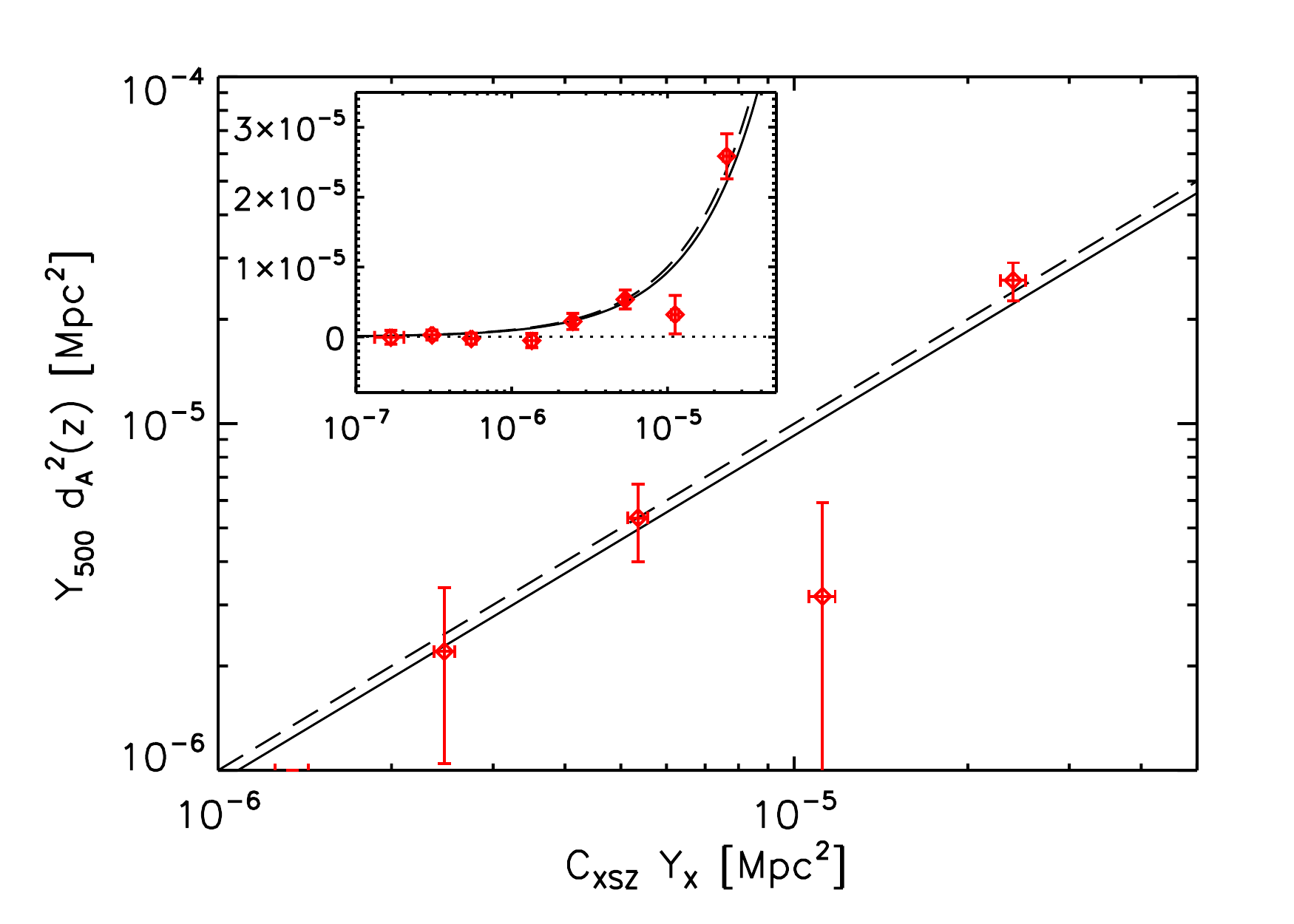}}
\caption{\emph{Planck} SZ flux $Y_{500}\, d_A^2(z)$ versus $Y_{X,500}$ for the \mbox{XXL-100-GC} sample (red diamonds). The expected value 0.924 ~\citep[Eq.~19 in][]{arnaud2010} is the solid black line and the unity is shown as the dashed line. The inset presents the same data points in the lin-log plane to show also the low $Y_{X,500}$ bins. The numerical values and error bars for the 8 points are given in Table~\ref{tab:binned_yx_ysz}.}
\label{fig:yx_ysz}
\end{figure}

The results presented in Table \ref{tab:binned_yx_ysz} are shown in Fig.~\ref{fig:yx_ysz}. In this Figure, we show the expected relation ($Y_{500}=0.924Y_{X,500}$) from the REXCESS sample~\citep[Eq.~19 of ][]{arnaud2010} and the one-to-one relation. We can see that the values of $Y_{500}$ measured by \emph{Planck} agree with the values of $Y_{X,500}$ obtained with \emph{XMM-Newton}. This provides an important additional test showing that X-ray gas mass measurements are not subject to significant systematic uncertainties. In light of these results, we conclude that the XXL galaxy cluster population does not appear to differ from other X-ray selected cluster samples. 

\subsubsection{Weak-lensing measurements}

An alternative possibility is that the weak-lensing calibrated $M-T$ relation used here and presented in Paper IV is biased high. For a discussion of the various systematic uncertainties affecting these measurements, we refer to Paper IV. We note, however, that our $M-T$ relation for XXL clusters agrees with the results of \citet{kettula15} combining CCCP, COSMOS and CFHTLS mass measurements. Therefore, any bias in our weak-lensing measurements would affect as well these other data sets, and the low gas fraction observed here should be a generic consequence of current weak-lensing mass calibrations. 

\subsection{A generalised hydrostatic bias?}
\label{sec:hse_bias}

The results presented in Fig. \ref{fig:fgas_mtot} clearly show a substantial tension between the gas fraction (and, in turn, baryon fraction) of the XXL-100-GC clusters with our weak-lensing mass calibration and the existing results based on hydrostatic masses. The straightforward interpretation is that the studies based on hydrostatic masses are biased low by the presence of a significant non-thermal pressure, which would result in an overestimated gas fraction. The effect of the mass bias on the $f_{\rm gas}-M$ plane is twofold. Obviously, when increasing the total mass the average gas fraction goes down by the same amount. On top of that, the increase in the total mass shifts the curve to the right, which given the gradient in the $f_{\rm gas}-M$ plane decreases the gas fraction at fixed mass further. On the other hand, the decrease of the gas fraction is mitigated by a slight increase in $M_{\rm gas,500}$ following from the increase in integration radius. This effect is however mild, since the slope of the gas density profile at $r_{\rm 500,MT}$ is steep\footnote{At $r_{500,MT}$, the mean slope of the density profiles is $d\log\rho/d\log R\sim-2$; therefore $M_{\rm gas,500}\propto r_{\rm 500,MT}\propto M_{\rm 500,MT}^{1/3}$}.

\begin{figure}
\centerline{\resizebox{\hsize}{!}{\includegraphics{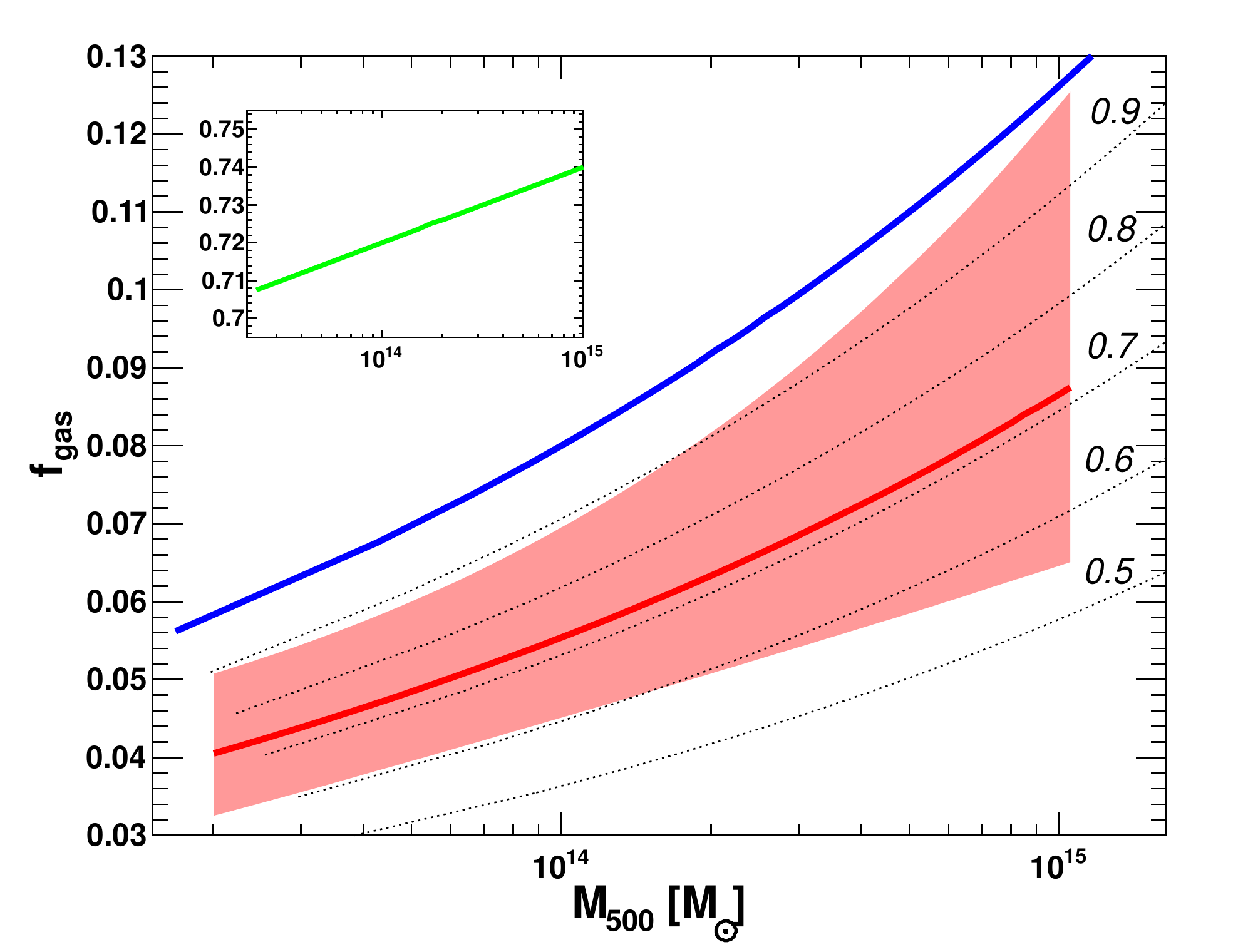}}}
\caption{$f_{\rm gas}-M$ relation within $r_{\rm 500,MT}$ obtained for XXL clusters (red line) compared to \citet[blue]{ettori15}. The green line in the inset shows the recovered hydrostatic bias $1-b=\frac{M_X}{M_{WL}}$ required to bring the two measurements in accord. The dashed lines show the expected true gas fraction curves for constant values of $1-b$ in the range [0.5-0.9].}
\label{fig:bias}
\end{figure}

By comparing the XXL-100-GC measurement with hydrostatic-based relations, we can work out the mean hydrostatic bias and its mass dependence required to match the two curves. To estimate the bias and its mass dependence, we took the measurement of \citet{ettori15} as a reference point, since it includes the largest sample of hydrostatic mass measurements, and expressed the bias as a power law,
\begin{equation}1-b=\frac{M_X}{M_{\rm WL}}=1-b_{0} +\alpha\log_{10}\left(\frac{M_{\rm WL}}{10^{14}M_\odot}\right).\end{equation}
In Fig. \ref{fig:bias} we show the two measurements and the bias required to reconcile the two measurements. By matching the two relations, we find $b_{0}=0.28_{-0.08}^{+0.07}$ and $\alpha=0.02_{-0.09}^{+0.10}$. 

Therefore, in case the discrepancy between weak-lensing and hydrostatic-based measurements comes from a generalised hydrostatic bias, we are observing a roughly mass-independent hydrostatic bias of 30\%, which is consistent with the offset between hydrostatic and lensing-based $M-T$ relations in the literature (see Fig. 8 of Paper IV). This result directly translates into significant differences in the inferred gas fraction. As a further check, we recomputed the gas mass and gas fraction using the $M-T$ relations of \citet{sun09} and \citet{lovisari15} and compared the resulting $f_{\rm gas}-M$ relation for XXL clusters with the results of these studies. The recovered gas fraction agrees well with the original measurements (see Appendix \ref{app:hydro}). Therefore, we conclude that the low gas fractions reported here are a direct consequence of the increased halo masses obtained through weak lensing.

\subsection{Comparison with previous measurements}
\label{sec:previous}

Numerical simulations predict that the bias in hydrostatic masses induced by the presence of non-thermal energy is of the order of $\sim15\%$ on average \citep[e.g.][]{rasia04,nagai07,nelson12}. The bias could be as high as 30\% in case the temperature structure in the ICM is strongly inhomogeneous \citep{rasia06}, although the mean temperature of local clusters was found to agree with the mass-weighted temperature \citep{frank13}. The bias $1-b=0.72_{-0.07}^{+0.08}$ recovered above is definitely on the high side of the expectations from numerical simulations. On the other hand, this value is in line with a number of recent weak-lensing measurements, such as Weighing the Giants \citep[WtG,][]{vdl14} and CCCP \citep{hoekstra15}. \citet{okabe15} presented weak-lensing mass measurements for the LoCuSS sample and found results $\sim10\%$ lower than WtG, but in good agreement with CCCP and CLASH \citep{donahue14}. While WtG and CCCP indicate a bias of the order of 20-30\% in the mass calibration adopted by the \emph{Planck} team \citep{planck_11_11}, \citet{smith15} showed that the LoCuSS data are consistent with no bias, depending on the method adopted to compute the sample average. No consensus has thus been reached on the true value of the hydrostatic bias \citep[see also][]{sereno15,applegate15}. We remark that the \emph{Planck} calibration, which serves as a benchmark for the measurement of the hydrostatic bias in the studies discussed here, is consistent with our calibration of the hydrostatic $f_{\rm gas}-M$ relation \citep[see Fig. 1 of ][]{ettori15}. 

As a word of caution, we note that systematic differences have been reported in the literature between hydrostatic masses extracted by different groups and different instruments \citep{rozo14,donahue14}. Since our calculation relies on a benchmark hydrostatic $f_{\rm gas}-M$ relation, the recovered hydrostatic bias certainly depends on the adopted hydrostatic relation. The relation derived by \citet{ettori15} is based on a large compilation of hydrostatic measurements (94 systems) obtained by several groups and several instruments. Therefore, it likely summarises our current knowledge of hydrostatic gas fraction measurements. Moreover, the XXL-100-GC sample is comprised mainly of galaxy groups and poor clusters in the temperature range 2-5 keV, for which the systematic uncertainties associated with the temperature measurements are much less important than for massive clusters \citep{nevalainen10,schellenberger15}. Nevertheless, systematic uncertainties in the adopted hydrostatic relation cannot be excluded.

Using the WtG weak-lensing mass calibration \citep{vdl14b}, \citet{mantz15} derived a mean gas fraction of $\sim0.11$ at $r_{500}$. This value is slightly lower than the hydrostatic-based calibrations discussed above, but higher than what is measured here. Given that the WtG clusters are more massive ($M_{500}>6\times10^{14}M_\odot$) than for XXL-100-GC, the two studies are broadly consistent, although less so than one would naively expect from the similarity of the $1-b$ values derived in this work and from the direct comparison of WtG and \emph{Planck} masses \citep{vdl14}.

\subsection{Comparison with numerical simulations}
\label{sec:agn}

\begin{figure*}
\resizebox{\hsize}{!}{\hbox{\includegraphics{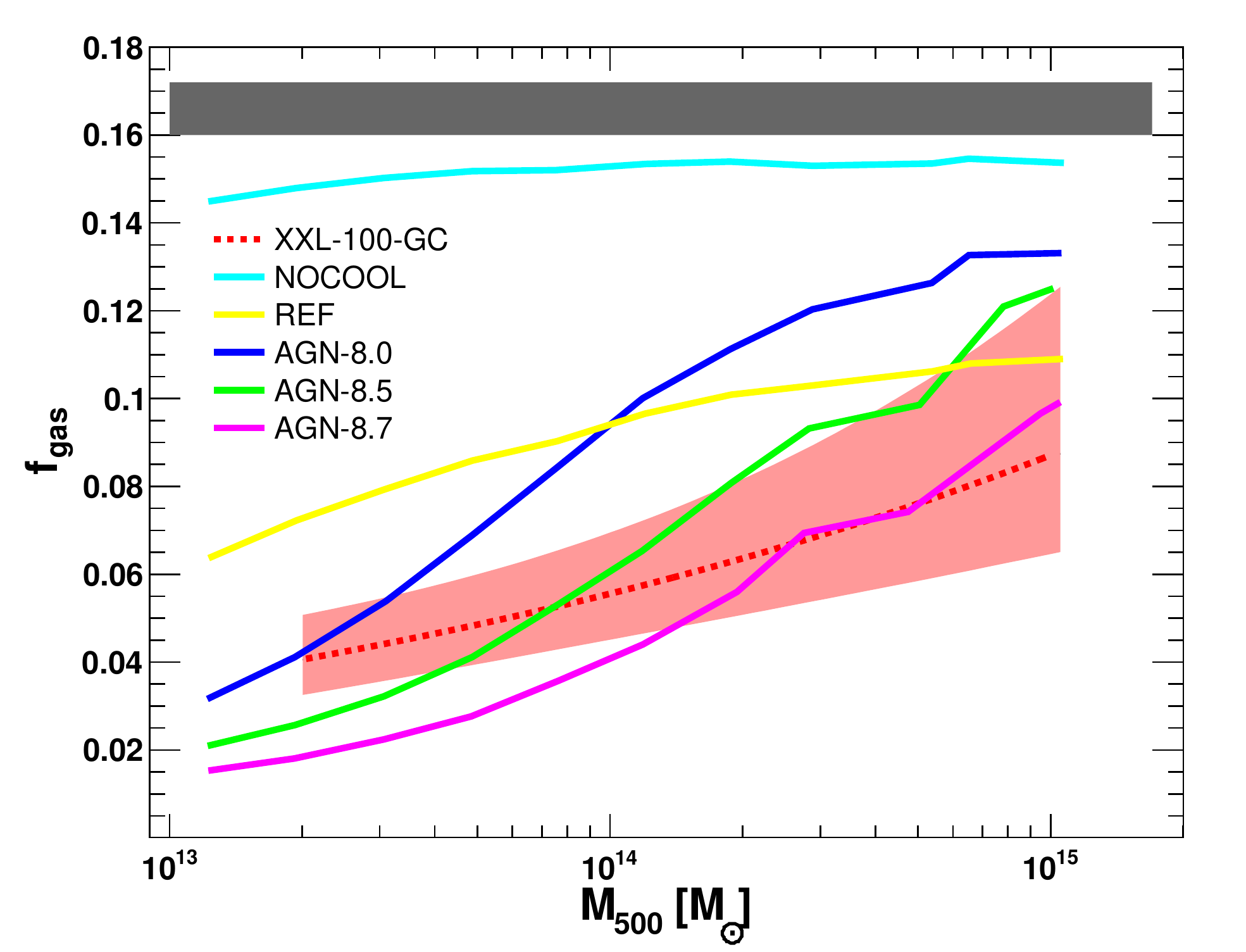}\includegraphics{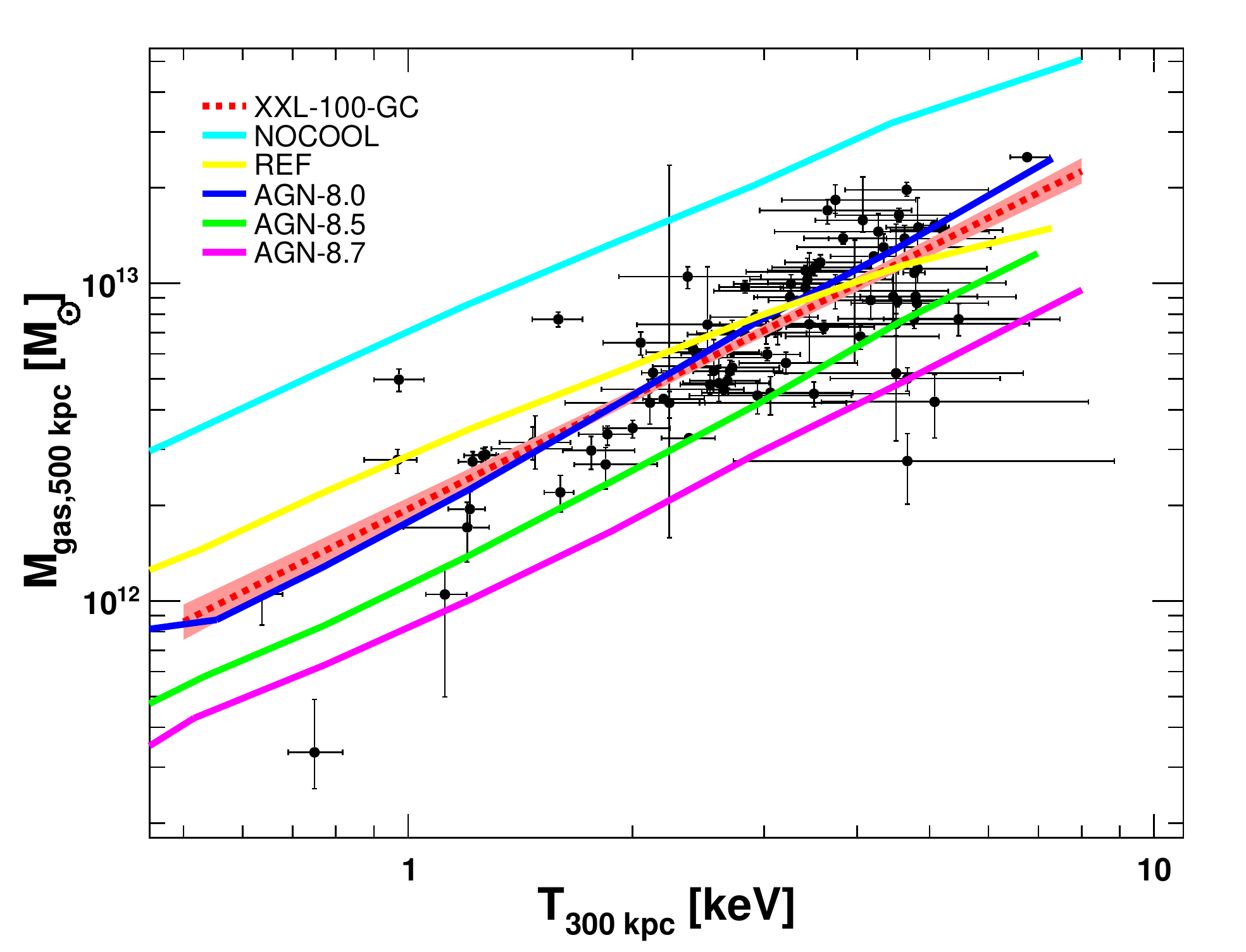}}}
\caption{\emph{Left panel:} Gas fraction of XXL-100-GC galaxy clusters (dashed red curve and red shaded area) compared to cosmo-OWLS simulations with different gas physics (non-radiative, NOCOOL; cooling and star formation, REF; AGN feedback with various energy injection, AGN-8.0, AGN-8.5, AGN-8.7). The grey shaded area shows the WMAP9 cosmic baryon fraction. \emph{Right panel:} Gas mass within 500 kpc as a function of the temperature inside 300 kpc for the XXL-100-GC sample (black points) compared to cosmo-OWLS simulations (same colour code). The red dashed curve and shaded area show the best fit to the data with a power law and its error envelope.}
\label{fig:fgas_sim}
\end{figure*}

The gas fraction of $5.5\%$ for $10^{14}M_\odot$ systems estimated here indicates that, once the stellar fraction is taken into account, even at the high-mass end the baryon fraction within $r_{\rm 500,MT}$ falls short of the Universal baryon fraction by a factor of two. As shown in the recent paper by \citet{sembolini15}, which presents the comparison between gas properties in a dozen different flavors of non-radiative numerical simulations, such a result cannot be explained by gravitational or hydrodynamical effects. Baryonic physics, and in particular feedback effects, must therefore be invoked to explain such a low gas fraction. 

We compared our results with the predictions of the cosmo-OWLS simulations \citep{lebrun14,mccarthy14}, which is a large suite of hydrodynamical simulations (an extension of the OverWhelmingly Large Simulations, OWLS, project of \citealt{schaye10}) utilising the smoothed particle hydrodynamics (SPH) code \texttt{GADGET-3} \citep{springel05}. This set of simulations includes several runs with different gas physics \citep[see Table 1 of ][for details]{lebrun14}: \emph{i)} a purely hydrodynamic run neglecting the effects of baryonic physics (hereafter NOCOOL); \emph{ii)} a run including gas cooling, star formation and feedback from supernovae (hereafter REF); \emph{iii)} three runs including AGN feedback. In the latter case, AGN feedback was modelled using the \citet{booth09} model, where a fraction of the accreted rest mass energy is used to increase the temperature of neighbouring gas particles by an amount $\Delta T_{\rm heat}$.  The black holes store up accretion energy in a reservoir until it is sufficient to heat neighbouring gas by $\Delta T_{\rm heat}$.  The three runs considered here include a deposited temperature $\Delta T_{\rm heat}$ of $10^{8}$ K (hereafter AGN-8.0), $10^{8.5}$ K (AGN-8.5), and $10^{8.7}$ K (AGN-8.7).  An increase in $\Delta T_{\rm heat}$ will in general lead to more bursty (and energetic) feedback, as more time is required between feedback events to store up sufficient energy to heat the gas to a higher value of $\Delta T_{\rm heat}$.

In the left-hand panel of Fig. \ref{fig:fgas_sim} we show the comparison in the $f_{\rm gas}-M$ plane between the XXL-100-GC curve and the results of the various cosmo-OWLS runs. As expected, the non-radiative run (NOCOOL) largely overestimates the observed gas fraction, since in this case the baryons condensate into dark-matter halos but do not form stars. The REF run is consistent with the observations at high mass but overestimates the gas fraction in the group regime. However, because of the absence of AGN feedback this run is affected by the usual ``cooling catastrophe'', which results in a stellar fraction much above the observed one. The curves providing the best match to the data are the AGN-8.5 and AGN-8.7 runs.  In these cases, AGN drive energetic outflows from the progenitor groups and clusters (typically at $z\sim2-3$, corresponding to the peak of the cosmic black hole accretion rate density) which are not efficiently recaptured later on, resulting in groups and clusters with lower than Universal baryon fractions within $r_{500}$ (see \citealt{mccarthy11}).  We note, however, that these models predict a steeper trend of increasing $f_{\rm gas}$ with halo mass compared to the data, which could be reproduced by invoking a mass dependence of the deposited heat $\Delta T_{\rm heat}$, although the different selection procedure for observed and simulated halos may play a role here. Therefore, if the results obtained in XXL-100-GC using weak-lensing calibration are to be trusted, our measurements favour the most extreme AGN feedback schemes tested in this work. 

Interestingly,  the AGN-8.0 model systematically overestimates the measured gas fraction. This run was found to provide the best match to hydrostatic-based gas fraction measurements and to X-ray-only proxies, such as the gas density and entropy profiles \citep[see][]{lebrun14}.  Conversely, in \citet{lebrun14} the AGN-8.5 and (particularly) AGN-8.7 models were strongly disfavoured by X-ray-only proxies, in particular the gas entropy. Indeed, a strong AGN feedback leads to a substantial entropy injection in the surrounding ICM \citep{gaspari15}, in excess of what is observed in local galaxy groups \citep[e.g.][]{ponman99,sun09}. As a further test, we made the comparison between the $M_{\rm gas}-T$ relation measured here and the results of the runs including various gas physics. In the right-hand panel of Fig. \ref{fig:fgas_sim} we show the gas mass measured within a fixed physical radius of 500 kpc as a function of the temperature inside a 300 kpc radius. The XXL-100-GC data points are compared to the simulation results obtained in the same physical regions. In agreement with \citet{lebrun14}, we find that the AGN-8.0 run closely reproduces the observed $M_{\rm gas}-T$ data, confirming that X-ray measurements prefer a mild entropy injection by AGN feedback. Conversely, the AGN-8.5 and AGN-8.7 schemes provide an excessive heating to the gas, which leads to an overestimate of the temperature at fixed gas mass compared to real systems. We emphasize that because of the use of fixed physical apertures this result is independent of any scaling relation. 

To summarise, we conclude that AGN feedback alone as is implemented in current numerical simulations cannot reproduce consistently the lensing-based $f_{\rm gas}-M$ relation and the thermodynamical properties of the ICM, which challenges our understanding of the ICM. This tension would be mitigated in case the weak-lensing mass calibration used here is biased high.

\subsection{Implications for cosmology}

Recently, the final results from the \emph{Planck} mission have been released, providing benchmark cosmological constraints through the CMB power spectrum \citep{planck15_13} and the number counts of SZ sources \citep{planck15_24}. The \emph{Planck} team noted a tension between the results obtained with the two techniques, cluster counts preferring systematically lower values of $\sigma_8$ and $\Omega_m$ than expected from the CMB power spectrum. To alleviate this issue, two possible interpretations were put forward. The two measurements could be reconciled for a neutrino mass in the range $\sum m_\nu\sim0.2$ eV since massive neutrinos induce an additional pressure term that would delay the formation of massive structures. Alternatively, this effect could be explained by a strong bias ($1-b=0.58\pm0.04$) in the mass calibration used by the \emph{Planck} team for the analysis of the SZ cluster counts, which is based on hydrostatic masses. 

The results presented here clearly highlight the tension that a strong hydrostatic bias implies on the baryon fraction (see Fig. \ref{fig:bias}). Since our reference hydrostatic masses agree with the \emph{Planck} mass calibration (see Sect. \ref{sec:previous}), a mass bias in the range $1-b=0.58$ would translate into an even lower baryon fraction than is measured here. Such a bias would imply $f_{\rm gas,500}\sim0.06$ in $10^{15}M_\odot$ halos, which would pose a considerable challenge for our understanding of cluster physics and evolution. Even the most extreme feedback schemes adopted in numerical simulations would fail to reproduce such characteristics (see Sect. \ref{sec:agn}), even though  such models have trouble  reproducing the global properties of galaxy clusters. Therefore, the results presented here argue against the strong hydrostatic bias  implied by \emph{Planck} primary CMB. A satisfactory solution to the \emph{Planck} CMB/SZ discrepancy must therefore be able to explain at the same time the baryon fraction of high-mass halos.

\section{Conclusion}

In this paper, we have presented a study of the baryon budget of dark-matter halos in the mass range $10^{13}-10^{15} M_\odot$ using the sample of the 100 brightest clusters discovered in the XXL Survey. Our main results can be summarised as follows:

\begin{itemize}
\item We developed a method to measure the gas mass from XXL Survey data (Sect. \ref{sec:method_mgas}). The method was calibrated using mock X-ray images drawn from cosmological simulations and was found to reproduce the true gas mass with excellent accuracy (see Fig. \ref{fig:mock}). The gas mass measurements for all the clusters in the XXL-100-GC sample are provided in Table \ref{tab:data}. We also provide measurements of $Y_{X,500}=T_{\rm 300 kpc}\times M_{\rm gas,500}$, which were found to agree well with the stacked \emph{Planck} SZ flux. \\

\item The scaling relation between the gas mass within $r_{\rm 500,MT}$ and the spectroscopic X-ray temperature is very tight and significantly steeper than the self-similar prediction  (see Fig. \ref{fig:mgas_tx}), in excellent agreement with previous measurements \citep{arnaud07}. We recovered the relation between hot gas fraction and halo mass within $r_{\rm 500,MT}$ by combining our $M_{\rm gas}-T$ relation with the $M_{\rm WL}-T$ relation based on an internal mass calibration using weak gravitational lensing (see Paper IV). The recovered relation reads $h_{70}^{-3/2}f_{\rm gas,500}=0.055_{-0.006}^{+0.007}\left(M_{\rm 500,MT}/10^{14}h_{70}^{-1}M_\odot\right)^{0.21_{-0.10}^{+0.11}}$.\\

\item We measured a mass-independent stellar fraction of $\sim1\%$ in XXL clusters, which agrees with previous measurements \citep{leauthaud12,coupon15}. Therefore, even when including the stellar mass fraction our baryon fraction measurement falls short of the cosmic value by about a factor of two ($f_{\rm bar, 500}=0.067\pm0.008$ in $10^{14}M_\odot$ halos). \\

\item Comparing our best-fit $f_{\rm gas}-M$ relation with results from the literature based on hydrostatic masses, we found that our weak-lensing based gas fraction is significantly lower than previous hydrostatic measurements (Fig. \ref{fig:fgas_mtot}). The tension is particularly important around $10^{14}M_\odot$, where the bulk of XXL-100-GC systems lies. The two methods can be reconciled by considering a roughly mass-independent hydrostatic bias $1-b=M_X/M_{\rm WL}=0.72_{-0.07}^{+0.08}$. This value is on the high side of the expectations of numerical simulations \citep{rasia04,nagai07}; however, it is consistent with recent studies such as WtG \citep{vdl14} and CCCP \citep{hoekstra15}. Therefore, the low gas fraction observed here directly follows from the higher masses obtained through weak lensing.\\

\item We compared our $f_{\rm gas}-M$ relation with the predictions of cosmological simulations using different gas physics \citep[cosmo-OWLS, ][]{lebrun14}. We found that our results favour extreme AGN feedback schemes in which a large fraction of the baryons is expelled from the potential well of dark matter halos. Such models are, however, in tension with X-ray-only proxies such as the gas density and entropy profiles 
\citep{lebrun14} and are not able to reproduce the relation between gas mass and temperature of XXL clusters. Therefore, the results presented here are challenging for current numerical simulations, and reconciling the observed gas fraction with the predictions would require that weak-lensing masses be systematically overestimated.\\

\item A mass bias $1-b=0.58\pm0.04$, which is required to reconcile \emph{Planck} cluster counts with primary CMB, would further exacerbate the tension between $f_{\rm bar}$ and the cosmological value, which would challenge our understanding of cluster physics. Therefore,  a satisfactory solution to the tension between CMB and cluster counts must also simultaneously explain the low baryon fraction measured here for massive halos.

\end{itemize}

\begin{acknowledgements}
XXL is an international project based on an XMM Very Large Program surveying two 25 deg$^2$ extragalactic fields at a depth of $\sim5\times10^{-15}$ ergs s$^{-1}$ cm$^{-2}$ in the [0.5-2] keV band. The XXL website is \texttt{http://irfu.cea.fr/xxl}. Multi-band information and spectroscopic follow-up of the X-ray sources are obtained through a number of survey programmes summarised at \texttt{http://xxlmultiwave.pbworks.com/}. DE acknowledges support from the Swiss National Research Foundation. AMCLB acknowledges support from an internally funded PhD studentship at the Astrophysics Research Institute of Liverpool John Moores University and from the French Agence Nationale de la Recherche under grant ANR-11-BD56-015. FP acknowledges support from the BMBF/DLR grant 50 OR 1117, the DFG grant RE 1462-6 and the DFG Transregio Programme TR33.

\end{acknowledgements}

\bibliographystyle{aa}
\bibliography{xxl_fgas}

\begin{thebibliography}{112}
\expandafter\ifx\csname natexlab\endcsname\relax\def\natexlab#1{#1}\fi

\bibitem[{{Allen} {et~al.}(2004){Allen}, {Schmidt}, {Ebeling}, {Fabian}, \&
  {van Speybroeck}}]{allen04}
{Allen}, S.~W., {Schmidt}, R.~W., {Ebeling}, H., {Fabian}, A.~C., \& {van
  Speybroeck}, L. 2004, \mnras, 353, 457

\bibitem[{{Anders} \& {Grevesse}(1989)}]{ag89}
{Anders}, E. \& {Grevesse}, N. 1989, \gca, 53, 197

\bibitem[{{Andreon}(2010)}]{andreon10}
{Andreon}, S. 2010, \mnras, 407, 263

\bibitem[{{Andreon}(2015)}]{andreon15}
{Andreon}, S. 2015, \aap, 575, A108

\bibitem[{{Applegate} {et~al.}(2015){Applegate}, {Mantz}, {Allen}, {von der
  Linden}, {Morris}, {Hilbert}, {Kelly}, {Burke}, {Ebeling}, {Rapetti}, \&
  {Schmidt}}]{applegate15}
{Applegate}, D.~E., {Mantz}, A., {Allen}, S.~W., {et~al.} 2015, \mnras,
  submitted [\eprint[arXiv]{1509.02162}]

\bibitem[{{Arnaud} \& {Evrard}(1999)}]{arnaudlxt}
{Arnaud}, M. \& {Evrard}, A.~E. 1999, \mnras, 305, 631

\bibitem[{{Arnaud} {et~al.}(2007){Arnaud}, {Pointecouteau}, \&
  {Pratt}}]{arnaud07}
{Arnaud}, M., {Pointecouteau}, E., \& {Pratt}, G.~W. 2007, \aap, 474, L37

\bibitem[{{Arnaud} {et~al.}(2010){Arnaud}, {Pratt}, {Piffaretti},
  {B{\"o}hringer}, {Croston}, \& {Pointecouteau}}]{arnaud2010}
{Arnaud}, M., {Pratt}, G.~W., {Piffaretti}, R., {et~al.} 2010, \aap, 517, A92

\bibitem[{{Arnouts} {et~al.}(2013){Arnouts}, {Le Floc'h}, {Chevallard},
  {Johnson}, {Ilbert}, {Treyer}, {Aussel}, {Capak}, {Sanders}, {Scoville},
  {McCracken}, {Milliard}, {Pozzetti}, \& {Salvato}}]{arnouts13}
{Arnouts}, S., {Le Floc'h}, E., {Chevallard}, J., {et~al.} 2013, \aap, 558, A67

\bibitem[{{Battaglia} {et~al.}(2013){Battaglia}, {Bond}, {Pfrommer}, \&
  {Sievers}}]{battaglia13}
{Battaglia}, N., {Bond}, J.~R., {Pfrommer}, C., \& {Sievers}, J.~L. 2013, \apj,
  777, 123

\bibitem[{{Behroozi} {et~al.}(2010){Behroozi}, {Conroy}, \&
  {Wechsler}}]{behroozi10}
{Behroozi}, P.~S., {Conroy}, C., \& {Wechsler}, R.~H. 2010, \apj, 717, 379

\bibitem[{{Booth} \& {Schaye}(2009)}]{booth09}
{Booth}, C.~M. \& {Schaye}, J. 2009, \mnras, 398, 53

\bibitem[{{Bruzual} \& {Charlot}(2003)}]{bruzual03}
{Bruzual}, G. \& {Charlot}, S. 2003, \mnras, 344, 1000

\bibitem[{{Bryan} \& {Norman}(1998)}]{bryan98}
{Bryan}, G.~L. \& {Norman}, M.~L. 1998, \apj, 495, 80

\bibitem[{{Budzynski} {et~al.}(2014){Budzynski}, {Koposov}, {McCarthy}, \&
  {Belokurov}}]{budzynski14}
{Budzynski}, J.~M., {Koposov}, S.~E., {McCarthy}, I.~G., \& {Belokurov}, V.
  2014, \mnras, 437, 1362

\bibitem[{{Buote} \& {Humphrey}(2012)}]{buote12b}
{Buote}, D.~A. \& {Humphrey}, P.~J. 2012, in Astrophysics and Space Science
  Library, Vol. 378, Astrophysics and Space Science Library, ed. D.-W. {Kim} \&
  S.~{Pellegrini}, 235

\bibitem[{{Cavaliere} \& {Fusco-Femiano}(1976)}]{cavaliere}
{Cavaliere}, A. \& {Fusco-Femiano}, R. 1976, \aap, 49, 137

\bibitem[{{Chabrier}(2003)}]{chabrier03}
{Chabrier}, G. 2003, \pasp, 115, 763

\bibitem[{{Coupon} {et~al.}(2015){Coupon}, {Arnouts}, {van Waerbeke},
  {Moutard}, {Ilbert}, {van Uitert}, {Erben}, {Garilli}, {Guzzo}, {Heymans},
  {Hildebrandt}, {Hoekstra}, {Kilbinger}, {Kitching}, {Mellier}, {Miller},
  {Scodeggio}, {Bonnett}, {Branchini}, {Davidzon}, {De Lucia}, {Fritz}, {Fu},
  {Hudelot}, {Hudson}, {Kuijken}, {Leauthaud}, {Le F{\`e}vre}, {McCracken},
  {Moscardini}, {Rowe}, {Schrabback}, {Semboloni}, \& {Velander}}]{coupon15}
{Coupon}, J., {Arnouts}, S., {van Waerbeke}, L., {et~al.} 2015, \mnras, 449,
  1352

\bibitem[{{Donahue} {et~al.}(2014){Donahue}, {Voit}, {Mahdavi}, {Umetsu},
  {Ettori}, {Merten}, {Postman}, {Hoffer}, {Baldi}, {Coe}, {Czakon},
  {Bartelmann}, {Benitez}, {Bouwens}, {Bradley}, {Broadhurst}, {Ford},
  {Gastaldello}, {Grillo}, {Infante}, {Jouvel}, {Koekemoer}, {Kelson}, {Lahav},
  {Lemze}, {Medezinski}, {Melchior}, {Meneghetti}, {Molino}, {Moustakas},
  {Moustakas}, {Nonino}, {Rosati}, {Sayers}, {Seitz}, {Van der Wel}, {Zheng},
  \& {Zitrin}}]{donahue14}
{Donahue}, M., {Voit}, G.~M., {Mahdavi}, A., {et~al.} 2014, \apj, 794, 136

\bibitem[{{Eckert} {et~al.}(2013){Eckert}, {Ettori}, {Molendi}, {Vazza}, \&
  {Paltani}}]{eckert13b}
{Eckert}, D., {Ettori}, S., {Molendi}, S., {Vazza}, F., \& {Paltani}, S. 2013,
  \aap, 551, A23

\bibitem[{{Eckert} {et~al.}(2011){Eckert}, {Molendi}, \& {Paltani}}]{ccbias1}
{Eckert}, D., {Molendi}, S., \& {Paltani}, S. 2011, \aap, 526, A79+

\bibitem[{{Eckert} {et~al.}(2012){Eckert}, {Vazza}, {Ettori}, {Molendi},
  {Nagai}, {Lau}, {Roncarelli}, {Rossetti}, {Snowden}, \& {Gastaldello}}]{e12}
{Eckert}, D., {Vazza}, F., {Ettori}, S., {et~al.} 2012, \aap, 541, A57

\bibitem[{{Eke} {et~al.}(1998){Eke}, {Navarro}, \& {Frenk}}]{eke98}
{Eke}, V.~R., {Navarro}, J.~F., \& {Frenk}, C.~S. 1998, \apj, 503, 569

\bibitem[{{Ettori}(2003)}]{ettori03b}
{Ettori}, S. 2003, \mnras, 344, L13

\bibitem[{{Ettori}(2015)}]{ettori15}
{Ettori}, S. 2015, \mnras, 446, 2629

\bibitem[{{Ettori} \& {Fabian}(1999)}]{ettori98}
{Ettori}, S. \& {Fabian}, A.~C. 1999, \mnras, 305, 834

\bibitem[{{Ettori} {et~al.}(2003){Ettori}, {Tozzi}, \& {Rosati}}]{ettori03}
{Ettori}, S., {Tozzi}, P., \& {Rosati}, P. 2003, \aap, 398, 879

\bibitem[{{Evrard}(1997)}]{evrard97}
{Evrard}, A.~E. 1997, \mnras, 292, 289

\bibitem[{{Fabian} {et~al.}(1981){Fabian}, {Hu}, {Cowie}, \&
  {Grindlay}}]{fabian81}
{Fabian}, A.~C., {Hu}, E.~M., {Cowie}, L.~L., \& {Grindlay}, J. 1981, \apj,
  248, 47

\bibitem[{{Foreman-Mackey} {et~al.}(2013){Foreman-Mackey}, {Hogg}, {Lang}, \&
  {Goodman}}]{foreman-mackey13}
{Foreman-Mackey}, D., {Hogg}, D.~W., {Lang}, D., \& {Goodman}, J. 2013, \pasp,
  125, 306

\bibitem[{{Frank} {et~al.}(2013){Frank}, {Peterson}, {Andersson}, {Fabian}, \&
  {Sanders}}]{frank13}
{Frank}, K.~A., {Peterson}, J.~R., {Andersson}, K., {Fabian}, A.~C., \&
  {Sanders}, J.~S. 2013, \apj, 764, 46

\bibitem[{{Gaspari}(2015)}]{gaspari15}
{Gaspari}, M. 2015, \mnras, 451, L60

\bibitem[{{George} {et~al.}(2011){George}, {Leauthaud}, {Bundy}, {Finoguenov},
  {Tinker}, {Lin}, {Mei}, {Kneib}, {Aussel}, {Behroozi}, {Busha}, {Capak},
  {Coccato}, {Covone}, {Faure}, {Fiorenza}, {Ilbert}, {Le Floc'h}, {Koekemoer},
  {Tanaka}, {Wechsler}, \& {Wolk}}]{george11}
{George}, M.~R., {Leauthaud}, A., {Bundy}, K., {et~al.} 2011, \apj, 742, 125

\bibitem[{{Giles} {et~al.}(2015){Giles}, {Maughan}, {Pacaud}, \& {et
  al.}}]{xxl3}
{Giles}, P., {Maughan}, B., {Pacaud}, F., \& {et al.} 2015, \aap, submitted
  (XXL Survey, III)

\bibitem[{{Giodini} {et~al.}(2009){Giodini}, {Pierini}, {Finoguenov}, {Pratt},
  {Boehringer}, {Leauthaud}, {Guzzo}, {Aussel}, {Bolzonella}, {Capak}, {Elvis},
  {Hasinger}, {Ilbert}, {Kartaltepe}, {Koekemoer}, {Lilly}, {Massey},
  {McCracken}, {Rhodes}, {Salvato}, {Sanders}, {Scoville}, {Sasaki}, {Smolcic},
  {Taniguchi}, {Thompson}, \& {COSMOS Collaboration}}]{giodini09}
{Giodini}, S., {Pierini}, D., {Finoguenov}, A., {et~al.} 2009, \apj, 703, 982

\bibitem[{{Gonzalez} {et~al.}(2007){Gonzalez}, {Zaritsky}, \&
  {Zabludoff}}]{gonzalez07}
{Gonzalez}, A.~H., {Zaritsky}, D., \& {Zabludoff}, A.~I. 2007, \apj, 666, 147

\bibitem[{{Guzzo} {et~al.}(2014){Guzzo}, {Scodeggio}, {Garilli}, {Granett},
  {Fritz}, {Abbas}, {Adami}, {Arnouts}, {Bel}, {Bolzonella}, {Bottini},
  {Branchini}, {Cappi}, {Coupon}, {Cucciati}, {Davidzon}, {De Lucia}, {de la
  Torre}, {Franzetti}, {Fumana}, {Hudelot}, {Ilbert}, {Iovino}, {Krywult}, {Le
  Brun}, {Le F{\`e}vre}, {Maccagni}, {Ma{\l}ek}, {Marulli}, {McCracken},
  {Paioro}, {Peacock}, {Polletta}, {Pollo}, {Schlagenhaufer}, {Tasca},
  {Tojeiro}, {Vergani}, {Zamorani}, {Zanichelli}, {Burden}, {Di Porto},
  {Marchetti}, {Marinoni}, {Mellier}, {Moscardini}, {Nichol}, {Percival},
  {Phleps}, \& {Wolk}}]{guzzo14}
{Guzzo}, L., {Scodeggio}, M., {Garilli}, B., {et~al.} 2014, \aap, 566, A108

\bibitem[{{Helsdon} \& {Ponman}(2000)}]{helsdon00}
{Helsdon}, S.~F. \& {Ponman}, T.~J. 2000, \mnras, 315, 356

\bibitem[{{Hinshaw} {et~al.}(2013){Hinshaw}, {Larson}, {Komatsu}, {Spergel},
  {Bennett}, {Dunkley}, {Nolta}, {Halpern}, {Hill}, {Odegard}, {Page}, {Smith},
  {Weiland}, {Gold}, {Jarosik}, {Kogut}, {Limon}, {Meyer}, {Tucker}, {Wollack},
  \& {Wright}}]{wmap9}
{Hinshaw}, G., {Larson}, D., {Komatsu}, E., {et~al.} 2013, \apjs, 208, 19

\bibitem[{{Hoekstra} {et~al.}(2015){Hoekstra}, {Herbonnet}, {Muzzin}, {Babul},
  {Mahdavi}, {Viola}, \& {Cacciato}}]{hoekstra15}
{Hoekstra}, H., {Herbonnet}, R., {Muzzin}, A., {et~al.} 2015, \mnras, 449, 685

\bibitem[{{Hudson} {et~al.}(2010){Hudson}, {Mittal}, {Reiprich}, {Nulsen},
  {Andernach}, \& {Sarazin}}]{hudson10}
{Hudson}, D.~S., {Mittal}, R., {Reiprich}, T.~H., {et~al.} 2010, \aap, 513,
  A37+

\bibitem[{{Kalberla} {et~al.}(2005){Kalberla}, {Burton}, {Hartmann}, {Arnal},
  {Bajaja}, {Morras}, \& {P{\"o}ppel}}]{kalberla}
{Kalberla}, P.~M.~W., {Burton}, W.~B., {Hartmann}, D., {et~al.} 2005, \aap,
  440, 775

\bibitem[{{Kelly}(2007)}]{kelly07}
{Kelly}, B.~C. 2007, \apj, 665, 1489

\bibitem[{{Kettula} {et~al.}(2013){Kettula}, {Finoguenov}, {Massey}, {Rhodes},
  {Hoekstra}, {Taylor}, {Spinelli}, {Tanaka}, {Ilbert}, {Capak}, {McCracken},
  \& {Koekemoer}}]{kettula13}
{Kettula}, K., {Finoguenov}, A., {Massey}, R., {et~al.} 2013, \apj, 778, 74

\bibitem[{{Kettula} {et~al.}(2015){Kettula}, {Giodini}, {van Uitert},
  {Hoekstra}, {Finoguenov}, {Lerchster}, {Erben}, {Heymans}, {Hildebrandt},
  {Kitching}, {Mahdavi}, {Mellier}, {Miller}, {Mirkazemi}, {Van Waerbeke},
  {Coupon}, {Egami}, {Fu}, {Hudson}, {Kneib}, {Kuijken}, {McCracken},
  {Pereira}, {Rowe}, {Schrabback}, {Tanaka}, \& {Velander}}]{kettula15}
{Kettula}, K., {Giodini}, S., {van Uitert}, E., {et~al.} 2015, \mnras, 451,
  1460

\bibitem[{{Kriss} {et~al.}(1983){Kriss}, {Cioffi}, \& {Canizares}}]{kriss}
{Kriss}, G.~A., {Cioffi}, D.~F., \& {Canizares}, C.~R. 1983, \apj, 272, 439

\bibitem[{{Kuntz} \& {Snowden}(2008)}]{kuntz08}
{Kuntz}, K.~D. \& {Snowden}, S.~L. 2008, \aap, 478, 575

\bibitem[{{Lagan{\'a}} {et~al.}(2013){Lagan{\'a}}, {Martinet}, {Durret}, {Lima
  Neto}, {Maughan}, \& {Zhang}}]{lagana13}
{Lagan{\'a}}, T.~F., {Martinet}, N., {Durret}, F., {et~al.} 2013, \aap, 555,
  A66

\bibitem[{{Landy} \& {Szalay}(1993)}]{landy93}
{Landy}, S.~D. \& {Szalay}, A.~S. 1993, \apj, 412, 64

\bibitem[{{Lavoie} {et~al.}(2015){Lavoie}, {Willis}, {Democles}, \& {et
  al.}}]{xxl15}
{Lavoie}, S., {Willis}, J.~P., {Democles}, J., \& {et al.} 2015, \mnras,
  submitted (XXL Survey, XV)

\bibitem[{{Le Brun} {et~al.}(2014){Le Brun}, {McCarthy}, {Schaye}, \&
  {Ponman}}]{lebrun14}
{Le Brun}, A.~M.~C., {McCarthy}, I.~G., {Schaye}, J., \& {Ponman}, T.~J. 2014,
  \mnras, 441, 1270

\bibitem[{{Le F{\`e}vre} {et~al.}(2005){Le F{\`e}vre}, {Guzzo}, {Meneux},
  {Pollo}, {Cappi}, {Colombi}, {Iovino}, {Marinoni}, {McCracken}, {Scaramella},
  {Bottini}, {Garilli}, {Le Brun}, {Maccagni}, {Picat}, {Scodeggio}, {Tresse},
  {Vettolani}, {Zanichelli}, {Adami}, {Arnaboldi}, {Arnouts}, {Bardelli},
  {Blaizot}, {Bolzonella}, {Charlot}, {Ciliegi}, {Contini}, {Foucaud},
  {Franzetti}, {Gavignaud}, {Ilbert}, {Marano}, {Mathez}, {Mazure}, {Merighi},
  {Paltani}, {Pell{\`o}}, {Pozzetti}, {Radovich}, {Zamorani}, {Zucca}, {Bondi},
  {Bongiorno}, {Busarello}, {Lamareille}, {Mellier}, {Merluzzi}, {Ripepi}, \&
  {Rizzo}}]{lefevre05}
{Le F{\`e}vre}, O., {Guzzo}, L., {Meneux}, B., {et~al.} 2005, \aap, 439, 877

\bibitem[{{Leauthaud} {et~al.}(2012){Leauthaud}, {Tinker}, {Bundy}, {Behroozi},
  {Massey}, {Rhodes}, {George}, {Kneib}, {Benson}, {Wechsler}, {Busha},
  {Capak}, {Cort{\^e}s}, {Ilbert}, {Koekemoer}, {Le F{\`e}vre}, {Lilly},
  {McCracken}, {Salvato}, {Schrabback}, {Scoville}, {Smith}, \&
  {Taylor}}]{leauthaud12}
{Leauthaud}, A., {Tinker}, J., {Bundy}, K., {et~al.} 2012, \apj, 744, 159

\bibitem[{{Leccardi} \& {Molendi}(2008{\natexlab{a}})}]{lm08b}
{Leccardi}, A. \& {Molendi}, S. 2008{\natexlab{a}}, \aap, 487, 461

\bibitem[{{Leccardi} \& {Molendi}(2008{\natexlab{b}})}]{lm08}
{Leccardi}, A. \& {Molendi}, S. 2008{\natexlab{b}}, \aap, 486, 359

\bibitem[{{Lieu} {et~al.}(2015){Lieu}, {Smith}, {Giles}, \& {et al.}}]{xxl4}
{Lieu}, M., {Smith}, G.~P., {Giles}, P., \& {et al.} 2015, \aap, submitted (XXL
  Survey, IV)

\bibitem[{{Lin} \& {Mohr}(2004)}]{lin04}
{Lin}, Y.-T. \& {Mohr}, J.~J. 2004, \apj, 617, 879

\bibitem[{{Lin} {et~al.}(2003){Lin}, {Mohr}, \& {Stanford}}]{lin03}
{Lin}, Y.-T., {Mohr}, J.~J., \& {Stanford}, S.~A. 2003, \apj, 591, 749

\bibitem[{{Lovisari} {et~al.}(2015){Lovisari}, {Reiprich}, \&
  {Schellenberger}}]{lovisari15}
{Lovisari}, L., {Reiprich}, T.~H., \& {Schellenberger}, G. 2015, \aap, 573,
  A118

\bibitem[{{Lucy}(1974)}]{lucy74}
{Lucy}, L.~B. 1974, \aj, 79, 745

\bibitem[{{Lucy}(1994)}]{lucy94}
{Lucy}, L.~B. 1994, \aap, 289, 983

\bibitem[{{Mahdavi} {et~al.}(2007){Mahdavi}, {Hoekstra}, {Babul}, {Sievers},
  {Myers}, \& {Henry}}]{mahdavi07}
{Mahdavi}, A., {Hoekstra}, H., {Babul}, A., {et~al.} 2007, \apj, 664, 162

\bibitem[{{Mantz} {et~al.}(2015){Mantz}, {von der Linden}, {Allen},
  {Applegate}, {Kelly}, {Morris}, {Rapetti}, {Schmidt}, {Adhikari}, {Allen},
  {Burchat}, {Burke}, {Cataneo}, {Donovan}, {Ebeling}, {Shandera}, \&
  {Wright}}]{mantz15}
{Mantz}, A.~B., {von der Linden}, A., {Allen}, S.~W., {et~al.} 2015, \mnras,
  446, 2205

\bibitem[{{McCarthy} {et~al.}(2014){McCarthy}, {Le Brun}, {Schaye}, \&
  {Holder}}]{mccarthy14}
{McCarthy}, I.~G., {Le Brun}, A.~M.~C., {Schaye}, J., \& {Holder}, G.~P. 2014,
  \mnras, 440, 3645

\bibitem[{{McCarthy} {et~al.}(2011){McCarthy}, {Schaye}, {Bower}, {Ponman},
  {Booth}, {Dalla Vecchia}, \& {Springel}}]{mccarthy11}
{McCarthy}, I.~G., {Schaye}, J., {Bower}, R.~G., {et~al.} 2011, \mnras, 412,
  1965

\bibitem[{{McCarthy} {et~al.}(2010){McCarthy}, {Schaye}, {Ponman}, {Bower},
  {Booth}, {Dalla Vecchia}, {Crain}, {Springel}, {Theuns}, \&
  {Wiersma}}]{mccarthy10}
{McCarthy}, I.~G., {Schaye}, J., {Ponman}, T.~J., {et~al.} 2010, \mnras, 406,
  822

\bibitem[{{Melin} {et~al.}(2006){Melin}, {Bartlett}, \&
  {Delabrouille}}]{melin2006}
{Melin}, J.-B., {Bartlett}, J.~G., \& {Delabrouille}, J. 2006, \aap, 459, 341

\bibitem[{{Mohr} {et~al.}(1999){Mohr}, {Mathiesen}, \& {Evrard}}]{mohr99}
{Mohr}, J.~J., {Mathiesen}, B., \& {Evrard}, A.~E. 1999, \apj, 517, 627

\bibitem[{{Morandi} {et~al.}(2007){Morandi}, {Ettori}, \&
  {Moscardini}}]{morandi07}
{Morandi}, A., {Ettori}, S., \& {Moscardini}, L. 2007, \mnras, 379, 518

\bibitem[{{Mulroy} {et~al.}(2014){Mulroy}, {Smith}, {Haines}, {Marrone},
  {Okabe}, {Pereira}, {Egami}, {Babul}, {Finoguenov}, \& {Martino}}]{mulroy14}
{Mulroy}, S.~L., {Smith}, G.~P., {Haines}, C.~P., {et~al.} 2014, \mnras, 443,
  3309

\bibitem[{{Nagai} {et~al.}(2007){Nagai}, {Vikhlinin}, \& {Kravtsov}}]{nagai07}
{Nagai}, D., {Vikhlinin}, A., \& {Kravtsov}, A.~V. 2007, \apj, 655, 98

\bibitem[{{Navarro} {et~al.}(1997){Navarro}, {Frenk}, \& {White}}]{nfw97}
{Navarro}, J.~F., {Frenk}, C.~S., \& {White}, S.~D.~M. 1997, \apj, 490, 493

\bibitem[{{Nelson} {et~al.}(2012){Nelson}, {Rudd}, {Shaw}, \&
  {Nagai}}]{nelson12}
{Nelson}, K., {Rudd}, D.~H., {Shaw}, L., \& {Nagai}, D. 2012, \apj, 751, 121

\bibitem[{{Nevalainen} {et~al.}(2010){Nevalainen}, {David}, \&
  {Guainazzi}}]{nevalainen10}
{Nevalainen}, J., {David}, L., \& {Guainazzi}, M. 2010, \aap, 523, A22

\bibitem[{{Okabe} \& {Smith}(2015)}]{okabe15}
{Okabe}, N. \& {Smith}, G.~P. 2015, \mnras, submitted
  [\eprint[arXiv]{1507.04493}]

\bibitem[{{Owers} {et~al.}(2009){Owers}, {Nulsen}, {Couch}, \&
  {Markevitch}}]{owers09}
{Owers}, M.~S., {Nulsen}, P.~E.~J., {Couch}, W.~J., \& {Markevitch}, M. 2009,
  \apj, 704, 1349

\bibitem[{{Pacaud} {et~al.}(2015){Pacaud}, {Clerc}, {Giles}, \& {et
  al.}}]{xxl2}
{Pacaud}, F., {Clerc}, N., {Giles}, P., \& {et al.} 2015, \aap, submitted (XXL
  Survey, II)

\bibitem[{{Pacaud} {et~al.}(2006){Pacaud}, {Pierre}, {Refregier}, {Gueguen},
  {Starck}, {Valtchanov}, {Read}, {Altieri}, {Chiappetti}, {Gandhi}, {Garcet},
  {Gosset}, {Ponman}, \& {Surdej}}]{pacaud06}
{Pacaud}, F., {Pierre}, M., {Refregier}, A., {et~al.} 2006, \mnras, 372, 578

\bibitem[{{Pierre} {et~al.}(2015){Pierre}, {Pacaud}, {Adami}, \& {et
  al.}}]{xxl1}
{Pierre}, M., {Pacaud}, F., {Adami}, C., \& {et al.} 2015, \aap, submitted (XXL
  Survey, I)

\bibitem[{{Planck Collaboration X}(2011)}]{PlanckMCXC}
{Planck Collaboration X}. 2011, \aap, 536, A10

\bibitem[{{Planck Collaboration XI}(2011)}]{planck_11_11}
{Planck Collaboration XI}. 2011, \aap, 536, A11

\bibitem[{{Planck Collaboration XI}(2013)}]{PlanckLBG}
{Planck Collaboration XI}. 2013, \aap, 557, A52

\bibitem[{{Planck Collaboration XII}(2011)}]{PlanckMaxBCG}
{Planck Collaboration XII}. 2011, \aap, 536, A12

\bibitem[{{Planck Collaboration XIII}(2015)}]{planck15_13}
{Planck Collaboration XIII}. 2015, \aap, submitted [\eprint[arXiv]{1502.01589}]

\bibitem[{{Planck Collaboration XXIV}(2015)}]{planck15_24}
{Planck Collaboration XXIV}. 2015, \aap, submitted [\eprint[arXiv]{1502.01597}]

\bibitem[{{Planelles} {et~al.}(2013){Planelles}, {Borgani}, {Dolag}, {Ettori},
  {Fabjan}, {Murante}, \& {Tornatore}}]{planelles13}
{Planelles}, S., {Borgani}, S., {Dolag}, K., {et~al.} 2013, \mnras, 431, 1487

\bibitem[{{Pointecouteau} {et~al.}(2004){Pointecouteau}, {Arnaud}, {Kaastra},
  \& {de Plaa}}]{pointe478}
{Pointecouteau}, E., {Arnaud}, M., {Kaastra}, J., \& {de Plaa}, J. 2004, \aap,
  423, 33

\bibitem[{{Ponman} {et~al.}(1999){Ponman}, {Cannon}, \& {Navarro}}]{ponman99}
{Ponman}, T.~J., {Cannon}, D.~B., \& {Navarro}, J.~F. 1999, \nat, 397, 135

\bibitem[{{Pratt} {et~al.}(2009){Pratt}, {Croston}, {Arnaud}, \&
  {B{\"o}hringer}}]{pratt09}
{Pratt}, G.~W., {Croston}, J.~H., {Arnaud}, M., \& {B{\"o}hringer}, H. 2009,
  \aap, 498, 361

\bibitem[{{Rasia} {et~al.}(2006){Rasia}, {Ettori}, {Moscardini}, {Mazzotta},
  {Borgani}, {Dolag}, {Tormen}, {Cheng}, \& {Diaferio}}]{rasia06}
{Rasia}, E., {Ettori}, S., {Moscardini}, L., {et~al.} 2006, \mnras, 369, 2013

\bibitem[{{Rasia} {et~al.}(2011){Rasia}, {Mazzotta}, {Evrard}, {Markevitch},
  {Dolag}, \& {Meneghetti}}]{rasia11}
{Rasia}, E., {Mazzotta}, P., {Evrard}, A., {et~al.} 2011, \apj, 729, 45

\bibitem[{{Rasia} {et~al.}(2004){Rasia}, {Tormen}, \& {Moscardini}}]{rasia04}
{Rasia}, E., {Tormen}, G., \& {Moscardini}, L. 2004, \mnras, 351, 237

\bibitem[{{Rossetti} {et~al.}(2013){Rossetti}, {Eckert}, {De Grandi},
  {Gastaldello}, {Ghizzardi}, {Roediger}, \& {Molendi}}]{rossetti13}
{Rossetti}, M., {Eckert}, D., {De Grandi}, S., {et~al.} 2013, \aap, 556, A44

\bibitem[{{Rozo} {et~al.}(2014){Rozo}, {Rykoff}, {Bartlett}, \&
  {Evrard}}]{rozo14}
{Rozo}, E., {Rykoff}, E.~S., {Bartlett}, J.~G., \& {Evrard}, A. 2014, \mnras,
  438, 49

\bibitem[{{Sarazin} \& {Bahcall}(1977)}]{sarazin77}
{Sarazin}, C.~L. \& {Bahcall}, J.~N. 1977, \apjs, 34, 451

\bibitem[{{Schaye} {et~al.}(2010){Schaye}, {Dalla Vecchia}, {Booth}, {Wiersma},
  {Theuns}, {Haas}, {Bertone}, {Duffy}, {McCarthy}, \& {van de
  Voort}}]{schaye10}
{Schaye}, J., {Dalla Vecchia}, C., {Booth}, C.~M., {et~al.} 2010, \mnras, 402,
  1536

\bibitem[{{Schellenberger} {et~al.}(2015){Schellenberger}, {Reiprich},
  {Lovisari}, {Nevalainen}, \& {David}}]{schellenberger15}
{Schellenberger}, G., {Reiprich}, T.~H., {Lovisari}, L., {Nevalainen}, J., \&
  {David}, L. 2015, \aap, 575, A30

\bibitem[{{Sembolini} {et~al.}(2015){Sembolini}, {Yepes}, {Pearce}, {Knebe},
  {Kay}, {Power}, {Cui}, {Beck}, {Borgani}, {Dalla Vecchia}, {Dav{\'e}}, {Jahan
  Elahi}, {February}, {Huang}, {Hobbs}, {Katz}, {Lau}, {McCarthy}, {Murante},
  {Nagai}, {Nelson}, {Newton}, {Puchwein}, {Read}, {Saro}, {Schaye}, \&
  {Thacker}}]{sembolini15}
{Sembolini}, F., {Yepes}, G., {Pearce}, F.~R., {et~al.} 2015, \mnras, submitted
  [\eprint[arXiv]{1503.06065}]

\bibitem[{{Sereno} \& {Ettori}(2015)}]{sereno15}
{Sereno}, M. \& {Ettori}, S. 2015, \mnras, 450, 3633

\bibitem[{{Smith} {et~al.}(2015){Smith}, {Mazzotta}, {Okabe}, {Ziparo},
  {Mulroy}, {Babul}, {Finoguenov}, {McCarthy}, {Lieu}, {Bahe}, {Bourdin},
  {Evrard}, {Futamase}, {Haines}, {Jauzac}, {Marrone}, {Martino}, {May},
  {Taylor}, \& {Umetsu}}]{smith15}
{Smith}, G.~P., {Mazzotta}, P., {Okabe}, N., {et~al.} 2015, \mnras, in press
  [\eprint[arXiv]{1511.01919}]

\bibitem[{{Smith} {et~al.}(2001){Smith}, {Brickhouse}, {Liedahl}, \&
  {Raymond}}]{apec}
{Smith}, R.~K., {Brickhouse}, N.~S., {Liedahl}, D.~A., \& {Raymond}, J.~C.
  2001, \apjl, 556, L91

\bibitem[{{Springel}(2005)}]{springel05}
{Springel}, V. 2005, \mnras, 364, 1105

\bibitem[{{Sun}(2012)}]{sun12}
{Sun}, M. 2012, New Journal of Physics, 14, 045004

\bibitem[{{Sun} {et~al.}(2009){Sun}, {Voit}, {Donahue}, {Jones}, {Forman}, \&
  {Vikhlinin}}]{sun09}
{Sun}, M., {Voit}, G.~M., {Donahue}, M., {et~al.} 2009, \apj, 693, 1142

\bibitem[{{Valtchanov} {et~al.}(2001){Valtchanov}, {Pierre}, \&
  {Gastaud}}]{valtchanov01}
{Valtchanov}, I., {Pierre}, M., \& {Gastaud}, R. 2001, \aap, 370, 689

\bibitem[{{van der Burg} {et~al.}(2014){van der Burg}, {Muzzin}, {Hoekstra},
  {Wilson}, {Lidman}, \& {Yee}}]{vdb14}
{van der Burg}, R.~F.~J., {Muzzin}, A., {Hoekstra}, H., {et~al.} 2014, \aap,
  561, A79

\bibitem[{{Vikhlinin} {et~al.}(2006){Vikhlinin}, {Kravtsov}, {Forman}, {Jones},
  {Markevitch}, {Murray}, \& {Van Speybroeck}}]{vikhlinin06}
{Vikhlinin}, A., {Kravtsov}, A., {Forman}, W., {et~al.} 2006, \apj, 640, 691

\bibitem[{{von der Linden} {et~al.}(2014{\natexlab{a}}){von der Linden},
  {Allen}, {Applegate}, {Kelly}, {Allen}, {Ebeling}, {Burchat}, {Burke},
  {Donovan}, {Morris}, {Blandford}, {Erben}, \& {Mantz}}]{vdl14b}
{von der Linden}, A., {Allen}, M.~T., {Applegate}, D.~E., {et~al.}
  2014{\natexlab{a}}, \mnras, 439, 2

\bibitem[{{von der Linden} {et~al.}(2014{\natexlab{b}}){von der Linden},
  {Mantz}, {Allen}, {Applegate}, {Kelly}, {Morris}, {Wright}, {Allen},
  {Burchat}, {Burke}, {Donovan}, \& {Ebeling}}]{vdl14}
{von der Linden}, A., {Mantz}, A., {Allen}, S.~W., {et~al.} 2014{\natexlab{b}},
  \mnras, 443, 1973

\bibitem[{{White} {et~al.}(1993){White}, {Navarro}, {Evrard}, \&
  {Frenk}}]{white93}
{White}, S.~D.~M., {Navarro}, J.~F., {Evrard}, A.~E., \& {Frenk}, C.~S. 1993,
  \nat, 366, 429

\bibitem[{{Zibetti} {et~al.}(2005){Zibetti}, {White}, {Schneider}, \&
  {Brinkmann}}]{zibetti05}
{Zibetti}, S., {White}, S.~D.~M., {Schneider}, D.~P., \& {Brinkmann}, J. 2005,
  \mnras, 358, 949

\end{thebibliography}

\appendix
\section{Master table}

\onecolumn
{\setstretch{1.3}
\begin{longtab}
\begin{longtable}{ccccccc}
\caption{\label{tab:data} Basic data}\\
\hline\hline
Cluster & $z$ & $T_{\rm 300 kpc}$  & $r_{\rm 500, MT}$ & $M_{\rm gas, 500}$  & $Y_{X, 500}$  & $f_{\rm gas, 500}$ \\
 & & [keV] & [kpc] & [$M_\odot$] & [keV $M_\odot$] & \\ \hline
\endfirsthead
\caption{continued.}\\
\hline\hline
Cluster & $z$ & $T_{\rm 300 kpc}$  & $r_{\rm 500, MT}$ & $M_{\rm gas, 500}$  & $Y_{X, 500}$  & $f_{\rm gas, 500}$ \\
 & & [keV] & [kpc] & [$M_\odot$] & [keV $M_\odot$] & \\ \hline
\endhead
\hline
\endfoot
XLSSC 001$^\star$ & 0.614 & $ 3.8_{-0.4}^{+0.6}$ & $ 777 \pm 118 $ & $(2.68_{-0.56}^{+0.56})\times10^{13}$ & $(1.04_{-0.24}^{+0.27})\times10^{14}$ & $0.106_{-0.055}^{+0.053}$ \\ 
XLSSC 003 & 0.836 & $ 3.4_{-0.6}^{+1.0}$ & $ 643 \pm 106 $ & $(1.09_{-0.19}^{+0.24})\times10^{13}$ & $(4.06_{-1.04}^{+1.42})\times10^{13}$ & $0.059_{-0.039}^{+0.032}$ \\ 
XLSSC 006$^\star$ & 0.429 & $ 4.8_{-0.5}^{+0.5}$ & $ 982 \pm 145 $ & $(4.09_{-0.83}^{+0.82})\times10^{13}$ & $(1.97_{-0.43}^{+0.46})\times10^{14}$ & $0.099_{-0.048}^{+0.048}$ \\ 
XLSSC 010 & 0.330 & $ 2.7_{-0.3}^{+0.5}$ & $ 751 \pm 113 $ & $(7.25_{-1.32}^{+1.60})\times10^{12}$ & $(2.04_{-0.44}^{+0.57})\times10^{13}$ & $0.044_{-0.024}^{+0.022}$ \\ 
XLSSC 011$^\star$ & 0.054 & $ 2.5_{-0.4}^{+0.5}$ & $ 831 \pm 136 $ & $(1.65_{-0.37}^{+0.52})\times10^{12}$ & $(4.26_{-1.14}^{+1.58})\times10^{12}$ & $0.010_{-0.006}^{+0.006}$ \\ 
XLSSC 022$^\star$ & 0.293 & $ 2.1_{-0.1}^{+0.1}$ & $ 671 \pm 94 $ & $(6.63_{-0.83}^{+0.90})\times10^{12}$ & $(1.43_{-0.19}^{+0.20})\times10^{13}$ & $0.058_{-0.026}^{+0.026}$ \\ 
XLSSC 023 & 0.328 & $ 2.1_{-0.2}^{+0.3}$ & $ 655 \pm 96 $ & $(5.20_{-1.15}^{+1.21})\times10^{12}$ & $(1.15_{-0.28}^{+0.32})\times10^{13}$ & $0.047_{-0.026}^{+0.023}$ \\ 
XLSSC 025$^\star$ & 0.265 & $ 2.5_{-0.2}^{+0.2}$ & $ 751 \pm 110 $ & $(9.69_{-2.45}^{+2.88})\times10^{12}$ & $(2.44_{-0.64}^{+0.76})\times10^{13}$ & $0.062_{-0.031}^{+0.033}$ \\ 
XLSSC 027$^\star$ & 0.295 & $ 2.7_{-0.3}^{+0.4}$ & $ 768 \pm 116 $ & $(8.10_{-1.16}^{+1.30})\times10^{12}$ & $(2.27_{-0.42}^{+0.49})\times10^{13}$ & $0.047_{-0.024}^{+0.023}$ \\ 
XLSSC 029 & 1.050 & $ 4.1_{-0.6}^{+1.1}$ & $ 626 \pm 98 $ & $(2.14_{-0.45}^{+0.74})\times10^{13}$ & $(9.48_{-2.54}^{+4.16})\times10^{13}$ & $0.098_{-0.062}^{+0.057}$ \\ 
XLSSC 036 & 0.492 & $ 3.6_{-0.4}^{+0.5}$ & $ 801 \pm 120 $ & $(2.02_{-0.36}^{+0.41})\times10^{13}$ & $(7.33_{-1.49}^{+1.80})\times10^{13}$ & $0.083_{-0.042}^{+0.041}$ \\ 
XLSSC 041$^\star$ & 0.142 & $ 1.9_{-0.2}^{+0.1}$ & $ 670 \pm 98 $ & $(4.86_{-0.98}^{+1.07})\times10^{12}$ & $(8.90_{-1.90}^{+2.15})\times10^{12}$ & $0.050_{-0.024}^{+0.024}$ \\ 
XLSSC 050 & 0.140 & $ 3.1_{-0.2}^{+0.2}$ & $ 897 \pm 127 $ & $(1.75_{-0.25}^{+0.20})\times10^{13}$ & $(5.45_{-0.82}^{+0.72})\times10^{13}$ & $0.075_{-0.034}^{+0.033}$ \\ 
XLSSC 052 & 0.056 & $ 0.6_{-0.1}^{+0.0}$ & $ 387 \pm 56 $ & $(6.19_{-1.75}^{+2.28})\times10^{11}$ & $(3.90_{-1.12}^{+1.48})\times10^{11}$ & $0.036_{-0.018}^{+0.020}$ \\ 
XLSSC 054$^\star$ & 0.054 & $ 2.0_{-0.1}^{+0.1}$ & $ 723 \pm 105 $ & $(4.52_{-1.03}^{+1.07})\times10^{12}$ & $(8.86_{-2.10}^{+2.26})\times10^{12}$ & $0.040_{-0.020}^{+0.020}$ \\ 
XLSSC 055$^\star$ & 0.232 & $ 3.0_{-0.3}^{+0.3}$ & $ 843 \pm 128 $ & $(9.60_{-1.43}^{+1.70})\times10^{12}$ & $(2.88_{-0.52}^{+0.63})\times10^{13}$ & $0.045_{-0.022}^{+0.022}$ \\ 
XLSSC 056$^\star$ & 0.348 & $ 3.2_{-0.3}^{+0.5}$ & $ 824 \pm 124 $ & $(1.45_{-0.19}^{+0.23})\times10^{13}$ & $(4.88_{-0.85}^{+1.03})\times10^{13}$ & $0.065_{-0.033}^{+0.031}$ \\ 
XLSSC 057$^\star$ & 0.153 & $ 2.2_{-0.1}^{+0.3}$ & $ 734 \pm 105 $ & $(7.05_{-1.13}^{+1.09})\times10^{12}$ & $(1.62_{-0.29}^{+0.31})\times10^{13}$ & $0.054_{-0.027}^{+0.025}$ \\ 
XLSSC 060$^\star$ & 0.139 & $ 4.8_{-0.2}^{+0.2}$ & $ 1136 \pm 158 $ & $(3.69_{-0.51}^{+0.41})\times10^{13}$ & $(1.76_{-0.25}^{+0.20})\times10^{14}$ & $0.078_{-0.034}^{+0.034}$ \\ 
XLSSC 061$^\star$ & 0.259 & $ 2.1_{-0.3}^{+0.5}$ & $ 678 \pm 106 $ & $(7.15_{-1.58}^{+1.75})\times10^{12}$ & $(1.58_{-0.41}^{+0.51})\times10^{13}$ & $0.063_{-0.038}^{+0.033}$ \\ 
XLSSC 062 & 0.059 & $ 0.8_{-0.1}^{+0.1}$ & $ 422 \pm 61 $ & $(3.24_{-0.68}^{+1.26})\times10^{11}$ & $(2.45_{-0.54}^{+0.97})\times10^{11}$ & $0.014_{-0.007}^{+0.008}$ \\ 
XLSSC 072 & 1.002 & $ 3.7_{-0.6}^{+1.1}$ & $ 613 \pm 99 $ & $(2.62_{-0.72}^{+0.85})\times10^{13}$ & $(1.05_{-0.33}^{+0.45})\times10^{14}$ & $0.134_{-0.092}^{+0.078}$ \\ 
XLSSC 083$^\star$ & 0.430 & $ 4.5_{-0.7}^{+1.1}$ & $ 943 \pm 154 $ & $(3.07_{-0.90}^{+1.01})\times10^{13}$ & $(1.45_{-0.47}^{+0.60})\times10^{14}$ & $0.084_{-0.055}^{+0.049}$ \\ 
XLSSC 084 & 0.430 & $ 4.5_{-1.3}^{+1.6}$ & $ 945 \pm 202 $ & $(2.33_{-0.85}^{+1.20})\times10^{13}$ & $(1.05_{-0.46}^{+0.71})\times10^{14}$ & $0.063_{-0.050}^{+0.052}$ \\ 
XLSSC 085 & 0.428 & $ 4.8_{-1.0}^{+2.0}$ & $ 976 \pm 176 $ & $(1.44_{-0.28}^{+0.31})\times10^{13}$ & $(7.71_{-2.33}^{+3.10})\times10^{13}$ & $0.035_{-0.029}^{+0.021}$ \\ 
XLSSC 087$^\star$ & 0.141 & $ 1.6_{-0.1}^{+0.1}$ & $ 619 \pm 87 $ & $(2.33_{-0.36}^{+0.47})\times10^{12}$ & $(3.71_{-0.60}^{+0.78})\times10^{12}$ & $0.030_{-0.014}^{+0.014}$ \\ 
XLSSC 088$^\star$ & 0.295 & $ 2.5_{-0.4}^{+0.6}$ & $ 726 \pm 120 $ & $(9.73_{-2.07}^{+2.05})\times10^{12}$ & $(2.47_{-0.65}^{+0.78})\times10^{13}$ & $0.067_{-0.041}^{+0.036}$ \\ 
XLSSC 089$^\star$ & 0.609 & $ 3.7_{-1.2}^{+1.6}$ & $ 769 \pm 171 $ & $(1.26_{-0.28}^{+0.41})\times10^{13}$ & $(5.00_{-1.90}^{+2.60})\times10^{13}$ & $0.052_{-0.043}^{+0.038}$ \\ 
XLSSC 090$^\star$ & 0.141 & $ 1.1_{-0.1}^{+0.1}$ & $ 507 \pm 72 $ & $(1.04_{-0.54}^{+0.24})\times10^{12}$ & $(1.17_{-0.60}^{+0.30})\times10^{12}$ & $0.025_{-0.017}^{+0.012}$ \\ 
XLSSC 091$^\star$ & 0.186 & $ 5.1_{-0.2}^{+0.2}$ & $ 1149 \pm 161 $ & $(5.00_{-0.83}^{+0.80})\times10^{13}$ & $(2.53_{-0.43}^{+0.42})\times10^{14}$ & $0.097_{-0.044}^{+0.044}$ \\ 
XLSSC 092$^\star$ & 0.432 & $ 3.1_{-0.6}^{+0.8}$ & $ 771 \pm 138 $ & $(1.39_{-0.37}^{+0.41})\times10^{13}$ & $(4.37_{-1.40}^{+1.76})\times10^{13}$ & $0.069_{-0.044}^{+0.042}$ \\ 
XLSSC 093$^\star$ & 0.429 & $ 3.4_{-0.4}^{+0.6}$ & $ 810 \pm 123 $ & $(2.48_{-0.58}^{+0.62})\times10^{13}$ & $(8.70_{-2.23}^{+2.59})\times10^{13}$ & $0.107_{-0.059}^{+0.055}$ \\ 
XLSSC 094 & 0.886 & $ 4.7_{-0.9}^{+1.2}$ & $ 742 \pm 129 $ & $(1.61_{-0.44}^{+0.74})\times10^{13}$ & $(7.94_{-2.57}^{+4.11})\times10^{13}$ & $0.053_{-0.035}^{+0.037}$ \\ 
XLSSC 095 & 0.138 & $ 0.9_{-0.1}^{+0.1}$ & $ 450 \pm 64 $ & $(3.28_{-0.82}^{+1.60})\times10^{11}$ & $(3.01_{-0.78}^{+1.47})\times10^{11}$ & $0.011_{-0.006}^{+0.007}$ \\ 
XLSSC 096 & 0.520 & $ 5.5_{-1.1}^{+2.0}$ & $ 1000 \pm 180 $ & $(2.02_{-0.64}^{+0.84})\times10^{13}$ & $(1.20_{-0.44}^{+0.66})\times10^{14}$ & $0.042_{-0.034}^{+0.028}$ \\ 
XLSSC 097$^\star$ & 0.760 & $ 4.6_{-1.0}^{+1.5}$ & $ 794 \pm 143 $ & $(2.65_{-0.73}^{+0.87})\times10^{13}$ & $(1.29_{-0.44}^{+0.59})\times10^{14}$ & $0.083_{-0.061}^{+0.052}$ \\ 
XLSSC 098$^\star$ & 0.297 & $ 2.9_{-0.6}^{+1.0}$ & $ 801 \pm 139 $ & $(1.03_{-0.32}^{+0.41})\times10^{13}$ & $(3.27_{-1.18}^{+1.67})\times10^{13}$ & $0.053_{-0.040}^{+0.035}$ \\ 
XLSSC 099$^\star$ & 0.391 & $ 5.1_{-1.5}^{+3.1}$ & $ 1032 \pm 221 $ & $(4.96_{-1.51}^{+3.07})\times10^{12}$ & $(3.06_{-1.30}^{+2.38})\times10^{13}$ & $0.011_{-0.012}^{+0.010}$ \\ 
XLSSC 100 & 0.915 & $ 4.3_{-1.2}^{+1.7}$ & $ 694 \pm 143 $ & $(1.73_{-0.44}^{+0.55})\times10^{13}$ & $(7.88_{-2.88}^{+3.95})\times10^{13}$ & $0.068_{-0.057}^{+0.047}$ \\ 
XLSSC 101 & 0.756 & $ 4.5_{-0.8}^{+0.8}$ & $ 788 \pm 134 $ & $(2.76_{-0.49}^{+0.55})\times10^{13}$ & $(1.24_{-0.30}^{+0.36})\times10^{14}$ & $0.089_{-0.047}^{+0.049}$ \\ 
XLSSC 102$^\star$ & 0.969 & $ 3.2_{-0.5}^{+0.8}$ & $ 574 \pm 96 $ & $(2.23_{-0.67}^{+0.75})\times10^{13}$ & $(7.46_{-2.51}^{+3.27})\times10^{13}$ & $0.145_{-0.097}^{+0.087}$ \\ 
XLSSC 103$^\star$ & 0.233 & $ 3.5_{-0.8}^{+1.2}$ & $ 913 \pm 172 $ & $(1.30_{-0.42}^{+0.51})\times10^{13}$ & $(4.77_{-1.82}^{+2.54})\times10^{13}$ & $0.048_{-0.037}^{+0.033}$ \\ 
XLSSC 104$^\star$ & 0.294 & $ 4.7_{-1.0}^{+1.6}$ & $ 1038 \pm 188 $ & $(8.08_{-1.68}^{+4.01})\times10^{12}$ & $(4.16_{-1.30}^{+2.24})\times10^{13}$ & $0.019_{-0.014}^{+0.014}$ \\ 
XLSSC 105 & 0.429 & $ 5.2_{-0.8}^{+1.1}$ & $ 1024 \pm 165 $ & $(2.90_{-0.35}^{+0.41})\times10^{13}$ & $(1.56_{-0.31}^{+0.37})\times10^{14}$ & $0.062_{-0.034}^{+0.031}$ \\ 
XLSSC 106$^\star$ & 0.300 & $ 3.3_{-0.3}^{+0.4}$ & $ 856 \pm 125 $ & $(2.15_{-0.48}^{+0.53})\times10^{13}$ & $(7.25_{-1.72}^{+1.98})\times10^{13}$ & $0.090_{-0.046}^{+0.045}$ \\ 
XLSSC 107$^\star$ & 0.436 & $ 2.7_{-0.3}^{+0.4}$ & $ 711 \pm 111 $ & $(1.14_{-0.32}^{+0.38})\times10^{13}$ & $(3.15_{-0.95}^{+1.21})\times10^{13}$ & $0.072_{-0.041}^{+0.041}$ \\ 
XLSSC 108$^\star$ & 0.254 & $ 2.2_{-0.1}^{+0.3}$ & $ 705 \pm 101 $ & $(5.18_{-0.90}^{+1.04})\times10^{12}$ & $(1.20_{-0.23}^{+0.27})\times10^{13}$ & $0.041_{-0.020}^{+0.019}$ \\ 
XLSSC 109$^\star$ & 0.491 & $ 3.5_{-0.8}^{+1.3}$ & $ 787 \pm 151 $ & $(1.46_{-0.58}^{+0.67})\times10^{13}$ & $(5.30_{-2.39}^{+3.29})\times10^{13}$ & $0.064_{-0.054}^{+0.047}$ \\ 
XLSSC 110 & 0.445 & $ 1.6_{-0.1}^{+0.1}$ & $ 525 \pm 76 $ & $(8.70_{-2.60}^{+2.99})\times10^{12}$ & $(1.38_{-0.42}^{+0.49})\times10^{13}$ & $0.135_{-0.071}^{+0.075}$ \\ 
XLSSC 111 & 0.299 & $ 4.5_{-0.5}^{+0.6}$ & $ 1017 \pm 152 $ & $(2.44_{-0.54}^{+0.60})\times10^{13}$ & $(1.12_{-0.27}^{+0.31})\times10^{14}$ & $0.061_{-0.032}^{+0.031}$ \\ 
XLSSC 112 & 0.139 & $ 1.8_{-0.1}^{+0.2}$ & $ 653 \pm 95 $ & $(4.17_{-1.00}^{+1.40})\times10^{12}$ & $(7.63_{-1.95}^{+2.76})\times10^{12}$ & $0.046_{-0.025}^{+0.026}$ \\ 
XLSSC 113$^\star$ & 0.050 & $ 1.2_{-0.1}^{+0.0}$ & $ 558 \pm 78 $ & $(1.57_{-0.25}^{+0.29})\times10^{12}$ & $(1.91_{-0.31}^{+0.36})\times10^{12}$ & $0.030_{-0.014}^{+0.014}$ \\ 
XLSSC 114$^\star$ & 0.234 & $ 4.7_{-1.9}^{+4.2}$ & $ 1071 \pm 288 $ & $(1.78_{-1.04}^{+1.63})\times10^{13}$ & $(9.99_{-6.59}^{+13.35})\times10^{13}$ & $0.041_{-0.068}^{+0.050}$ \\ 
XLSSC 115$^\star$ & 0.043 & $ 2.0_{-0.2}^{+0.6}$ & $ 743 \pm 113 $ & $(5.36_{-1.63}^{+2.51})\times10^{11}$ & $(1.25_{-0.41}^{+0.68})\times10^{12}$ & $0.004_{-0.003}^{+0.003}$ \\ 
XLSSC 501 & 0.333 & $ 2.8_{-0.4}^{+0.6}$ & $ 768 \pm 121 $ & $(1.21_{-0.26}^{+0.29})\times10^{13}$ & $(3.58_{-0.91}^{+1.13})\times10^{13}$ & $0.068_{-0.040}^{+0.036}$ \\ 
XLSSC 502 & 0.141 & $ 1.2_{-0.1}^{+0.0}$ & $ 532 \pm 75 $ & $(3.03_{-0.75}^{+0.91})\times10^{12}$ & $(3.67_{-0.92}^{+1.11})\times10^{12}$ & $0.062_{-0.030}^{+0.032}$ \\ 
XLSSC 503 & 0.336 & $ 2.0_{-0.2}^{+0.3}$ & $ 642 \pm 97 $ & $(1.06_{-0.30}^{+0.34})\times10^{13}$ & $(2.21_{-0.65}^{+0.78})\times10^{13}$ & $0.101_{-0.056}^{+0.056}$ \\ 
XLSSC 504 & 0.243 & $ 13.9_{-5.4}^{+13.9}$ & $ 1953 \pm 503 $ & $(4.36_{-2.36}^{+17.19})\times10^{12}$ & $(8.67_{-5.38}^{+31.27})\times10^{13}$ & $0.002_{-0.003}^{+0.007}$ \\ 
XLSSC 505 & 0.055 & $ 1.7_{-0.1}^{+0.2}$ & $ 661 \pm 92 $ & $(1.50_{-0.27}^{+0.34})\times10^{12}$ & $(2.64_{-0.49}^{+0.64})\times10^{12}$ & $0.017_{-0.008}^{+0.008}$ \\ 
XLSSC 506 & 0.717 & $ 4.5_{-1.5}^{+2.1}$ & $ 798 \pm 188 $ & $(1.07_{-0.29}^{+0.41})\times10^{13}$ & $(5.06_{-2.09}^{+3.07})\times10^{13}$ & $0.035_{-0.032}^{+0.028}$ \\ 
XLSSC 507 & 0.566 & $ 2.4_{-0.5}^{+0.6}$ & $ 612 \pm 106 $ & $(1.39_{-0.34}^{+0.38})\times10^{13}$ & $(3.37_{-1.03}^{+1.30})\times10^{13}$ & $0.118_{-0.076}^{+0.069}$ \\ 
XLSSC 508 & 0.539 & $ 3.3_{-0.5}^{+0.7}$ & $ 742 \pm 123 $ & $(2.37_{-0.75}^{+0.86})\times10^{13}$ & $(7.91_{-2.74}^{+3.49})\times10^{13}$ & $0.117_{-0.075}^{+0.072}$ \\ 
XLSSC 509 & 0.633 & $ 4.2_{-0.8}^{+1.1}$ & $ 806 \pm 143 $ & $(1.07_{-0.21}^{+0.31})\times10^{13}$ & $(4.68_{-1.33}^{+1.89})\times10^{13}$ & $0.037_{-0.024}^{+0.022}$ \\ 
XLSSC 510 & 0.395 & $ 2.6_{-0.3}^{+0.4}$ & $ 711 \pm 106 $ & $(6.66_{-1.58}^{+2.26})\times10^{12}$ & $(1.82_{-0.48}^{+0.68})\times10^{13}$ & $0.044_{-0.024}^{+0.025}$ \\ 
XLSSC 511 & 0.130 & $ 1.3_{-0.1}^{+0.1}$ & $ 545 \pm 77 $ & $(3.25_{-0.69}^{+0.68})\times10^{12}$ & $(4.06_{-0.88}^{+0.89})\times10^{12}$ & $0.063_{-0.030}^{+0.030}$ \\ 
XLSSC 512 & 0.402 & $ 3.6_{-0.4}^{+0.6}$ & $ 848 \pm 128 $ & $(1.72_{-0.39}^{+0.41})\times10^{13}$ & $(6.39_{-1.59}^{+1.85})\times10^{13}$ & $0.066_{-0.036}^{+0.034}$ \\ 
XLSSC 513 & 0.378 & $ 4.2_{-0.5}^{+0.8}$ & $ 936 \pm 144 $ & $(3.85_{-0.90}^{+0.87})\times10^{13}$ & $(1.68_{-0.44}^{+0.50})\times10^{14}$ & $0.114_{-0.065}^{+0.059}$ \\ 
XLSSC 514 & 0.169 & $ 1.5_{-0.1}^{+0.2}$ & $ 582 \pm 86 $ & $(4.23_{-1.26}^{+1.53})\times10^{12}$ & $(6.31_{-1.94}^{+2.44})\times10^{12}$ & $0.065_{-0.035}^{+0.037}$ \\ 
XLSSC 515 & 0.101 & $ 1.2_{-0.1}^{+0.1}$ & $ 540 \pm 77 $ & $(2.05_{-0.37}^{+0.46})\times10^{12}$ & $(2.45_{-0.46}^{+0.58})\times10^{12}$ & $0.042_{-0.019}^{+0.020}$ \\ 
XLSSC 516 & 0.866 & $ 4.8_{-0.8}^{+1.0}$ & $ 695 \pm 114 $ & $(1.07_{-0.33}^{+0.56})\times10^{13}$ & $(5.31_{-1.76}^{+3.06})\times10^{13}$ & $0.037_{-0.023}^{+0.026}$ \\ 
XLSSC 517 & 0.699 & $ 3.5_{-0.6}^{+1.1}$ & $ 698 \pm 118 $ & $(1.46_{-0.37}^{+0.48})\times10^{13}$ & $(5.47_{-1.70}^{+2.53})\times10^{13}$ & $0.072_{-0.052}^{+0.044}$ \\ 
XLSSC 518 & 0.177 & $ 1.3_{-0.0}^{+0.0}$ & $ 535 \pm 75 $ & $(3.07_{-0.53}^{+0.56})\times10^{12}$ & $(3.90_{-0.68}^{+0.72})\times10^{12}$ & $0.060_{-0.027}^{+0.027}$ \\ 
XLSSC 519 & 0.270 & $ 1.5_{-0.2}^{+0.2}$ & $ 555 \pm 86 $ & $(3.27_{-0.62}^{+1.20})\times10^{12}$ & $(4.84_{-1.12}^{+1.81})\times10^{12}$ & $0.052_{-0.026}^{+0.031}$ \\ 
XLSSC 520 & 0.175 & $ 2.6_{-0.1}^{+0.2}$ & $ 805 \pm 113 $ & $(9.30_{-2.00}^{+2.32})\times10^{12}$ & $(2.49_{-0.55}^{+0.64})\times10^{13}$ & $0.053_{-0.026}^{+0.026}$ \\ 
XLSSC 521 & 0.807 & $ 4.7_{-0.8}^{+1.3}$ & $ 775 \pm 131 $ & $(3.53_{-0.71}^{+0.78})\times10^{13}$ & $(1.75_{-0.48}^{+0.63})\times10^{14}$ & $0.112_{-0.075}^{+0.062}$ \\ 
XLSSC 522 & 0.395 & $ 2.6_{-0.3}^{+0.4}$ & $ 711 \pm 108 $ & $(9.45_{-2.14}^{+2.23})\times10^{12}$ & $(2.49_{-0.62}^{+0.70})\times10^{13}$ & $0.062_{-0.033}^{+0.032}$ \\ 
XLSSC 523 & 0.343 & $ 2.9_{-0.4}^{+0.6}$ & $ 779 \pm 121 $ & $(1.41_{-0.25}^{+0.27})\times10^{13}$ & $(4.32_{-0.99}^{+1.18})\times10^{13}$ & $0.075_{-0.043}^{+0.038}$ \\ 
XLSSC 524 & 0.270 & $ 2.6_{-0.4}^{+0.5}$ & $ 754 \pm 121 $ & $(1.16_{-0.32}^{+0.38})\times10^{13}$ & $(3.07_{-0.94}^{+1.23})\times10^{13}$ & $0.074_{-0.044}^{+0.043}$ \\ 
XLSSC 525 & 0.379 & $ 3.4_{-0.2}^{+0.3}$ & $ 832 \pm 120 $ & $(2.23_{-0.53}^{+0.60})\times10^{13}$ & $(7.57_{-1.84}^{+2.15})\times10^{13}$ & $0.093_{-0.046}^{+0.047}$ \\ 
XLSSC 526 & 0.273 & $ 2.8_{-0.2}^{+0.4}$ & $ 794 \pm 115 $ & $(1.59_{-0.24}^{+0.27})\times10^{13}$ & $(4.65_{-0.81}^{+0.97})\times10^{13}$ & $0.086_{-0.042}^{+0.040}$ \\ 
XLSSC 527 & 0.076 & $ 3.1_{-1.0}^{+2.8}$ & $ 926 \pm 211 $ & $(7.70_{-4.56}^{+12.86})\times10^{11}$ & $(3.20_{-2.06}^{+6.23})\times10^{12}$ & $0.003_{-0.005}^{+0.006}$ \\ 
XLSSC 528 & 0.302 & $ 3.2_{-0.4}^{+0.8}$ & $ 839 \pm 132 $ & $(1.32_{-0.32}^{+0.37})\times10^{13}$ & $(4.52_{-1.27}^{+1.61})\times10^{13}$ & $0.059_{-0.037}^{+0.032}$ \\ 
XLSSC 529 & 0.547 & $ 3.5_{-0.4}^{+0.7}$ & $ 769 \pm 119 $ & $(1.79_{-0.29}^{+0.33})\times10^{13}$ & $(6.55_{-1.37}^{+1.68})\times10^{13}$ & $0.079_{-0.043}^{+0.039}$ \\ 
XLSSC 530 & 0.182 & $ 2.0_{-0.2}^{+0.2}$ & $ 686 \pm 100 $ & $(5.24_{-1.02}^{+1.07})\times10^{12}$ & $(1.07_{-0.23}^{+0.25})\times10^{13}$ & $0.048_{-0.024}^{+0.023}$ \\ 
XLSSC 531 & 0.391 & $ 4.5_{-1.4}^{+2.2}$ & $ 966 \pm 214 $ & $(8.52_{-4.84}^{+9.90})\times10^{12}$ & $(4.05_{-2.39}^{+5.77})\times10^{13}$ & $0.023_{-0.024}^{+0.030}$ \\ 
XLSSC 532 & 0.392 & $ 3.0_{-0.5}^{+0.6}$ & $ 772 \pm 126 $ & $(9.49_{-1.59}^{+1.99})\times10^{12}$ & $(2.92_{-0.66}^{+0.85})\times10^{13}$ & $0.049_{-0.027}^{+0.026}$ \\ 
XLSSC 533 & 0.107 & $ 2.4_{-0.1}^{+0.1}$ & $ 789 \pm 111 $ & $(1.33_{-0.29}^{+0.31})\times10^{13}$ & $(3.20_{-0.71}^{+0.77})\times10^{13}$ & $0.086_{-0.041}^{+0.042}$ \\ 
XLSSC 534 & 0.853 & $ 4.3_{-1.0}^{+1.7}$ & $ 725 \pm 138 $ & $(2.39_{-0.82}^{+1.04})\times10^{13}$ & $(1.11_{-0.45}^{+0.69})\times10^{14}$ & $0.088_{-0.074}^{+0.063}$ \\ 
XLSSC 535 & 0.172 & $ 2.4_{-0.2}^{+0.3}$ & $ 756 \pm 112 $ & $(1.08_{-0.27}^{+0.31})\times10^{13}$ & $(2.57_{-0.67}^{+0.81})\times10^{13}$ & $0.075_{-0.039}^{+0.040}$ \\ 
XLSSC 536 & 0.170 & $ 1.8_{-0.2}^{+0.3}$ & $ 659 \pm 97 $ & $(4.70_{-1.53}^{+1.89})\times10^{12}$ & $(9.12_{-3.08}^{+3.97})\times10^{12}$ & $0.049_{-0.030}^{+0.030}$ \\ 
XLSSC 537 & 0.515 & $ 4.8_{-0.9}^{+1.1}$ & $ 934 \pm 159 $ & $(2.04_{-0.55}^{+1.09})\times10^{13}$ & $(1.02_{-0.33}^{+0.63})\times10^{14}$ & $0.052_{-0.033}^{+0.038}$ \\ 
XLSSC 538 & 0.332 & $ 3.1_{-0.6}^{+0.9}$ & $ 804 \pm 140 $ & $(6.69_{-1.85}^{+2.23})\times10^{12}$ & $(2.15_{-0.72}^{+0.98})\times10^{13}$ & $0.033_{-0.023}^{+0.020}$ \\ 
XLSSC 539 & 0.184 & $ 1.2_{-0.2}^{+0.1}$ & $ 520 \pm 89 $ & $(1.87_{-0.51}^{+0.69})\times10^{12}$ & $(2.03_{-0.61}^{+0.83})\times10^{12}$ & $0.040_{-0.020}^{+0.025}$ \\ 
XLSSC 540 & 0.414 & $ 3.1_{-0.3}^{+0.4}$ & $ 776 \pm 118 $ & $(1.07_{-0.21}^{+0.24})\times10^{13}$ & $(3.36_{-0.75}^{+0.89})\times10^{13}$ & $0.053_{-0.027}^{+0.027}$ \\ 
XLSSC 541 & 0.187 & $ 2.7_{-0.3}^{+0.3}$ & $ 805 \pm 121 $ & $(8.81_{-1.46}^{+1.53})\times10^{12}$ & $(2.34_{-0.45}^{+0.50})\times10^{13}$ & $0.050_{-0.024}^{+0.024}$ \\ 
XLSSC 542 & 0.402 & $ 6.8_{-0.3}^{+0.5}$ & $ 1202 \pm 170 $ & $(8.43_{-1.46}^{+1.49})\times10^{13}$ & $(5.78_{-1.04}^{+1.10})\times10^{14}$ & $0.114_{-0.053}^{+0.052}$ \\ 
XLSSC 543 & 0.381 & $ 2.4_{-0.3}^{+0.5}$ & $ 689 \pm 109 $ & $(1.12_{-0.32}^{+0.36})\times10^{13}$ & $(2.82_{-0.88}^{+1.10})\times10^{13}$ & $0.082_{-0.051}^{+0.047}$ \\ 
XLSSC 544 & 0.095 & $ 2.4_{-0.2}^{+0.2}$ & $ 788 \pm 114 $ & $(6.17_{-1.08}^{+1.07})\times10^{12}$ & $(1.47_{-0.28}^{+0.29})\times10^{13}$ & $0.041_{-0.019}^{+0.019}$ \\ 
XLSSC 545 & 0.353 & $ 2.2_{-0.6}^{+1.6}$ & $ 668 \pm 138 $ & $(5.79_{-4.14}^{+20.24})\times10^{12}$ & $(1.60_{-1.18}^{+5.89})\times10^{13}$ & $0.048_{-0.071}^{+0.171}$ \\ 
XLSSC 546 & 0.792 & $ 3.5_{-0.6}^{+0.7}$ & $ 668 \pm 110 $ & $(1.10_{-0.20}^{+0.25})\times10^{13}$ & $(3.95_{-0.95}^{+1.20})\times10^{13}$ & $0.056_{-0.032}^{+0.030}$ \\ 
XLSSC 547 & 0.371 & $ 4.0_{-0.8}^{+1.1}$ & $ 920 \pm 165 $ & $(9.50_{-1.58}^{+2.15})\times10^{12}$ & $(4.00_{-1.05}^{+1.38})\times10^{13}$ & $0.030_{-0.019}^{+0.017}$ \\ 
XLSSC 548 & 0.321 & $ 1.0_{-0.1}^{+0.1}$ & $ 427 \pm 62 $ & $(3.69_{-1.11}^{+1.38})\times10^{12}$ & $(3.59_{-1.10}^{+1.41})\times10^{12}$ & $0.121_{-0.064}^{+0.069}$ \\ 
XLSSC 549 & 0.808 & $ 4.0_{-0.9}^{+2.4}$ & $ 709 \pm 135 $ & $(1.14_{-0.56}^{+0.68})\times10^{13}$ & $(5.40_{-2.88}^{+4.70})\times10^{13}$ & $0.048_{-0.056}^{+0.039}$ \\ 
XLSSC 550 & 0.109 & $ 1.0_{-0.1}^{+0.1}$ & $ 475 \pm 71 $ & $(2.52_{-0.66}^{+0.74})\times10^{12}$ & $(2.36_{-0.64}^{+0.73})\times10^{12}$ & $0.075_{-0.038}^{+0.040}$ \\ 
XLSSC 551 & 0.475 & $ 2.5_{-0.5}^{+0.7}$ & $ 667 \pm 117 $ & $(8.72_{-2.09}^{+5.05})\times10^{12}$ & $(2.36_{-0.74}^{+1.45})\times10^{13}$ & $0.064_{-0.043}^{+0.050}$ \\ 

\end{longtable}
\noindent \textbf{Column description:} 1. Cluster name. The clusters marked with an asterisk were used for the stellar fraction analysis (see Sect. \ref{sec:hod}); 2. Cluster redshift (Paper II); 3. X-ray temperature within 300 kpc aperture (Paper III); 4. $r_{\rm 500,MT}$ estimated using the XXL $M_{\rm WL}-T$ relation (Paper IV); 5. Gas mass within $r_{\rm 500,MT}$ (this work); 6. Integrated Compton parameter $Y_{X,500}$ within $r_{\rm 500,MT}$ (this work); 7. Hot gas fraction within $r_{\rm 500,MT}$ (this work).
\end{longtab}
}
\twocolumn

\section{Multiscale forward-fitting deprojection}
\label{app:deproj}

If the geometry of the emitting region is assumed to be known, the intrinsic emissivity profiles can in principle be recovered from the observed projected emission-measure profiles by inverting the projection kernel \citep[e.g.][]{kriss}. In practice, the problem is rendered complicated by the presence of noise in the original data. As for all inverse problems, the projection kernel smoothes small-scale fluctuations, thus the inverse transformation has the opposite effect and the noise can be greatly amplified \citep[see][]{lucy74,lucy94}. This effect is particularly important in the low signal-to-noise regime. 

Two main approaches usually exist to solve this problem: direct geometrical deprojection \citep[e.g.][]{fabian81,kriss,morandi07} and forward fitting using a parametric form \citep[e.g.][]{cavaliere,sarazin77,pointe478,mahdavi07}. The advantage of the former is that it makes no assumption on the shape of the intrinsic profile, thus it is in principle the most general. However, this method is very sensitive to measurement uncertainties, since small variations in the projected profile can be greatly magnified; therefore, the resulting profile is generally not smooth. Moreover, error bars obtained through this technique are not statistically meaningful, given that the result depends on the adopted binning. Conversely, the latter can provide rigorously determined error bars, but requires strong assumptions on the shape of the density profile since the chosen functional form (usually an extended version of a beta model) must reproduce the observed shape accurately. For a review of the various existing deprojection methods, we refer to \citet{buote12b}.

In this work, we propose a different approach to combine the rigorous results provided by forward fitting with minimal priors on the shape of the density profile. We suggest  decomposing the density profile into a sum of analytical multiscale functions which can be independently deprojected since the projection kernel is linear. Namely, the projected emission-measure profiles were fit with a sum of $N$ King functions,

\begin{equation}EM(s)=\sum_{i=1}^N A_i\left(1+\left(\frac{s}{r_{c,i}}\right)^2\right)^{-3\beta_i+0.5} ,\label{eq:multiscale}\end{equation}

\noindent where the core radii $r_{c,i}$ are fixed adaptively to reproduce the shape of the profile on a given scale and $s$ is the projected cluster-centric distance. The normalisations and slopes are left free to vary while fitting. As is known from the beta model \citep{cavaliere}, these functional forms have the good property that the corresponding deprojected function is analytical. Given that the projection kernel is linear, this property is preserved for any linear combination of King functions, i.e.

\begin{equation}n_{\rm gas}^2(r)=\sum_{i=1}^N n_{0,i}^2 \left(1+\left(\frac{r}{r_{c,i}}\right)^2\right)^{-3\beta_i}. \end{equation}

\noindent Here, $n_{0,i}^2$ is proportional to $A_i$. Thus, the deprojected gas density can be easily obtained from the fitted parameters. In this particular case, the normalisations of the deprojected profile $n_{0,i}^2$ can be evaluated directly from the projected normalisations \citep[see e.g. Appendix A of][]{hudson10},

\begin{equation} n_{0,i}^2 = \frac{\Gamma(3\beta)}{\Gamma(3\beta-0.5)\sqrt{\pi}r_{c,i}}A_i,\label{eq:multi_norm}\end{equation}

\noindent where $r_{c,i}$ is expressed in physical units. 

Naturally, this method can be generalised to any type of base functions, given the problem of interest. In the particular case of galaxy cluster emissivity profiles, the choice of King functions is easily justified.

\begin{figure}
\resizebox{\hsize}{!}{\includegraphics{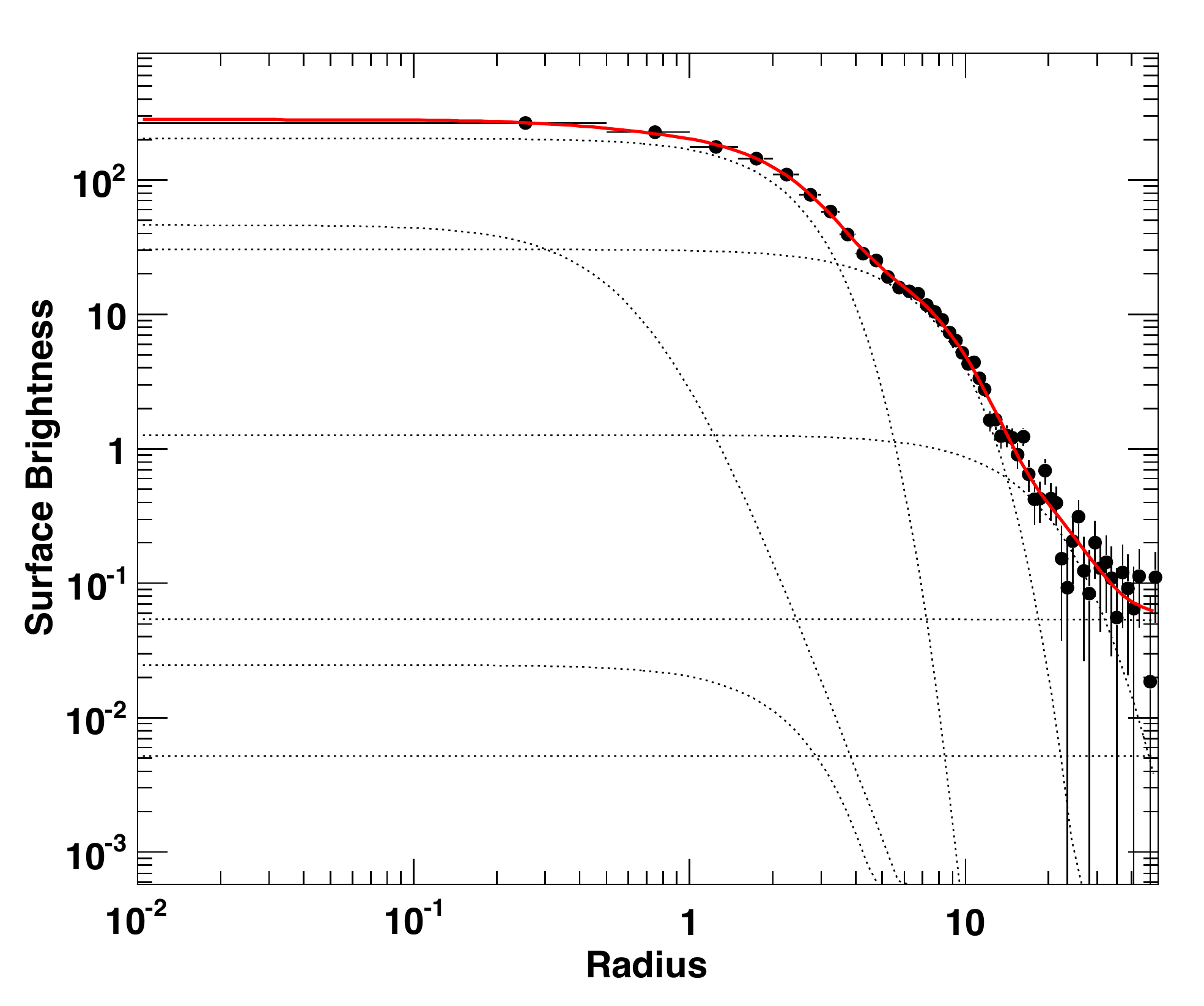}}
\caption{Example of application of the multiscale forward-fitting technique to simulated data (Eq. \ref{eq:multiscale}). The red curve shows the total model, while the dashed lines represent the contribution of each individual component. The units on both axes are arbitrary.}
\label{fig:multiscale}
\end{figure}

In Fig. \ref{fig:multiscale} we show a test of this method on simulated data. Specifically, we simulated an \emph{XMM-Newton} pointing with an extended source with a random radius-dependent slope and extracted the surface-brightness profile from the image. The simulated surface-brightness profile was then fit with Eq. \ref{eq:multiscale} (red curve). The dashed curves indicate the contribution of each component to the model.

We applied the technique described here by including one King component for each set of four data points. The total model thus has $N/2$ parameters, where $N$ is the number of data points in the emission-measure profile. The core radius of each component was fixed to the median radius of the corresponding block of data points. This choice allows a very large freedom to the fitted function and is sufficient to fit adequately any surface-brightness profile, provided that the following priors are verified: \emph{i)} the gas density is smooth and decreases monotonically with radius and \emph{ii)} the gas density is always $\geq0$. The former condition should be valid if the point sources and neighbouring clusters are properly excised, while the latter is always valid as long as the subtraction of the background is correct. Therefore, this approach combines the rigorous approach of forward fitting with weak priors on the cluster density profile.

\section{Modelling the \emph{XMM-Newton} PSF}
\label{app:psf}

Since high-redshift clusters are only slightly extended for \emph{XMM-Newton}, an accurate modelling of the instrument's point-spread function (PSF) is required to give a realistic estimate of the gas density profile, and hence gas mass. The observed image is given by the 2D convolution of the original image with the instrumental PSF. Performing such a convolution directly is time-consuming since it requires computing a triple integral on the fly. Thus, fitting a surface-brightness model accounting for PSF convolution is lengthy and numerically difficult.

To alleviate this problem, we take advantage of the finite binning of the surface-brightness profile to turn the continuous problem into a discrete one. For a given radial binning $\{r_i\}_{i=1}^N$ we assume that in each bin the surface brightness is approximately constant, which is a reasonable approximation for a bin width of 8 arcsec. Then we define a convolution matrix $P$ such that

\begin{equation} P_{i,j}=\mbox{Prob}(j\rightarrow i), \end{equation}

\noindent i.e. $P_{i,j}$ is the probability that a photon originating from bin $j$ is detected in bin $i$. The matrix is normalised such that $\sum_{i=1}^N P_{i,j}=1, \forall j$. Then from the model $EM(r_i)$ (see Eq. \ref{eq:multiscale}), the convolved model $\tilde {EM}(r_i)$ can be written as

\begin{equation} \tilde{EM}_i = \sum_{j=1}^N P_{i,j} EM(r_j) .\end{equation}

\noindent The convolved model $\tilde{EM}$ is then fit to the observed surface-brightness profile.

\begin{figure}
\centerline{\resizebox{0.8\hsize}{!}{\includegraphics{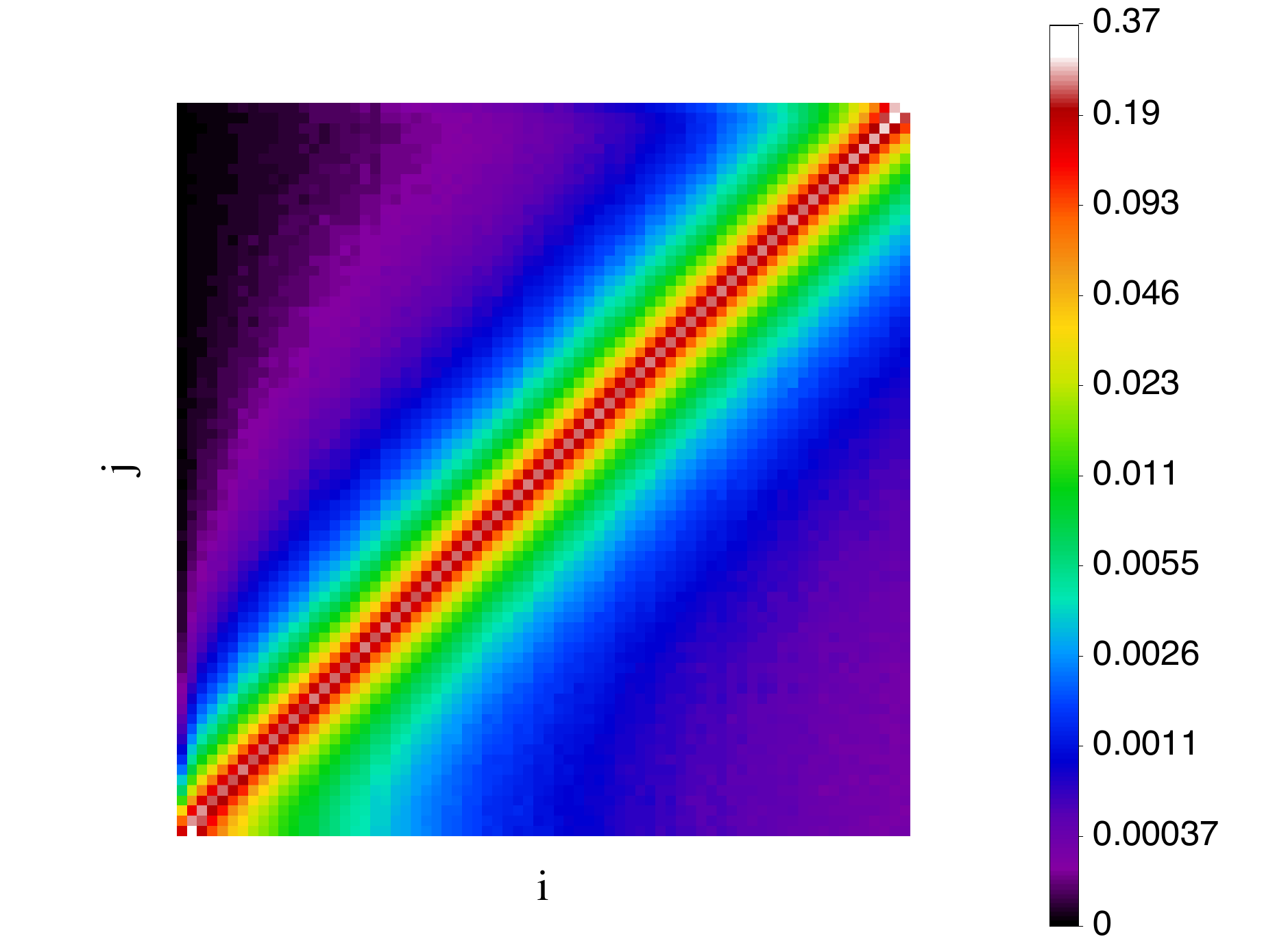}}}
\caption{PSF convolution matrix $P$ computed from ray tracing (see text). The value $P_{i,j}$ gives the probability that a photon originating from bin $j$ is detected in bin $i$.}
\label{psf}
\end{figure}

To compute the convolution matrix $P$, we adopt a ray tracing approach. Namely, for each radial bin, we simulate $10^6$ ray-tracing photons, taking vignetting effects and CCD gaps into accounts, and randomise the position of each simulated photon according to the properties of the \emph{XMM-Newton} PSF. For each off-axis angle, we model the PSF as a King profile with the parameters given in the latest \emph{XMM-Newton} calibration files. We then compute the fraction of photons falling into each radial bin. An example of convolution matrix is shown in Fig. \ref{psf}.

To validate this method, we applied it to the NW cold front of A2142 \citep[see][]{rossetti13} and compared our results with the values obtained from higher resolution \emph{Chandra} data \citep{owers09}. Without PSF convolution, the broken power-law model gives a poor fit to the data ($\chi^2=88.5$ for 55 degrees of freedom (d.o.f.), $P_{null}=0.0028$), and converges to a density jump $n_{in}/n_{out}=1.69\pm0.03$, which is inconsistent with the \emph{Chandra} value ($2.0\pm0.1$). Modelling the PSF using the method described above, we obtain a very good fit ($\chi^2=50.1/55$ d.o.f., $P_{null}=0.66$) and a density jump $n_{in}/n_{out}=2.0\pm0.04$, in agreement with the \emph{Chandra} measurement. This demonstrates the ability of our method to take the \emph{XMM-Newton} PSF into account.

\section{Comparing the $M-T$ relation with the weak lensing data}
\label{app:r500}

\begin{figure}
\centerline{\resizebox{\hsize}{!}{\includegraphics{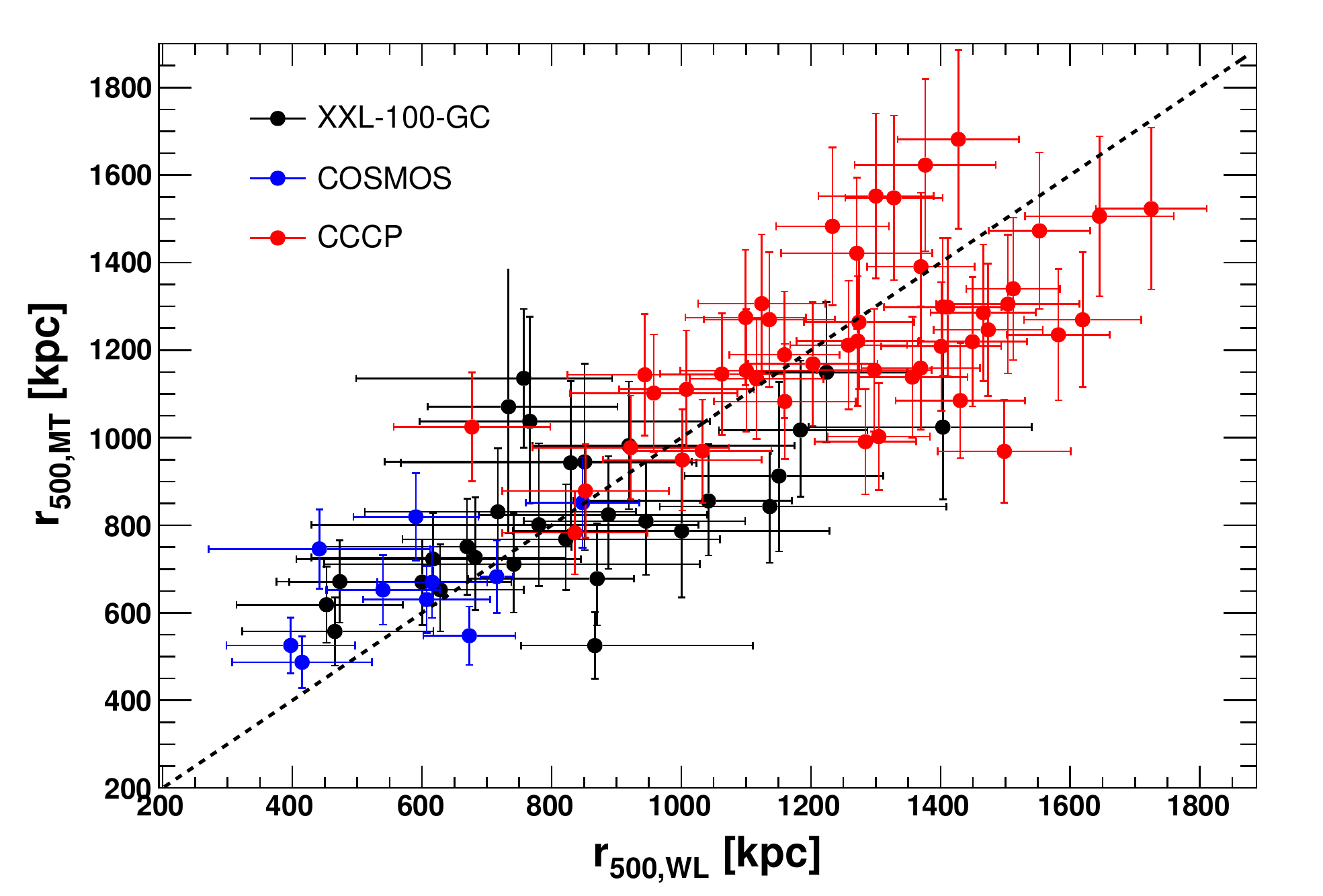}}}
\caption{Comparison between the values of $r_{500}$ obtained by using the $M-T$ relation of Paper IV and the values measured through directly using weak lensing for the XXL-100-GC (black), CCCP (red), and COSMOS (blue) samples. The dashed black line represents unity.}
\label{fig:verif}
\end{figure}

Since weak-lensing masses are available only for a fraction of XXL-100-GC systems, our analysis rests on the assumption that on average the masses and values of $r_{\rm 500,MT}$ for our systems are  consistent with the values obtained through weak lensing. To test this assumption, we used the weak-lensing data from XXL-100-GC, CCCP, and COSMOS used in Paper IV and applied our $M-T$ relation to the same systems. In the XXL-100-GC case, only the clusters for which a significant mass measurement could be obtained are considered here. In Fig. \ref{fig:verif} we show the comparison between the values of $r_{500}$ obtained from the $M-T$ relation and the values measured directly through weak lensing. The two sets of values are consistent, leading to  a mean ratio $r_{\rm 500,MT}/r_{\rm 500,WL}=1.022\pm0.021$. Therefore, there is no significant difference between the values of $r_{500,MT}$ and the ones expected from weak lensing.

\section{Gas fraction of the XXL sample using hydrostatic $M-T$ relations}
\label{app:hydro}

\begin{figure}
\centerline{\resizebox{\hsize}{!}{\includegraphics{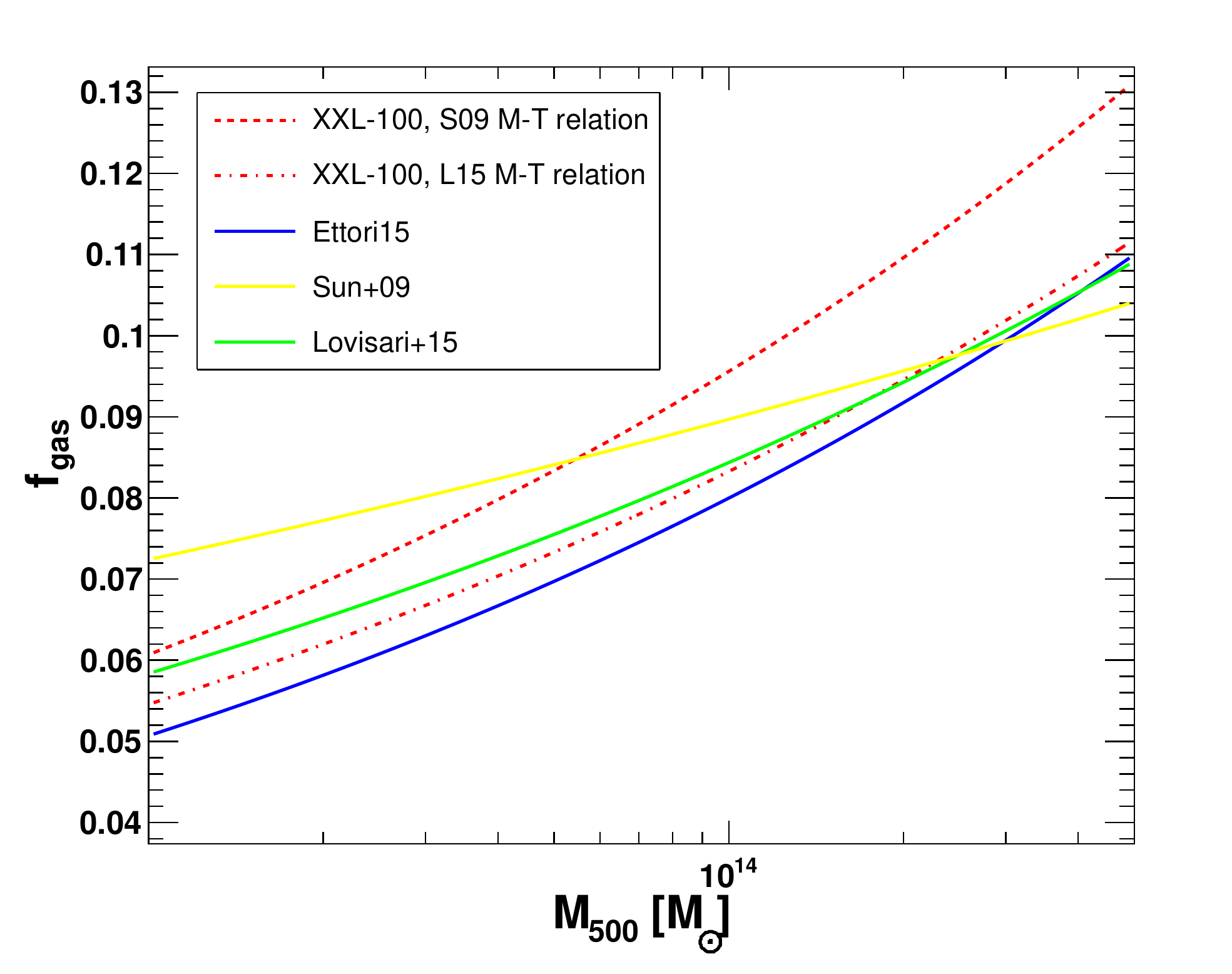}}}
\caption{Gas fraction curves for the XXL-100-GC sample estimated assuming the hydrostatic $M-T$ relations of \citet[dotted red curve]{sun09} and \citet[dash-dotted red cuve]{lovisari15}, compared to literature measurements based on hydrostatic masses.}
\label{fig:hydro}
\end{figure}

As a further check of the quality of the X-ray analysis for the XXL-100-GC sample, we   calculated the cluster mass and $r_{\rm 500,MT}$ using the hydrostatic-based mass-temperature relations of \citet[Tier1+2+clusters]{sun09} and \citet{lovisari15}. In both cases, we recomputed the gas mass within the corresponding aperture and calculated the gas fraction by fitting again the $M_{\rm gas}-T$ relation using the gas masses measured within the modified aperture and combining it with the adopted $M-T$ relation (see Sect. \ref{sec:fgas_mtot}). In Fig. \ref{fig:hydro} we show the resulting curves in the $f_{\rm gas}-M$ plane compared to the reference hydrostatic measurements. The results obtained for the XXL-100-GC sample agree well with the literature measurements. Above $10^{14}M_\odot$, the gas fraction of XXL-100-GC clusters estimated using the \citet{sun09} relation slightly exceeds that measured by \citet{sun09}. Differences of this order are  to be expected at the high-mass end, however, given that our sample only contains a small number of systems beyond $\sim3\times10^{14}M_\odot$.

\end{document}